\documentstyle[a4wide,12pt]{article}
\begin{document}

\def\d{\partial }
\def\f{\frac }
\def\be{\begin{eqnarray*}}
\def\ee{\end{eqnarray*}}
\def\u{\underline}

\font\qq=cmr10 at 6pt
\def\permille{$\,\,\!$\raisebox{1.2mm}{\qq 0}$\!$/$\!_{\raisebox{0mm}
{\qq 0}}\,\!_{\raisebox{0mm}{\qq 0}}$}

\def\thebibliography#1{\section*{References\markboth
 {REFERENCES}{REFERENCES}}\list
 {}{\setlength\labelwidth{1.4em}\leftmargin\labelwidth
 \setlength\parsep{0pt}\setlength\itemsep{0pt}
 \setlength{\itemindent}{-\leftmargin}
 \usecounter{enumi}}
 \def\newblock{\hskip .11em plus .33em minus -.07em}
 \sloppy
 \sfcode`\.=1000\relax}
\let\endthebibliography=\endlist


\def\@cite#1#2{{#1\if@tempswa , #2\fi}}
\def\@biblabel#1{}


\def\ssim{\setbox0=\hbox{$\sim$}%
\setbox1=\hbox{$<$}\dimen0=\ht1%
\advance\dimen0by-1.2pt\,\lower.6\dimen0%
\copy0\kern-\wd0\raise.4\dimen0\copy1 \,}


\def\gsim{\setbox0=\hbox{$\sim$}%
\setbox1=\hbox{$>$}\dimen0=\ht1%
\advance\dimen0by-1.2pt\,\lower.6\dimen0%
\copy0\kern-\wd0\raise.4\dimen0\copy1\,}


\def\lambdab{\lambda\mkern-9mu\lower1.2pt\hbox{$\mathchar'26$}}%

\def\apj{{\it ApJ}, }             
\def\aj{{\it AJ}, }             
\def\apjs{{\it ApJS}, }     
\def\apjl{{\it Astrophys.~J.~(Letters)}, }  
\def\pasp{{\it PASP}, }       
\def\mnras{{\it MNRAS}, }       
\def\mn{{\it MNRAS}, }
\def\aa{{\it A\&A}, }               
\def\aasup{{\it A\&AS}, }     

\def\aar{{\it Astron.~Astrophys.~Review}, }   
\def\baas{{\it Bull.~Am.~Astron.~Soc.}, }          
\def\assl{{\it Astrophys.~Sp.~Sc.~library.}, }      
\def\aspcs{{\it PASPC}, }      

\def\ub#1{\ifmmode\mbox{\boldmath{$#1$}}\else{\boldmath{$#1$}}\fi}
\def\vect#1{\ifmmode\mbox{\boldmath{$#1$}}\else{\boldmath{$#1$}}\fi}
\def\hi{\noindent \hangindent=2.5em}
\def\ls{\vskip 12.045pt}   
\def\ni{\noindent}        
\def\et{{\it et\thinspace al.}\ }    
\def\kms{km\thinspace s$^{-1}$ }     
\def\amm{\AA\thinspace mm$^{-1}$ }     
\def\solar{\ifmmode_{\mathord\odot}\else$_{\mathord\odot}$\fi} 
\def\Mo{\ifmmode M_{\odot} ~ \else $M_{\odot}$~ \fi}


{\noindent \LARGE Evidences of secular dynamical evolution }

\vspace{.5cm}

{\noindent \LARGE in disk galaxies}

\vspace{1.5cm}

{\noindent \large Louis Martinet}

\vspace{.5cm}

{\noindent\it Geneva Observatory, 51 chemin des Maillettes, CH-1290
Sauverny, Switzerland}

\vspace{1.5cm}

{\noindent \bf Abstract}

\vspace{0.5cm}

After a recall of fundamental concepts used in galactic dynamics, we
review observational facts as well as results of orbit theory and
numerical simulations which suggest long-term evolution of galaxies.
Dynamical interactions between galaxy constituents (bulges, disks,
bars, haloes, dark matter) are discussed and effects of external
perturbations on internal structures are examined. We report recent
developments on the connection between dynamical interaction processes
and efficiency of star formation in galaxies of various morphological
types. The Hubble sequence as an evolution sequence is revisited.

\vspace{0.5cm}

\newpage
\tableofcontents
\newpage

\section{Introduction}          

The purpose of the present review is not to extensively discuss all
the recent developments on galaxy formation and evolution. As it can
be seen below, the literature on the subject is abundant if not
redundant.  Our attention will be centered on clues of secular
dynamical evolution in disk galaxies which can be detected by
observations or suggested by numerical simulations or theoretical
approaches. We will emphasize some fundamental physical processes at
work in this context as resonances, torques, energy dissipation,
dynamical friction, loss and gain of angular momentum, accretion, etc.

The most significant progresses in the field come from considerations
on facts still neglected or not well understood in the past as the
role of gas, the interaction effects, not only between galaxies but
also between various components of a given system, the necessity to
take account of 3-D structures, the star formation in competition with
dynamical processes.

Recent observations as well as numerical experiments or theoretical
developments suggest that disk galaxies are the seat of evolutionary
processes on time scales of the order of the Hubble time or smaller.
We will not deal with galactic formation, since another review in the
same journal will be devoted to this phase of evolution.

We assume that the reader has had an introduction course in galactic
dynamics and is familiar with classical principles of stellar dynamics
as well as hydrodynamics. For those readers who lack this basic
knowledge, we warmly recommend ``Galactic Dynamics'' by Binney and
Tremaine (1987) and a general review on hydrodynamics by Monaghan
(1992).

We begin (section 2) by recalling the observed global properties of
galaxies as well as correlations between typical features along the well
known Hubble sequence.

In section 3, we review the fundamental tools of the galactic dynamics
and the different approaches such as orbit theory, analytical
developments and numerical simulations with their qualities and
failures. The use of action-angle variables will be reintroduced in
view of applications in other sections.

In section 4, we indicate more specifically what we can learn from
stellar orbit theory on galactic evolution, emphasizing concepts newly
introduced in the field, in particular in three dimensional problems,
such as complex instability or effects of a compact central mass.
Chaotic orbital behavior will be mentioned as far as it can play a
role in the secular evolution of the systems.

The problem of local and global instabilities of disks taking account
of stars and gas will be discussed in section 5. In this context
arguments for suggesting cold gas as a main constituent of the dark
matter present in galaxies will be given.

In section 6, we explore the interactions between various components
of disk galaxies (disks, spheroids, bars) and their consequences on
the long term evolution.

The dissipative evolution of disks due to the presence of gas is be
discussed in section 7 on the basis of recent simulations using an
N-body code coupled either with sticky particle code or with smooth
particle hydrodynamic code for gas behavior description. Spiral
activity and gas fueling of galactic nuclei driven by bars will be
studied.

Galaxies evolve in various environments. Interactions are able to
modify their morphologies. In section 8, the consequences of satellite
accretions are examined as well as the result of strong encounters of
disks. A correct treatment of dynamical friction in numerical
simulations is discussed.

In section 9, we examine how the star formation can contribute to
modify the ideas on the long-term evolution of galaxies.  This part is
still exploratory since a rigorous theory of star formation taking
account of the implied complex physical processes is not yet available
and one is constrained to introduce ad hoc assumptions for this
purpose.

By way of conclusion, we give in section 10 some arguments in favor of a
possible evolution of galaxies along the Hubble sequence from late to
early types, due to the dynamical processes previously described.

Since our purpose is essentially to emphasize the clues of
evolutionary dynamical processes during the life of disk galaxies, it
is useful to close this introduction by referring the reader to books
which deal with the structure and evolution of galaxies in general or
which describe in more details some specific subjects which we will
not be able to tackle here.

Amongst very numerous books and reviews, we propose the following (not
exhaustive) list.

\begin{enumerate}

\item In Annual Review of Astronomy and Astrophysics:

\begin{itemize}

\item[-] E. Athanassoula and A. Bosma (1985) ``Shells and rings around
galaxies'', {\bf 23}, 147

\item[-] J. Sellwood (1987)  ``The art of $N$-body building'', {\bf 25}, 151

\item[-] C. M. Telesco (1988)  ``Enhanced star formation and IR emission in
the centers of galaxies'', {\bf 26}, 343

\item[-] J. J. Binney (1992)  ``Warps'', {\bf 30}, 51

\item[-] J. E. Barnes and L. E. Hernquist (1992)  ``Dynamics of interacting
galaxies'', {\bf 30}, 705

\item[-] S. R. Majewski (1993)  ``Galactic structure surveys and the
evolution of the Milky Way'', {\bf 31}, 575

\item[-] M. S. Roberts and M. P. Haynes (1994)
	 ``Physical parameters along the Hubble sequence''
	 {\bf 32}, 115
\end{itemize}

\item In the series of International Astronomical Union Symposia
(Reidel Dordrecht, eds.) various reviews can be found. Let us quote in:

\begin{itemize}
\item[-] N$^\circ$ 144  ``The interstellar disk-halo connection in
galaxies'' H. Bloemen ed.  (1990)

\item[-] N$^\circ$ 146  ``Dynamics and their molecular cloud distribution''
F. Combes and  F. Casoli eds.  (1991)

\item[-] N$^\circ$ 149  ``The stellar populations of galaxies'' B. Barbuy and
A. Renzini eds.  (1992)

\item[-] N$^\circ$ 153  ``Galactic bulges'' H. Dejonghe and H. J. Habing''
eds. (1993)

\end{itemize}

\item Amongst the most recent books containing proceedings of other
conferences, we mention:

\begin{itemize}

\item[-]  ``Windows in galaxies'' (1990) \assl Vol. 160, G. Fabbiano, J. S.
Gallagher, A. Renzini eds. Kluwer, Dordrecht

\item[-]  ``Evolution of the Universe of galaxies'' (1990) \aspcs Vol. 10,
R. G. Kron eds.

\item[-]  ``Galactic models'' (1990), Annals of the New York Academy of
Sciences
 Vol. 596 J.R. Buchler ed.

\item[-]``Baryonic dark matter'' (1990),{\it NATO ASI Series C} no 306, D.
Lynden-Bell, G. Gilmore eds.  Kluwer, Dordrecht

\item[-]  ``Chemical and dynamical evolution of galaxies'' (1990), F.
Ferrini, J. Franco, F. Matteucci  eds. ETS Editrice Pisa

\item[-]  ``The interstellar medium in galaxies'' (1990), \assl Vol. 161,
H. A. Thronson, J. M. Shull  eds.  Kluwer, Dordrecht

\item[-] ``Dynamics and interactions of galaxies'' (1990), R. Wielen (ed.)
Springer, Berlin

\item[-]  `Evolution of interstellar matter and dynamics of galaxies''
(1990), J. Palous, W. B. Burton, Lindblad P. O. eds. Cambridge Univ.
Press, Cambridge

\item[-]  ``Dynamics of disk galaxies'' (1991), {\it Proceedings of Varberg
Conference}, B. Sundelius eds. G\"{o}teborg, Sweden

\item[-]  ``Physics of nearby galaxies, nature or nurture ?'' (1992), T. X.
Thuan, C. Balkowski, D. T. T. Van eds. Editions Fronti\`eres, Gifs s/ Yvette

\item[-]  ``Morphological and physical classification of galaxies'' (1992),
 OAC Fifth International Workshop, G. Busarello, M. Capaccioli, G. Longo
eds.   Kluwer, Dordrecht

\item[-]  ``Star formation in stellar systems'' (1992), G. Tenorio-Tagle,
M. Prieto and F. Sanchez  eds. Cambridge Univ. Press, Cambridge

\end{itemize}

\item In the series of Saas-Fee Courses of the Swiss Society of
Astrophysics and Astronomy:

\begin{itemize}

\item[-]  ``The galactic interstellar medium'' (1991),  D.
Pfenniger and P. Bartholdi  eds. Springer, Berlin

\end{itemize}

\item Specific monographies are also recommended

\begin{itemize}

\item[-]  ``Dynamics of barred galaxies'' (1993), J. Sellwood and A.
Wilkinson, in {\it Rep. Prog. Phys.}, {\bf 56}, 173

\item[-]  ``Orbits in barred galaxies'' (1989), G. Contopoulos and P.
Grosb\o l, \aar  {\bf 1}, 261

\item[-] ``Galaxy formation'' (1993), J. Silk and R. Wyse, in {\it Physics
Reports}, {\bf 231}, 293

\end{itemize}

\end{enumerate}

\newpage

\section{Global observational properties of spiral galaxies}

Whatever the origin of galaxy morphologies may be (initial conditions
or (and) result of secular evolution), the structural properties of
different Hubble types correlated with physical features have to be
explained. It is useful, in our present context, to draw up a list of
these properties. One must pay attention to the fact that the basic
morphological elements of the Hubble sequence of spirals have been set
forth by Sandage for bright galaxies.  Moreover the morphological
classification highly depends on the used passbands.  Recent J, H, K
photometry of galaxies confirms this evidence.

\begin{enumerate}

\item Systematics of the galactic bulge/disk ratios, carefully studied
by Simien and de Vaucouleurs (1986) confirms previous results according
to which the mean fractional luminosity of the bulge is a monotonic
decreasing function of the Hubble type from early (lenticular) types to
late-types, typically from 4 for Sa galaxies to 0.1 for Sc-d galaxies.

\item Kennicutt (1981) gave measurements of shapes and pitch angles
$\bar{\imath} $ of spiral arms in Sa-Sc galaxies. A clear dependence of
$\bar{\imath} $  on the absolute blue luminosity and V$_{rot}$ appears.
The result is in agreement with the increase of theoretical pitch
angles from early to late-type spirals inferred from the density wave
theory by Roberts et al. (1975).

\item Strong bars (strong in the sense of an important deviation to
axisymmetry) are present in $\sim$1/3 of spirals and weak bars (or ovals)
represent another third.

\begin{table}[h]
\caption{Percentage of barred galaxies of various Hubble types and
different inclination. $(b/a)_{25} = $ isophote axis ratio at $R_{25}$.}
\vspace{3mm}
\hrule

\[ 0 < (b/a)_{25} \leq 0.6 \left\{ \begin{array}{lr}
                                   \mbox{SBa, SBab} & 23\% \\
                                   \mbox{SBb, SBbc} & 29\% \\
                                   \mbox{SBc, SBcd} & 15\%
                                   \end{array}
                            \right. \]

\[ 0.6 < (b/a)_{25} \leq 1 \left\{ \begin{array}{lr}
                                   \mbox{SBa, SBab} & 53\% \\
                                   \mbox{SBb, SBbc} & 56\% \\
                                   \mbox{SBc, SBcd} & 30\%
                                   \end{array}
                              \right. \]
\hrule
\end{table}
 Bars are ubiquitous and certainly play an essential role in disk
evolution as we will see later. It is interesting to note that these
percentages do not vary very much in various environments. A maximum
percentage of barred galaxies seems to exist in Sb-Sbc types but we
must be very careful with statistics concerning the morphological
types. It is obviously easier to detect bars in face-on or nearly
face-on galaxies. We separately give in Table 1 the fractions SB/(SA +
SB) as a function of the inclination of the galactic disks with
respect to the line of sight.  The de Vaucouleurs parameter
$(b/a)_{25}$ is equal to 1 for face-on objects and 0 for edge-on. The
intermediate class SAB (or X in the de Vaucouleurs notation) is
excluded from this statistics. We can say that the majority of
galaxies must be barred or present an oval in the center if we add the
fact that photometric data in IR reveal bars which were not previously
observed in bluer band (see for ex. Elmegreen 1981).

Early type galaxies seem to have larger bars relative to the galactic
disk size than late type galaxies. The correlation between the size of
bulges and the length of bars found by Athanassoula and Martinet (1980)
is confirmed  by Baumgart and Peterson (1986).

\item Hogg et al. (1993) found a significant tendency to an
increase with the Hubble type from Sa's to Sd's of the cool gas (HI,
H$_2$, dust).

Fig. 10 of an old paper by Roberts (1969) already displayed the same
tendency for $M_{HI}/M_{TOT}$ from less than 0.05 for Sa-Sab's to
0.10--0.15 for Sbc--Sd's.

\item From a series  of papers by Rubin and collaborators (see a
synthesis of results in Rubin et al. (1985)) we have a relation
between the optical maximum rotation velocity $V_{max}$, as well as
the blue absolute magnitude, and the Hubble type. This relation is
confirmed by using HI rotation curves as shown by Zaritsky (1993). The
tendency is a $ \langle V_{max} \rangle$ decline from early to late
types.

\item The Tully-Fisher direct relation between the absolute luminosity
and the
maximum rotation velocity (Tully-Fisher, 1977) shown in fig. 5 is another
tendency in spiral galaxies. However, this relation seems to be
dependent on the pass-band, as confirmed by Gavazzi (1993). In
infra-red, there is less type-dependence contrarily to visible bands.

\item The morphology of rotation curves may change according to the
luminosity. In the context of the conspiracy problem between optical and
dark matter in spirals (see section 5.3), Casertano and  van Gorkom (1991)
distinguished three kinds of rotation curves as seen in one of their figures
reproduced here (fig. 1).

Spiral galaxies occupy a limited band in the plane $(V_{max}, h)$,
where $h$ is the disk exponential scale length. We have suggested that
systems with $ (V_{max}, h)$ outside this band would be dynamically
strongly unstable (see Martinet, 1988).

\item The central velocity dispersion $\sigma_0$ is also function of
Hubble type (Dressler and Sandage, 1983): $<\sigma_0>$ varies from
$\sim$ 115  $kms^{-1}$ for Sc + Sbc's galaxies to 170 $kms^{-1} $ for Sa
and 220  $kms^{-1}$ for S0.

\item The parameter $V_{max}/\sigma_0$ indicates the ratio of systematic
rotational motion to central random motions in galaxies. Whereas it is
smaller than one for elliptical galaxies, it is larger than 1 for
spirals. But we do not find any significant variation with the spiral
Hubble type. Typical values are 1.8 for Sa's, 1.6 for Sb's, 1.7 for Sc's.

\item The distribution of integrated $H_\alpha + [NII]$ equivalent width
of normal galaxies, which can be considered as an estimate of the  current
star formation rate (SFR) scaled to the total red luminosity, shows a
strong trend with Hubble type. The total SFR ranges from 0.1 to 1
$M_\odot yr^{-1}$ in S0-a galaxies to 10 $M_\odot yr^{-1}$ in Sc-Irr
galaxies (Kennicutt, 1983). The range of variation is larger if we take
account of  extreme dwarfs (0.001 $M_\odot yr^{-1}$) and starbursts
(till more than 100 $M_\odot yr^{-1}$).

\item The history of star formation is globally described by the
parameter $b= SFR \cdot \tau_d/M_d$, where $\tau_d$ is the age of the disk and
$M_d$ its total mass. $b$ measures the ratio of the current star
formation rate to the average past rate. It increases from 0.05 in
Sa-Sab's to 1.0 in Sc's (Kennicutt, 1993).

\item We will come back (section 10) on the problem of dark matter for
which some arguments suggest that its quantity increases from Sa's to
Sc-d's.

\item Zaritsky et al. (1994) suggest that Sb-Sc galaxies may have
steeper metal abundances gradients than either earlier or later spirals.
Furthermore, barred galaxies seem to generally have flatter gradients
than unbarred galaxies. But these results must be confirmed on larger samples.

\end{enumerate}

Summarizing this section, we retain numerous tendencies  of physical
features with the Hubble type. The bulge to disk ratio, the maximum
rotation velocity, the core velocity dispersion decrease from Sa's to
Sc's.
But it is important to emphasize a large or even very large scatter at
each type or subtype. The mass of cool gas, the pitch
angle of spiral arms, the current star formation increase from Sa's to
Sc's. Here also, we must account for large scatter inside each type or
subtype.

As mentioned above, these tendencies have been observed on samples of
giant and supergiant galaxies: the used catalogues are dominated by
objects of high luminosity. We must not lose sight of the fact that
Sa-Sbc types are much less frequent among dwarfs than for all
Shapley-Ames objects. Amongst dwarfs, Scd-Im galaxies represent 29\% of
the sample compared to only 5\% for all Shapley-Ames galaxies. At present
time we observe objects from mostly unevolved disks (Malin 1 presents a
HI disk size of $0.5 h^{-1}$ Mpc !) to disks which have exhausted more than
99\% of their gas. In brief, the whole galactic evolution story goes
beyond the frame of the giant galaxy correlations.

\newpage

\section{Stellar dynamics in galaxies}

For ten to twenty years, studies in galactic dynamics mainly
concentrated on the following topics:

\begin{enumerate}

\item \underline{Dark matter}, on which numerous questions remain
unsolved: what? how much? where?

\item \underline{Triaxiality} of elliptical galaxies, bulges of spirals
and bars, which is tractable in the frame of dynamical systems with
three degrees of freedom.

\item \underline{Swing amplification} of spiral waves, described by
Toomre (1981) which was presented as one possible solution to the
problem of persistence of the observed spiral structures.

\item \underline{Interaction of galaxies}, the most spectacular effect of
which being warps, polar rings, shells, bridges, tails and mergers.

\item \underline{Dissipative behavior of gas}, the importance of which
is now recognized in the evolution problems.
\end{enumerate}

With respect to the simplifications generally inherent in classical
purely ``stellar dynamics'' analytical treatments of galactic dynamics, the
topics mentioned above introduce various levels of complexity. Dark
matter is not necessarily conservative, as suggested in section 5.3,
triaxiality introduces new phenomena in orbital behavior (section 4),
swing amplification and interactions between galaxies as well as between
components of a given galaxy imply non-stationnarity, asymmetries,
coupling of dissipative gaseous and conservative stellar components,
hence coupling of hydrodynamics and stellar dynamics. It is clear that
one can try to succeed in understanding such a complexity only by using
numerical calculations, as we will show in the next sections.
Nevertheless it is not useless to recall here some of the fundamentals
concepts of classical galactic dynamics.

\subsection{Fundamentals of classical galactic dynamics}

An analytical description of a galaxy can be tried by using a set of
relations between the stellar density distribution $\rho(\vect{x},t) \geq
$ 0, the potential $\Phi(\vect{x},t)$ and the phase density
$ F(\vect{x},\vect{v},t) \geq 0$, where $\vect{x}$ and  $\vect{v}$ are
respectively the position and velocity vectors. These relations are
implicitly described by the Boltzmann equation ($\cal B$) and the Poisson
equation ($\cal P$):

\begin{eqnarray*}
({\cal B}) & \displaystyle \frac{D F}{Dt} = \frac{\d F}{\d t} +
\frac{\d F}{\d \vect{x}} \vect{v} - \vect{\nabla} \Phi \frac{\d F}{\d
\vect{v}} = 0 \\[5mm] ({\cal P}) & \displaystyle
\nabla^2\Phi(\vect{x},t) = 4\pi G \rho(\vect{x},t)
\end{eqnarray*}

Moreover

$$ \rho(\vect{x},t) = \int_{\Gamma_v} F(\vect{x},\vect{v},t) d\vect{v} $$
where $\Gamma_v$ is the velocity space.

If $\rho(\vect{x},t)$ is the observed density (or an imposed density
$\rho_{imp}$), ($\cal P $) gives the corresponding imposed potential,
($\cal B $) gives in principle the phase density response and the
third equation the density response $\rho_{resp}$. The
self-consistency condition implies $\rho_{imp} = \rho_{resp}$.

In ($\cal B $) such as written, the evolution due to stellar
encounters $(\d F/\d t)_{enc}$ is ignored. It is generally admitted
that in galactic dynamics, the influence of encounters implying stars
is negligible, taking account the fact that the 2-body relaxation time
$t_{rel} \sim V^3/Nm_1m_2$ (H\'enon, 1973) would be much larger than
the age of the system ($V$ is the star typical relative velocity,
$m_1$ and $m_2$ the respective masses of test and field objects).
However, encounters of giant molecular clouds with stars could be
effective to heat disks of spirals and contribute to $(\d F/\d
t)_{enc} \neq 0$, so that ($\cal B $) without second member would be
only justified for time scales less than some $10^8$ years.
Furthermore, even if $(\d F/\d t)_{enc} $ is ignored, a system could
have a collisionless evolution which is revealed by numerical
simulations as well as by the observations. The collective relaxation
certainly plays a role in the secular evolution of disks in
particular.

Exact analytical solutions of the integro-differential system  ($\cal B
$) $+$ ($\cal P $) have been found only in particular stationary cases
of symmetries such as spherical systems with $F(E)$ or $F(E,J^2)$, or
infinitely thin axisymmetric disks with $F(E,J_Z)$. Actually, disks are
generally not axisymmetric, they are perturbed by spiral or bar
components and they are not very thin. Then

$$ \rho = \rho(R,\phi,z,t) ,  \ \ \ \  \Phi = \Phi(R,\phi,z,t) ,
\ \ \ \ F=F(R,\phi,z,v_R,v_\phi,v_z,t)
$$
with a stationary zero order possible solution

$$ \rho_0(R,z), \ \ \ \ \Phi_0(R,z), \ \ \ \ F_0(R,z,v_R,v_{\phi},v_z)
$$

If the disk is thick or if we look for a solution for a axisymmetric
galaxy including a halo, we have the additional problem of the third
integral: $F(E,J_z)$, where $E=\frac{1}{2}(v^2_R + v^2_z +
\frac{J^2_z}{r^2}) + \Phi(R,z)$ and $J_z = rv_\phi$, is not compatible
with the velocity distribution in the solar neighborhood where the
observations suggest that the velocity dispersion ratio
$\sigma_R/\sigma_z \simeq 2$.

In principle, the phase density should be written as a function of a
set of isolating distinct integrals (Jeans Theorem).  Literature is
rich in discussions on the difficulty to realize this end beyond the
classical integrals of energy $E$ and angular momentum $J_z$ in the
axisymmetric case. Only by having recourse numerically to the
Poincar\'e method of surface of section, it was possible to partially
clear up the problem of integrability of specific potentials. An
exemplary case was explored by H\'enon and Heiles (1964). In general,
dynamical models such as those which are able to represent real
galaxies are non-integrable, in the sense that there exists always a
subspace of the phase space occupied by chaotic orbits in such systems
(see also section 4).

For an up-to-date review of the integrability of galaxy models see
{\it i.e.} de Zeeuw (1988).

Because of the difficulties to solve the system ($\cal B $) $+$ ($\cal
P $) or to find explicit form for $ F(\vect{x}, \vect{v})$ in terms of
integrals, one resorted for a long time to the Jeans equations
(Stellar hydrodynamic equations) obtained by multiplying ($\cal B$) by
$v_i^k$ with $k=0$ and $1$ and integrating on the velocity space. Some
useful relations between the density of a stellar population, the
moments of the velocity distribution and the radial or perpendicular
forces $K_R$ and $K_z$ have been obtained. For various applications to
problems of galactic structure such as the asymmetrical drift, the
motions perpendicular to the galactic plane and the determination of
the galactic mass from the kinematics of globular clusters in our
Galaxy or the relation between the azimuthal systematic motions and
the shape of spheroidal systems, cf. Binney and Tremaine (1987). The
drawback of the method consists in the impossibility to close the
system of equations without introducing arbitrary hypothesis with
regard to the moments of $F$. But see a recent method suggested by
Amendt and Cuddeford (1991) to overcome this difficulty.

The introduction of perturbations in the fundamental equations of the
self-consistent problem mentioned above very quickly gives cumbersome
expressions, even if one makes use of the fact that apparently spiral
perturbations, for instance, are small with respect to the
axisymmetric field. The linearisation of the equations in this case
has been considered as an adequate approximation of the first order.

We have

$$ \Phi = \Phi_0 +  \Phi_1 + ...\;  \; , \; \;  F = F_0 + F_1 + ...
\; \; , \; \;  \rho = \rho_0 + \rho_1 + ...
$$
where $\Phi_0$ is an imposed axisymmetric potential and $\Phi_1$ equal an
imposed spiral or weak bar perturbation $\Phi_1 \ll \Phi_0$, $ \rho_1 \ll
\rho_0$, $ F_1 \ll F_0$.

$(\cal B)$ becomes $D(F_0,\Phi_0) + D(F_1,\Phi_0) + D(F_0,\Phi_1) = 0$

A zero order solution is given by solving  $D(F_0, \Phi_0) = 0$. The
first order solution is obtained from the equation:

$$ D(F_1,\Phi_0) = -  D(F_0,\Phi_1) $$

If $\Phi_1$ is known, for ex.

$$
\Phi_1(R,\phi,t) = \sum_{m} \Phi_{1m}(R)
exp [i(\omega t-m\phi)]
$$
where m represents the  symmetry of components (for
a spiral, $ m = $ number of arms), we will formally have $ F_1 = - \int
D(F_0,\Phi_1) dt$, where the integration is performed along the
unperturbed orbits. $D(F_1,\Phi)$ is non-linear. Non-linear effects in
spiral cases are weak in general except for resonance regions which will
be described right below.

Calculations lead to denominators in $F_1$ in the form of

$$\sin(\omega \tau_0-m\phi_0)
$$
where $\tau_0$ = half epicyclic period $= \pi/\kappa_0$, $\phi_0 =$
angle between the apocenter and pericenter of the orbit, $\omega = m
\Omega_p$, where $\Omega_p$ is the perturbation pattern angular velocity.
This denominator is $0$ if

$$
\frac{\Omega_p - \Omega}{\kappa} = \pm \frac{n}{m}  =  \mbox{rational
number}
$$

For $m=2$, if $n = 0$, $\Omega_p = \Omega$ (corotation resonance) and
if $n = 1$ $\Omega_p = \Omega-\kappa/2$ or $\Omega + \kappa/2$
respectively corresponding to the Inner (ILR) and Outer (OLR) Lindblad
Resonances. They are the most important resonances in the galactic
disks.  They play a significant role when a perturbation to the
axisymmetric case is present. A star at resonance meets the
perturbation pattern always at the same point on its orbit so that a
cumulative perturbation effect is to expect. It is necessary to stress
that these resonances are defined in the frame of a theory of small
motions around circular orbits.  In the presence of a strong bar, the
frequencies $\Omega(R)$ and $\kappa(R)$ as well as the ``vertical''
frequency $\nu_z(R)$ do not rigorously represent the actual orbit
frequencies inside the bar.

Strong bars (axis ratio larger than 1.5 - 2) cannot be treated in the
frame of the linearized equations. Non-linear effects are important.
Orbits present large deviations with respect to the familiar epicyclic
orbits of the axisymmetric case. Fig. 2 shows some typical resonant
periodic orbits in perturbed systems, numerically calculated. For the
description of non-linear theory at the ILR, the reader is referred to a
review by Contopoulos and Grosb\o l (1989).

The previous discussion shows that the analytical approach of the
self-gravitating problem in disk galaxies is attended by large
difficulties. Recently two numerical methods have been proposed to
obtain a self-consistent solution for spiral or barred galaxies. The first
one (Contopoulos and Grosb\o l (1989))is based on the existence and the
shape of periodic orbits in spiral galaxies. The other, used by Pfenniger
(1984b) for the construction of 2-D self-gravitating barred galaxies, is a
variant of the Schwarzschild (1979) linear programming approach to
obtain equilibrium model of elliptical galaxies. Both methods consider
various kinds of possible orbits and the time they spend in various
regions of the system.

A fortiori the treatment of dynamical evolution of the galaxies with
time is out of reach by analytical methods. This justifies the use of
numerical simulations on computers of galaxy evolution with the help of
N-body and hydrodynamical codes. This approach already allowed to treat
a lot of problems concerning the equilibrium, the stability and the
evolution of disk galaxies such as those described in section 5 and the
next ones.

\subsection{Angle-action variables}

The formalism of angle-action variables was introduced by Born (1927).
It appeared in galactic dynamics in papers by Lynden-Bell and Kalnajs
(1972) and Lynden-Bell (1973). It has been proved to work in various
problems connected to the disk evolution: a) orbital behavior of stars
belonging to the spheroidal component of a galaxy during the slow
formation of the disk (Binney and May, (1986), b) trapping of orbits in
the potential sink of a bar (Lynden-Bell, 1973 and 1979), c) dynamics of
resonances in spirals and bars (Contopoulos, 1988), d) evolution of
barred galaxies by dynamical friction (Weinberg, 1985), e) the exchange
of angular momentum between a bar and a disk (Little and Carlberg, 1991),
f) the dynamical interaction between a bar and a spheroid (Hernquist and
Weinberg, 1992). In some cases, it is cumbersome to go beyond a linear
treatment but this formalism gives some useful insight on the reality of
the mentioned processes. We will come back in next sections on some of
the most recent applications of the formalism.

A suitable set of angle-action variables $(I_1, w_1, I_2, w_2)$ for
galactic disks is defined as follows:

With $R_h =$ radius of the circular orbit corresponding to the angular
momentum $h$,

$R_1 = R - R_h = a sin(w_1)$, $w_1$ is the radial oscillation
phase.

$I_1 = 1/2 \kappa a^2 = \left(E_R - E(h)\right) /\kappa $ is a function of the
radial amplitude.

$w_2 $ is the galactocentric angle to an epicenter in uniform
motion.

$I_2 = h = $ angular momentum.

For a thick disk, a third action is introduced: $I_3$ is a function of
the perpendicular amplitude of motion $ = E_z/\nu_z$, where  $\nu_z$ is
the frequency of the perpendicular motion, to which corresponds $w_z = $
latitudinal angle $=$ perpendicular oscillation phase.

For a perturbed system, $H(\ub{I},\ub{w} ) =
H_0(\ub{I}) + \epsilon H_1(\ub{I},\ub{w} )$,   the
equations of motion are

$$
\left\{ \begin{array}{rcl}
   \dot{\ub{I}}   & = & - \epsilon  \ub{\nabla}\!_w
H_1(\ub{I},\ub{w} ) \\[4.5mm]
   \dot{\ub{w}}  & = &
  \mbox{\boldmath{{$\nabla$}}}\!_I H_0(\ub{I}) +  \epsilon
 \mbox{\boldmath{{$\nabla$}}}\!_I H_1(\ub{I},\ub{w} )
        \end{array} \right.
$$

\noindent
the Hamiltonian being explicitly $H = \Phi_0 + \epsilon \Phi_1 -
\Omega_p h$  where $\Omega_p$ is the angular velocity of the spiral or
bar perturbation.

For example $\epsilon \Phi_1 = \sum_{m}\sum_{n}[ \epsilon_{mn}
\cos(mw_1-nw_2) + \epsilon'_{mn} \sin(mw_1-nw_2)]$

A useful property of the actions is their adiabatic invariance in cases
of slow time variation of potentials.

This consideration has been exploited for instance by Lynden-Bell (1979)
in an very elegant treatment of the trapping of quasi-resonant orbits by
a bar.

\newpage
\section{What can we learn from studies of stellar orbits}

\subsection{Introduction}

{}From simplicity to complexity, studies of stellar orbits in galaxies can
be tackled according to the three following approaches:

\begin{enumerate}

\item \underline{Study of orbits in a given potential}

The aim is to obtain a classification of main types of orbits existing
in a given system, in particular the periodic orbits. This approach is
able to bring constraints on the real existence of such a system and that,
in so far as the shape of some orbits is not always compatible with the
geometry of the imposed density profile as explained below.
In this approach, the collective effects are neglected so that some results
need confirmation by N-body simulations, for instance the problem of the disc
instabilities perpendicular to the galactic plane (section 6.4).

\item \underline{Response of a given population to any perturbation}

This approach implies the calculation of the time spent by the orbits in
given regions and allows to study mechanisms which explain some
observed structures as rings (resulting from trapping of matter at some
resonances) or thickening of disks.

\item If in the frame of the previous approach, we succeed to find a
response in density equivalent to the imposed density, for some
particular  weighting of the various kinds of possible orbits, we can
claim to construct a \underline{self-gravitating equilibrium model}, as
already mentioned in section 3.1.

\end{enumerate}

In this section we will show that the study of orbital behaviors in
given systems is a complementary approach to other ways of analyzing the
structure and the evolution of the galaxies. In particular the question
of the importance of chaotic orbits in galactic potentials has been the
subject of numerous investigations in the recent past: what is the
permissible percentage of such orbits in various morphological types of
galaxies if they have to be considered as self-gravitating equilibrium
systems?

Let us begin by a question of terminology. Roughly spoken, a system is
called ergodic if any trajectory (except for a set of null measures)
fills densely its energy surface. The sequence of intersection points in
a space of section (Poincar\'e,1899) corresponding to such an ergodic
trajectory would fill densely this space of section. In spite of the
widely-held use of the term of ``ergodicity''  in works dealing with
galactic dynamics problems, ergodic orbits strictly as such, are not
found in smooth ``realistic''  as well as in noisy models of galaxies.
Apart from very peculiar cases, at the very most we may speak about
chaotic behavior in {\it some} regions of phase space. Some other
adjectives are also used in literature: irregular, wild, erratic,
semi-ergodic... This behavior implies that in a space of section the
sequence of points corresponding to the trajectory jumps more or less
randomly in the fraction of space left open by the invariant curves
which correspond in contrast to a regular behavior. In fact, as often as
not, a chaotic behavior is characterized by the presence of {\it
cantori} which, by any means, are not to be confused with ergodicity.

See Percival (1989) for an introduction to the concept of cantori and
H\'enon (1981) for a clear description of the surface of section method
and applications to various dynamical systems.

\subsection{Interaction of resonances. Heteroclinic orbits}

Gerhard (1985) looked for the perturbations $\epsilon H_1$ of integrable
Hamiltonians $H_0$ which are consistent with observations of early-type
 galaxies
as well as approximately preserving the regular orbital structure of
integrable potentials. Limiting his treatment to small perturbations and
to homoclinic orbits in order to be able to use
the Melnikov integral technique, the author finds that only perturbations such
 as cos $m\phi (m =$ 0, 2,
4), modest ellipticity gradients or small figure rotation are working in
this context. However, it seems to be clear that in realistic systems, a
chaotic behavior can be amplified by resonance interactions and the
presence of heteroclinic orbits. Such a situation was described in the
inner regions of an axisymmetric model of our Galaxy for stars with
small angular momentum (Martinet, 1974).

Evidence for chaos triggered by resonance interactions in triaxials
models of galaxies, has been given by Martinet \& Udry (1990) in
connection with the adopted morphology for these systems. Orbits  were
systematically studied in slowly rotating modified Hubble profile models
of various axis ratios $(a : b : c)$: I) a nearly spherical one, II) a
Schwarzschild model (1 : 0.625 : 0.5), III) a strongly triaxial one $(1:
0.6 : 0.15)$  and IV) a bar $(1: 0.25 : 0.125)$. Surfaces of section as
well as the rotation number ``rot''  attached to invariant curves have
been obtained for different values of the Hamiltonian. An example is
given in Fig. 3 for model II. Rational values of ``rot''  correspond to
resonant periodic orbits. When the orbits are not regular, it is
impossible to define ``rot''  (the invariant curves are dissolved)
without ambiguity (bottom figure) and discontinuities in ``rot'' $(x)$
appear, indicating the range of resonances in interaction responsible of
chaotic behavior apparent in the surface of section (top figure).

The main result of this investigation is that only models I and II show
moderate chaotic regions (cantori) with not really detected resonance
interactions, possibly confined to a very narrow range of rational
numbers. On the contrary, models III and IV develop important chaotic
regions: for model IV for ex., resonances in the range 1/5 $<$ ``rot'' $<$
 2/3 are implied in the set-up of chaos. Figure 4 shows that in an
existence diagram  of axis ratios, our models I and II are located in
the region occupied by real triaxial galaxies or bulges (various
symbols) according to predictions inferred from observations. On the
contrary, the highly triaxial models III and IV are in a region devoided
of such real systems. This could be a sketch of constraints on the
possible shape of triaxial galaxies. We must notice that $N$-body
equilibrium figures obtained by gravitational collapse of initial
rotating anisotropic bodies have $b/a$ from 0.5 to 1 and $c/a > $ 0.45
(Udry, 1992) also in agreement with observational predictions. It is the
same for Barnes (1992) products of mergers. Too highly triaxial objects
might not exist because real equilibrium systems  would not tolerate too
much chaos!

\subsection{Effects of asymmetries or noise on the orbital behavior in
triaxial systems}

The gravitational attraction of a co-rotating nearby body is able to
deform the shape of orbits in the given systems defined above. In particular,
instead of having a main stable orbit $x_1$ which bifurcates into two
stable branches (for ex. Martinet \& Zeeuw, 1988), the asymmetry
triggered by an eccentric Plummer sphere leads to a continuous deviation
of $x_1$ from the plane, gradually becoming a banana-shaped orbit (Udry
1991). We will come back below to the problems rising from the existence
of this kind of centrophobic orbits. The addition of a high frequency
sinusoidal function to the given potential to represent some noise
locally modifies the isodensity contours. Udry suggested the following
form

$$
\Phi = \Phi_0 \left[ 1+ \epsilon \sin \left( \sum_i  k_i x_i \right) \right]
$$
for a noisy 3-D potential. $\Phi_0 $ is the potential defined in the
previous section (Hubble profile model). As shown in Fig. 5 displaying
surfaces of section for several cases of the amplitude $\epsilon$  and
the frequencies $k_i $  of the noise in the 2-D case, chaotic behavior
is favored by such perturbations. As $\epsilon$ and (or) $k_i$
increase, the invariant curves become thicker and are then progressively
destroyed. See also Pfenniger and Friedli (1991).

\subsection{Complex instability in triaxial systems}

This is one of the new orbital behavior which appear in dynamical
system with three degrees of freedom (Magnenat and Martinet, 1983). It
consists of the fact that the Jacobian matrix associated with the
linearized transformation describing the motion close to a periodic
orbit has all its eigenvalues complex and outside the unit circle. Real
eigenvectors do not exist. This behavior was mentioned by Broucke
(1969) in the 3-body problem and emphasized by Magnenat (1982) for the
case of cubic potential. Important zones of complex instabilities have
been found in rather academic potentials (Contopoulos and Magnenat
(1985)). Pfenniger (1985) found complex unstable tridimensional periodic orbits
in a large range of energy and parameters in a model of barred galaxy.
It is not clear in which situations  complex instability would
produce a great amount of chaotic motions.

We mention here a case which could be interesting for the equilibrium
and the stability of real galaxies. Ordinarily, four main sequences of
periodic orbits are considered for a specific triaxial slowly rotating
system: the stable anomalous orbits tipped relatively to the equatorial
plane and circling the long axis, the unstable anomalous inclined orbits
which circle the intermediate axis, the normal stable retrograde orbits
in the equatorial plane and the z-axis orbits. The z-axis orbits are
stable (s) in the inner part, then become simply unstable (u) against
perturbations parallel to the intermediate axis and still further out
become doubly unstable (du) against additional perturbations parallel to
the long axis. There four sequences are interconnected in a way
described by fig. 1 of the paper by Heissler et al. (1982). Martinet and
Pfenniger (1987) studied the stability of the z-axis orbit when the
figure rotation and the shape of the system change. The sequence of
bifurcations mentioned above is modified and the onset of complex
instability at high energies on the family of z-axis orbits is fairly
general for rotating triaxial stellar systems. For a figure rotation
large enough, the direct transition of stability to complex instability
prevents the stable anomalous orbits from joining the z-axis. Let us
consider a 3-D barred potential (axisymmetric background plus a Ferrer's
bar such as used by  Pfenniger (1984a), to which a Plummer
sphere is added to represent a condensed massive object at the center of
the system). It is observed that even a very small mass concentration
$M_p$ in the core amplifies the complex instability: the critical value
of the Hamiltonian at which the transition stable $ \longrightarrow $
complex unstable occurs is lowered. For a value of $ M_p/M_{tot} \geq
0.0006 $, the $z$-axis orbit is practically fully complex unstable from
$z = 0$. The effect of such an instability is described  in Fig. 6: the
importance of stochastic diffusion of orbits starting near the z-axis is
increasing with the central Plummer sphere mass. The diffusion time by
this process could be shorter than the Hubble time if the central mass
makes the most of the z-axis unstable. It seems possible that by the
association of a central small mass and a rotating bar, which both
create a large number of stochastic orbits, bulges may grow secularly
through the enhanced diffusion of stochastic orbits. Of course, other
stochasticity-producing events, like mergers, may happen in the life of
a galaxy. The exact diffusion speed is sensitively dependent on small
perturbations. Since real galaxies are much less smooth, symmetric and
steady than the potentials above, we can expect that the stochastic
diffusion is probably even faster in real galaxies than in the present
models. However, in the approach described here as in all calculations
where orbits are studied in given fixed potentials, the collective
processes are not taken into account so that only $N$-body simulations are
able to give a more specific idea of the mentioned effects.

\subsection{Barred galaxies}

Detailed numerical calculations of stellar orbits in given barred
potentials have been a cornerstone for understanding the dynamics of SB
galaxies, which has recently been extensively reviewed by Sellwood \&
Wilkinson (1993). Here we summarize again some important conclusions
inferred from recent works about the relation between the shape of bars
and the onset of chaos.

The essential point is that if the axis ratio of the bar in the plane of
the disk is larger than 3 to 4, and/or the mass of the bar is larger
than 1/4 of the total mass inside corotation, an extended chaotic region
can occur in the inner parts of the galaxy and that the corresponding
orbits cannot enhance the bar anymore. The stability of the main
periodic orbit $x_1$ along the bar is necessary to maintain the barred
structure (for details of the 2-D case see for ex. Athanassoula et al.
(1983) or Contopoulos and Papayannopoulos (1980)). These results suggest
that models with the properties just mentioned should be excluded for a
self-gravitating barred galaxy. As indicated in the introduction, such
predictions need to be confirmed taking account of collective effects.

The effect of a compact mass at the center of a galaxy has been reported
by Hasan \& Norman (1990, and references therein). The percentage of
phase space volume occupied by direct orbits along the bar decreases
from 50\% in absence of such a compact mass or if the axis ratio $a/b$
of the bar is $\sim$ 2 to 10--15\%, for instance, if the compact mass
is one tenth of the total mass or if $a/b \sim$ 4.

At the present time, the only equilibrium model existing for a 2-D
barred galaxy is the numerical one constructed by Pfenniger (1984b). In
this work it is shown that the percentage of semi-ergotic stars may be
as large as 30\% but more probably below 10\% if the axis ratio is 4 and
the mass of the bar is 1/5 of the total mass. Locally, however, around
the Lagrangian points at the end of the bar, they can be 100\%.

For a 3-D barred galaxy the estimation of the percentage of chaotic
orbits is more complicated. From Pfenniger (1984a), it appears that
semi-ergodicity is favored by instability strips perpendicular to the
galactic plane. The bar growth is limited if axis ratios $a/b$ and
$a/c$ are too large. A too thin bar ($a/c \sim 10$) induces a lot of
chaos. In a 3-D strongly barred galaxy, the resonant family 4 : 4 : 1
has been proved to have a complex unstable part (Pfenniger, 1985). The
resulting orbital diffusion could be a possibility to populate the
inner halo (see 4.6). Much work is necessary to estimate the
permissible percentage of chaotic orbits in this case. 3-D equilibrium
models do not exist at present time, in particular because vertical
density structure of bars is weakly constrained by observations.

\subsection{Boxlets}

For a wide class of non-rotating triaxial potentials, most of the phase
space is occupied by four well known major families: box, inner and
outer long axis tubes, short axis tubes. The $x$-axis periodic family is
generic for the boxes provided that it is stable. Boxes appear essential
for the construction of equilibrium models of triaxial systems. However,
if there is sharp variation of the density profile near the center,
$x$-axis orbits can be unstable. Centrophilic boxes could be replaced by
centrophobic boxlets such as bananas (Miralda-Escud\'e \& Schwarzschild 1989).

Two points about boxlets need further investigations: firstly, the fact
that this shape is not always compatible with the shape of the imposed
density.
This may cause a problem for the construction for equilibrium models.
Secondly, if the boxlets are unstable, they can participate to diffusion
of chaotic behavior. The question concerning what percentage of such
orbits is permissible for the construction of self-gravitating systems
is still open. This question is still more complicated for rotating
systems (Martinet \& Udry  1990).

\subsection{Concluding remarks}

In conclusion, we can summarize what we can learn from orbit calculation
in imposed galactic potentials
\begin{itemize}

\item[-] Existing morphologies of galaxies seem to be incompatible with too
high percentages of semi-ergodic orbits. Slowly rotating triaxial
dynamical models with a major to minor axis $a/c$ ratio larger than 2.5
or fast rotating barred systems with axis ratio $a/b$  larger than 3 or
4 display important chaotic behaviors. Systems with such morphological
features are in fact apparently not observed!

\item[-] Association of a compact central mass and a rotating bar, which
creates a large number of chaotic orbits, can trigger a secular growth
of bulges by enhanced diffusion of there orbits.

\item[-] 3-D $N$-body bars in disks have axis ratios $a/b < 4$. That
corresponds to systems for which the orbital calculations predict a
moderate percentage of semi-ergodic orbits.

\item[-] Quantitative estimations of percentages of semi-ergodic orbits in
3-D systems compatible with observed morphologies of barred galaxies is
an open question.

\item[-] Response to a given potential is not sufficient for answering the
question of the permissible percentage of chaotic orbits in various
morphological types. Furthermore galaxies evolve, bars may grow, then
dissolve. Ellipticals may accrete. Gas and its various interactions
with stars (star formation, gaseous response, stellar winds etc.)
contribute to modify the structure of the systems. Does the secular
evolution lead such systems to a state close to an integrable
potential with modest percentage of chaotic orbits, as suggested by
Gerhard (1985)?

\end{itemize}

\newpage
\section{Stability of disks}

In  the present section, we will remind some considerations on the local
and the global stability of stellar disks. Then we will examine how the
stability criteria are altered by the presence of gas. That will serve
as a preamble to the description of the effects of gas inflow towards
the central regions in section 7.

\subsection{Local stability of stellar disks against axisymmetric
\protect\newline
perturbations in terms of radial distance to the center}

The well known criterion introduced by Toomre (1964) indicates that
stellar disks are stable against local radial perturbations if

\be
  Q = \f{\sigma_R(R)}{(\sigma_R)_{min}} = \f{\sigma_R(R) \kappa(R)}{k G
\Sigma(R)} \geq 1
\ee

\noindent
where $\kappa$ is the epicyclic frequency, $\Sigma(R)$, the projected
surface density and $k$, a constant which is equal to 3.36 for a
infinitely thin stellar sheet, 2.6 for a not infinitely thin disk with
$\sigma_z/\sigma_R$ = .6 (Vandervoort, 1970) and 2.9 if the galaxy
contains $\sim$ 10\% (in mass) of gas (Toomre, 1974). For some
galaxies, $k$ could be higher so that we maintain $k$ as a free
parameter for the moment but later we will often refer to the standard
value of 3.36.

As soon as $Q$ = 1, the Jeans instability modes are suppressed. The
criterion has been applied frequently to particular galaxies (i.e.
Kormendy, 1984, van der Kruit and Freeman, 1986, Bottema, 1988). The
influence of various factors on the radial behavior $Q(R)$ in disks
has been discussed by Martinet (1988) and Bottema (1993). We discuss
below some of their conclusions. But let us remind without delay that
even if $Q \geq 1$, the disk can be violently unstable against
non-axisymmetric global perturbations and develop massive bars. It is
a result of numerous numerical simulations. In fact, a really
satisfactory criterion of global stability against the formation of
bars does not exist at the present time in spite of several tentatives
(Ostriker and Peebles, 1974; Efstathiou et al., 1982). Although dark
halos have often been used to explain the mechanism of stabilization,
some more recent numerical experiments by way of a $N$-body code
(Athanassoula and Sellwood, 1986) have shown that halos do not
represent the ``cure-all'' in this context and that a disk which is
sufficiently hot in its central parts could prevent the growth of a
bar in some cases. The result of their $N$-body model seems to
indicate that $Q\geq$ 2 - 2.5 might be a global stability criterion.
The $\langle Q \rangle$ values given in recent literature for
individual galaxies (see Bottema, 1993, and reference therein) could
certainly suffer from more or less uncertainties due to badly
determined quantities, such as the mass-to-light ratio in the disc or
the central luminosity density $L_{0,0}$. Nevertheless, the parameter
$Q$ remains a powerful tool for the estimate of the disc's
``temperature'' and it is therefore important to understand its
behavior under various structural conditions specified below.
Furthermore, a better appreciation of the possible values of $Q$ as a
function of the distance to the center $R$ can be helpful.  Let us
consider the following models for the density distribution
perpendicular to the galactic plane:

\be
\rho(z) = 2^{-2/n} \rho_e \mbox{sech}^{2/n}(nz/z_e)
\ee

Two extreme cases are $n =$1 (isothermal case) and $n = \infty$
(exponential case). For the isothermal case, the surface density
at distance $R$ from the center is

\be
\Sigma(R) & = & 2 \int_0^\infty \rho(R,0) \mbox{sech}^2(z/z_0) dz \\[4mm]
        & = & 2 z_0 \rho(R,0)
\ee

\noindent
with $ z_0 = 2 z_e$ and $\rho_0 = \rho_e/4$.

In such a case the velocity dispersion in the $z$-direction is given by

\be
\sigma^2_z(R) & = & 2\pi G  \rho(R,0) z_0^2\\
              & = & \pi G \Sigma(R) z_0
\ee

In these formulas, $z_0$ is an estimate of the constant height of the
disk. The distribution of light in the old disk is (van der Kruit and Searle,
1981)

\be
L(R,z) = L_{0,0} e^{-R/h} \mbox{sech}^2(z/z_0)
\ee

Furthermore, the following expression for $V(R) $ can fit reasonably
well many observed rotation curves:

\begin{eqnarray*}
V(R) = V_{max} \sqrt{\f{R^2}{R^2+R^2_m}}
\end{eqnarray*}

In this case
\be
\kappa(R) = \sqrt{2}  V_{max} \f{\sqrt{R^2+ 2R^2_m}}{R^2+R^2_m}
\ee

Taking account of these relations, we can rewrite the equation for $Q$
in the form

\be Q = .264 \f{\alpha}{k} \cdot \f{V_{max} }{ (L_{0,0})^{1/2} (M/L)^{1/2}}
\cdot \f{ e^{R/2h}  }{R^2+R^2_m }\sqrt{R^2+ 2R^2_m}
\ee

\noindent
with $V_{max}$ expressed in $km s^{-1}$, $L_{0,0}$ in 10$^{-2} L_\odot
pc^{-3}$, $R$ in kiloparsecs. $\alpha = \sigma_R/\sigma_z \simeq $ 2.

This formula is valid in the region where the model is applicable, i.e.
between $R = h$ and $R = 3h$ for late-type spirals (without large
bulges).

Figure 7 compares the behavior of $F(R, R_m, h)$, factor $Q$ containing
$R$, for two cases ($R_m = $ 0 and $R_m = $  h), assuming $\alpha $ as well
as $(M/L)_D$ constant. It is generally admitted that $M/L$
is constant in the whole disc on the basis of some data on the colors.
However, we cannot exclude a variation of  $(M/L)_D$ with $R$. If the
central parts are redder than the outer ones,  $M/L$ could decrease
considerably from the center to the edge so that $Q$ could increase more
strongly with $R$. Further, if we suppose $Q = $ const as it has often
been assumed,   $(M/L)_D$ behaves like

\begin{eqnarray*}
\f{exp(R/h)}{(R^2 + R^2_m)^2} (R^2 + 2R^2_m)
\end{eqnarray*}

One can also write (Bottema, 1993)

\be
Q = 0.10 \left( \f{\langle \sigma_R^2 \rangle ^{1/2}_{R=0}}{\mbox{km s}^{-1}}
\right)
 \left( \f{ \Sigma_0}{M_\odot \mbox{pc}^{-2}}\right)^{-1}
 \left(  \f{V_{max}}{ \mbox{km s}^{-1}    }\right)
         \f{F(R, R_m, h)}{\mbox{kpc}^{-1}}
\ee

\noindent
with

\begin{eqnarray*}
F(R, R_m, h) = e^{R/2h}  \left(  \f{\sqrt{R^2+ 2R^2_{m}}}{R^2+ R^2_{m}}
\right)
\end{eqnarray*}

Adopting a value of $1.4$ $kpc$ for $R_m$  considered as independent of
the brightness of a galaxy, this author introduces the relation

$$ h = 4.95 \times 10^{-6} \times v_{max}^{2.6} $$

\noindent
inferred from the Tully-Fisher relation and $\log R_{25} = -3.9 - 0.245
M_B$ for Sb and Sc galaxies by Rubin et al. (1985) with $R_{25} \simeq
3.1 \times h$ (Freeman's law).

Replacing $\Sigma_0$ by $\mu_0^B (\f{M}{L})_B $ where $\mu_0^B = 136
\; L_\odot pc^{-2}$, the expression for $Q$ leads to

\be
\sigma_R(R=h)  =  \f{816 Q (M/L)_B}{ V_{max} F( V_{max}, R=h)}
\ee

The relation between the velocity dispersion at $R = h$ and $ V_{max}$
is plotted in fig 8,  a) for various $M/L$, adopting $Q = $1.7 for all the
galaxies which is the value suggested be Sellwood and Carlberg (1984)
from their numerical simulations and b) for $Q =$ 2 or 3, adopting $M/L
= $ 2.0 for all the galaxies. This figure reproduced from Fig 11. of
Bottema's (1993) paper, shows a very good agreement between the relation
adopted above and the observational values for $(M/L)_B \simeq $ 2.8 or
for $Q$ between 2 and 2.5. Assuming that the central velocity dispersion
of disks is not very different from values of $\sigma_0$ observed by
Whitmore at al. (1985) for the bulges and using values of $ V_{max}$
given by Corradi and Capaccioli (1991) for about thirty galaxies, we
have constructed a similar diagram which gives the same agreement.

For the model previously defined we can estimate the ratio of the radial
($\kappa$) to perpendicular ($\nu_z$) frequencies as a function of $R$.

The force perpendicular to the galactic plane $K_z$, is, for a
self-gravitating isothermal sheet ($n=1$),

\be
K_z = - \f{2\sigma_z^2}{z_0} \cdot \mbox{th}(z/z_0)
\ee

Then

\be
\nu^2_z = -\left(  \f{\d K_z}{\d \dot{z} } \right)_{z = 0} = 4 \pi G
L_{0,0} \left( \f{M}{L} \right) \exp \left( - \f{R}{h}   \right)
\ee
and
\be
\f{\kappa}{\nu_z} = \f{1}{\sqrt{2} \pi} \cdot \f{k}{\alpha} \cdot Q
\ee

For example, with $\alpha = $ 2 and $k = $ 3.36

\be
\f{\kappa}{\nu_z} = .378 \times Q
\ee

This result is independent of the $ (M/L)_B$ ratio as well as of
$L_{0,0}$. The behavior of $\kappa/\nu_z$ versus $R$ is easily deduced from
that of $Q$.

In a flattened galaxy we expect that $\f{\kappa}{\nu_z} < 1$ on a large range
of $R$ except for the central parts. Therefore $Q$ ought to be smaller
that 2.65 in order to satisfy this condition. If $k = $ 2.9 instead of
3.36, the condition becomes $Q < $ 3. If $k = 2.6$, $Q < $ 3.4.

The case $Q \sim $ 2.65 is interesting because the resulting
$\kappa/\nu_z = $ 1/1 is a case of resonance coupling between the
perpendicular and the plane motions for eccentric orbits in an
axisymmetric or triaxial galaxy (Binney, 1981) which is able to trigger
a secular evolution of the system by way of an excursion of orbits out
of the galactic plane. However, the importance of the process can only
be tested by $N$-body simulations which take the collective effects into
account.

It is easy to show and interesting to note that if we replace the
isothermal model by the exponential one ($ n = \infty $) or
better by an intermediate one ($n =$ 2), the last relation between
$\kappa/\nu_z$ and $Q$ is unchanged.

It remains to explain how the stellar velocity dispersion in disks can
reach values which lead to $Q > $ 1 to 2. Our Galaxy gives the unique
opportunity to study in details the kinematic properties of stars
(systematic and residual motions), especially in the solar neighborhood:
in the external galaxies, kinematical data are restricted to rotation
curves, central l.o.s. velocity dispersions and, for several objects
(Bottema, 1993), the variation of the radial or perpendicular velocity
dispersion in terms of the distance to the center. We have no
simultaneous access to the components in the usual preferential
directions $\sigma_R, \sigma_\phi, \sigma_z$, as it is the case in our
Galaxy. Now these data are important to constrain the possible processes
responsible of the disk heating.

\subsection{Disk heating in our Galaxy}

The relation age-kinematics  for stars in the solar neighborhood which
is an undisputed evidence for the disk heating has been repeatedly
discussed in the past. An extensive review has been published by Lacey
(1991). From it, we will retain recent observational data and briefly
discuss some of the scenarios suggested to explain the increase of
velocity dispersion with age in the galactic disk.

The estimation of stellar ages put difficulties for oldest stars (age
higher than 2 billions years). The selection of stellar groups near the
main sequence according to the color index or the spectral type gives
only  very rough results. The chromospheric activity concerns essentially
K stars. At present the most effective method is based on the ($u, b, v, y$)
Str\"{o}mgren photometry added to isochrones of stellar evolution models
for B to F spectral types, but the squeezing of the isochrones on a
range of color index corresponding to an age range between 3 and 10
billions years induces large uncertainties on the individual ages.

An age-kinematics relation has been obtained by Carlberg et al. (1985)
from a sample of 225 F dwarfs and by Str\"{o}mgren (1987) from 558
population I dwarfs and 1294 stars of any heavy element abundance.
Clearly there is a disagreement between these recent relations and the
older ones (Wielen, 1977) for ages larger than 2. 10$^9$ years. A part of
this disagreement comes from the choice of the galactic age (15. 10$^9$
years for recent determination, 10. 10$^9$ years for the older). The law of
variation of the velocity dispersion $\sigma$ in term of age is

$$
\sigma(\tau) = (\sigma_0^{1/p} + C\tau)^p
$$

\noindent
with $p = $ 0.5 for the old data and 0.2 - 0.3 for the new ones. As for
the velocity dispersion ratios in terms of age, it is not indicated to
use them as strong constraints for physical process models because of their
high error bars. Average values are  $\sigma_R : \sigma_\phi : \sigma_z
= 1 : 0.6 (\pm 0.03) : .32 (\pm 0.03)$. A third group of data should be
explained by a correct model of disk heating: it is the radial
dependence of   $\sigma_R$ or  $\sigma_z$. At the present time only the
local kinematics has been used to test the various mechanisms suggested,
which we will now enumerate. Of course, these mechanisms concern any
galactic disk but are only constrained by local observations.

\begin{itemize}

\item[a)] \underline{Diffusion by massive gas clouds}

Analytical investigations and numerical simulations lead to an estimate
of orbital diffusion time by 2-body star--cloud encounters

$$
\tau_{diff}  \div  \f{\sigma^3_\ast}{G^2 \rho_c M_c \mbox{ln}\Lambda }
\sim 10 \times 10^9 \mbox{ years}
$$

with $\sigma_\ast \sim $ 30 $kms^{-1}$, $\rho_c \sim $ 0.1 $M_\odot
pc^{-3}$, $M_c = $ cloud average mass $ \approx $ 10$^6 M_\odot$ and
$\Lambda$, the ``Coulomb logarithm''.  There exists a population of
molecular clouds with these properties.  Because the diffusion
transfers the energy between radial and perpendicular oscillations and
increases on average corresponding energies to the detriment of the
systematic circular motions, this process implies a velocity
isotropisation which contradicts the observed values
$\sigma_z/\sigma_R = $ 0.5. It also predicts $\sigma$ ($10^{10}$
years) $\sim $ 20 $kms^{-1}$ whereas the old disk has $\sigma \simeq $
50 $kms^{-1}$.

\item[b)] \underline{Diffusion by spiral density waves}

This process, proposed by Toomre (1964) and numerically investigated by
various authors (for ex. Sellwood and Carlberg (1984) and references
therein), is exposed to a difficulty due to the fact that the stability
of disks is very sensitive to the velocity dispersion. If $\sigma_R$ is
too weak, spiral waves (or bars) develop, but because of the field
fluctuations, $\sigma$ increases and $Q$ becomes of the order of 2 after
10$^9$ years. Then the disk is too hot to maintain the spiral structure.
The dissipation in gas or a continuous accretion of matter are able to
reduce this effect but one must cope with the problem of maintaining the
spiral arms for a long time. Furthermore spiral modes found by
analytical or numerical investigation seem to be unable to heat the
whole disk.  The suggestion of recurrent instabilities by Sellwood and
Lin (1989) could be a solution. However, Carlberg (1987) found that
vertical heating by the process invoked does not work by lack of
conspiration between the radial and the perpendicular frequencies. In
general $\mid m(\Omega - \Omega_p) \mid \leq \kappa$ and
$\nu_z/\kappa \sim $ 2-3. But see sect. 6 for barred galaxies.

\item[c)] \underline{Combined effect of spirals and clouds}

The respective failures of the previous processes led Carlberg (1987) and
Jenkins and Binney (1990) to consider the combined effect of both. A
stochastic acceleration, described by Fokker - Planck diffusion
equation, is produced. For a clear setting-up of this equation, cf. H\'enon
(1973). The ratio  $\sigma_z/\sigma_R $   depends on a parameter $\beta$
which represents the ratio of the heating actions by spiral wave and
cloud diffusion (  $\beta = $ 0  when the later acts alone). A good
agreement with observations is obtained for    $\beta = $ 90, which
represents a spiral perturbation of the potential $\Delta \Phi \approx
$ 11   $(kms^{-1})^2$ . Then  $\sigma_R \div t^\alpha$ with $\alpha \sim
$ 0.5 and  $\sigma_z \simeq  t^\alpha$ with  $ \alpha \simeq $ 0.3.

The resulting  $F(v_z)$ is Gaussian. However better observational data
with more accurate age calibration are necessary, in particular to
specify the value of  $\alpha $. Let us note that in fact, molecular
clouds are a small scale limit of a chaotic spiral structure.
Finally, the self-gravitation of the disk should be taken into account:
 Heating induces the disk thickening, then density decreases what
induces an adiabatic cooling whereas  matter accretion on the disk
can trigger the formation of new stars which increases density,
inducing an adiabatic heating.

\item[d)] \underline{Heating of disks by satellite accretions}

This process will be examined in section 8, devoted to environmental
effects on the internal dynamical evolution of disks.

\item[e)] \underline{Heating by vertical resonances}

This process will be described in section 6.4.
\end{itemize}

\subsection{Stability of a two-fluid system against axisymmetric and
 non--axisymmetric perturbations.}

A real galaxy consisting of stars and gas, it is important to study the
physical effects of gas inclusion on the stability.

Local gravitational instabilities in a two-component system (gas and
stars) obviously depends on the parameters of both gas and stellar
fluids. A condition of neutral stability has been given by Jog and
Solomon (1984) in the form of the dispersion relation

\be
 \f{2 \pi G k \Sigma_\ast}{\kappa^2 + k^2\sigma^2_\ast} + \f{2 \pi G
k \Sigma_g}{\kappa^2 + k^2\sigma^2_g} - 1 = 0
\ee

\noindent
the first and the second terms corresponding respectively to stars and
gas with $k = $ wave number, $\sigma_g =$ sound speed in the gas,
$\sigma_\ast = $ velocity dispersion of stars. Even a small gaseous
fraction (less than 10 - 20\%) of the total disk density significantly
decreases the stability of disks against axisymmetric perturbations. An
important extension of this work has been recently published by Jog
(1992) who studied the growth of non-axisymmetric perturbations in disks
represented as two-fluid systems. The underlying phenomenon previously
studied for one-component cases and called ``swing amplification''  by
Toomre (1981 and references therein) results from a conspiration of
three effects: shearing, due do the differential rotation, shaking due
to the epicyclic motions and self-gravity of the matter involved. Tagger
et al. (1994) add a fourth factor, the thickness of the disk.

In her treatment, Jog assumes that the stellar ($s$) and gaseous ($g$)
components are two isothermal fluids with surface density $\mu_i$ ($i =
s$ or $g$) and sound velocity or 1-D velocity dispersion  $c_i$ ($ c_g
\ll c_s$). $\mu_0$ and $\Phi_0$ are the unperturbed density and potential
in an infinitely thin disk supported by the differential rotation and
residual motions of both fluids. The fluid representation is considered
as relevant for local analysis far from resonances if the shearing is
not too important. The mathematical treatment is simplified. However in
a more rigorous approach, the isothermal assumption for the stellar
velocity distribution has to be dropped and the viscosity of the gas
has to be taken into account.

We quote here the main points of the formalism used by Jog. The Euler
equation in an uniformly rotating frame are

\be
\f{\d V_i}{\d t} + (V_i \cdot \nabla)V_i = -
\f{c_i^2}{\mu_{0i}}\nabla\mu_{i} - \nabla(\Phi_s + \Phi_g) - 2\Omega \times V_i
+
\Omega^2 r
\ee

\noindent
where $ V_i $ is the 2-D fluid velocity with respect to rotating axes.

It is a matter of studying the dynamics of a local region around a point
($r_0,\phi_0$)  corotating in the defined frame, when a perturbation
$\delta \mu, \delta \phi$ is introduced. The Euler, continuity  and
Poisson equations can successively be written, by using sheared comoving
axes (coordinates $x', y', z', t'$), as

\be
\f{\d v_{xi}}{\d\tau} - \f{\Omega}{A} v_{yi} = -i  \f{k_y}{2A} \tau
\left[ - (\delta \Phi_s + \delta \Phi_g) - \f{c_i^2}{\mu_{0i}} (\delta
\mu_i) \right]
\ee

\be
\f{\d v_{yi}}{\d\tau} + \f{B}{A} v_{xi} = i  \f{k_y}{2A}
\left[ - (\delta \Phi_s + \delta \Phi_g) - \f{c_i^2}{\mu_{0i}} (\delta
\mu_i)  \right]
\ee

\be
\f{\d }{\d\tau} (\delta \mu_i) -i  \f{k_y}{2A} \tau \mu_{0i} v_{xi} +
i  \f{k_y}{2A} \mu_{0i} v_{yi} = 0
\ee

\be
\left[ -k_y^2(1+\tau^2) + \f{\d '^2 }{\d z^2} \right]  (\delta \Phi_s + \delta
\Phi_g) = 4\pi G  (\delta \mu_s + \delta \mu_g)  \delta (z')
\ee

\be
  (\delta \Phi_s + \delta \Phi_g) = - \left[ \f{2\pi
G}{k_y(1+\tau^2)^{1/2}} \right]  (\delta \mu_s + \delta \mu_g)
\ee

Here $\tau = 2At -\f{k_x}{k_y}$ is a dimensionless measure of time,
$k_x$ and $ k_y$ the wave numbers in $x$, $y$ directions, $x$ being
along the initial outward direction. $A$ and $B$ are the Oort
constants. A trial solution has been introduced, proportional to
$\exp[i(k_x x' + k_y y')]$. The equation must be solved to obtain
$\delta \mu_s$ and $\delta \mu_g$ as functions of time, given their
initial values.

It is illuminating to have Euler's equations rewritten for $\Theta_i =
\delta \mu_i / \mu_{0i}$    and after substitution of Poisson equation
into them,

\be
\left( \f{d^2\Theta_i}{d\tau^2}  \right) - \left(
\f{d\Theta_i}{d\tau}  \right)  \left( \f{2\tau}{1 + \tau^2}  \right) +
\Theta_i \left[ \f{\kappa^2}{4A^2} +  \f{2B/A}{1 + \tau^2} + \f{k_y^2}{4A^2}
(1 + \tau^2)c^2_i \right] & \nonumber \\[4mm]
 =  (\mu_{0s}\Theta_s + \mu_{0g}\Theta_g)
\left(\f{\pi G k_y}{2A^2}\right)(1+\tau^2)^{1/2}&
\ee

We clearly see in the brackets the terms respectively corresponding to
the epicyclic motion, the shear and the fluid pressure whereas the right
hand side corresponds to the self-gravity of the $s - g$ two-fluid system.

{}From this equation, following a mode evolution from a leading feature
($\tau < 0$) to a trailing one ($\tau > 0$), it is easy to infer: 1) At
large $|\tau|$, the pressure term dominates, $\Theta_i$ is oscillatory
with constant amplitude and a frequency proportional to $c_i\tau$ ($c_s
>c_g$)  and the $s$ and $g$ equations are weakly coupled. 2) For $\tau
\rightarrow 0$, epicyclic and shearing terms become important,
eventually with cancellation for a flat rotation curve. If the
self-gravity dominates, we have swing amplification, which is temporary
because for $\tau \gg 0$, we find back an oscillatory solution. 3) The
coupling between both fluids is higher at low $\mid \tau \mid$. It is
noted that if $k_y \rightarrow 0$ (purely radial perturbation), the last
equation, after Fourier analysis, is reduced to the dispersion relation
for the axisymmetric case obtained by Jog and Solomon (1984)

\be
 (\omega^2 - \kappa^2 -k^2_xc^2_s +2\pi G k_x \mu_{0s})  (\omega^2 -
\kappa^2 -k^2_xc^2_g +2\pi G k_x \mu_{0g}) \nonumber \\[4mm]
 - (2\pi G k_x \mu_{0s}) (2\pi G k_x \mu_{0g}) = 0
\ee
$\omega$ being the perturbation frequency in the trial solution $\Theta_i
=\Theta_{i0} exp(i\omega t)$

The next step is to look for dependence of the solution on
various crucial parameters such as $Q_s$ and $Q_g$ (``Toomre's
parameters'' ), the fraction of mass $\epsilon$ in form of gas, the rate of
shearing $\eta = 2A/\Omega_0$ and $X$, the wavelength of perturbation in
terms of the critical wavelength for growth of instabilities

$$
X = \f{\lambda_y}{\lambda_{crit}} = \f{2\pi r}{m} \cdot
\f{\kappa^2}{4\pi^2 G \mu_0^2}
$$

\noindent
where $m$ is the arm number, $\mu_0 = \mu_{0s} + \mu_{0g}$.

It is to note that in the present context (the gas is the colder
component)
$$
\f{Q_g \epsilon}{Q_s(1- \epsilon)} < 1
$$

Jog gives the new form of the differential equation for $\Theta_i$ when
these parameters are introduced  ($ \xi^2 = \kappa^2/4A^2 =
2(2-\eta)/\eta^2$)

\be
\left( \f{d^2\Theta_s}{d\tau^2}  \right) & - & \left(
\f{d\Theta_s}{d\tau}  \right)  \left( \f{2\tau}{1 + \tau^2}  \right) \nonumber
\\[4mm]
& + & \Theta_s \left[ \xi^2 + \f{2(\eta - 2)}{\eta(1+\tau^2)} +
     \f{(1+\tau^2)Q^2_s(1-\epsilon)^2\xi^2}{4X^2} \right] \nonumber \\[4mm]
& &  = \;
 \f{\xi^2}{X} (1+\tau^2)^{1/2} [\Theta_s(1-\epsilon) + \Theta_g \epsilon]
\ee

\be
\left( \f{d^2\Theta_g}{d\tau^2}  \right)& -& \left(
\f{d\Theta_g}{d\tau}  \right)  \left( \f{2\tau}{1 + \tau^2}  \right) \nonumber
\\[4mm]
& + & \Theta_g \left[ \xi^2 + \f{2(\eta - 2)}{\eta(1+\tau^2)} +
     \f{(1+\tau^2)Q^2_g \epsilon^2\xi^2}{4X^2} \right] \nonumber \\[4mm]
& &  = \;
 \f{\xi^2}{X} (1+\tau^2)^{1/2} [\Theta_s(1-\epsilon) + \Theta_g \epsilon]
\ee

A typical case is illustrated in fig. 9. The choice of parameters is
$Q_s = $1.5, $Q_g = $1.5, $\epsilon = $ 0.1, $X = $1, $\eta = $1. The
two-fluid system is stable to axisymmetric perturbations. The
amplification is higher in the gas with a more highly wound spiral feature.

Other cases, shown in Jog's paper, indicate the dependence of the
solution on the mentioned parameters and lead to the following important
conclusion:
Growth of non axisymmetric perturbations in a real galaxy may  occur even
if the system is stable against axisymmetric perturbations and/or if
either fluid component is stable against non-axisymmetric component.

We add that, according to the fact that the process is more effective
when the gas fraction is high, it must be more important in late-type
galaxies.

\subsection{Global stability of disks and the dark matter problem}

The essential of what we know on the large scale instabilities (or
stability!) of disks has been obtained through $N$-body simulations.
In the early seventies, two kinds of related results appeared
simultaneously: whereas the first flat rotation curves were published
suggesting the presence of hidden forms of matter in isolated
galaxies, seminal papers by Miller and Prendergast (1968) and Hohl and
Hockney (1969) presented $N$-body simulations of disk galaxies in view
of explaining the formation and the maintenance of spiral structures
in the disks. At that time, the galactic astronomers were focused on
the linear density wave theory (Lin and Shu, 1964; Goldreich and
Lynden-Bell, 1965) concerning tightly wound spirals.

The central problem for people working on $N$-body numerical experiments
was to suppress these strong bars which form spontaneously in the
simulations, in order to obtain the well defined grand design spiral
structure, ``so often observed''  and ``well explained by the density wave
theory''. In fact, in the seventies, bars were not much studied and
practically not observed in details with the noticeable exception of de
Vaucouleurs who insisted many times on their importance in the spiral
classification. A further motivation to ignore them was that bars could
hardly be included in the density wave theory.

Essentially, two solutions were often considered to prevent the
formation of bars: hot disks or massive halos (Ostriker and Peebles
1973; Einasto et al. 1974). But in the seventies no data could indicate
that the inner parts of the galactic disks were hot enough. Discs were
considered as consisting of cold populations by extrapolation of the
solar neighborhood data. Now, more recent observations indicate that the
velocity dispersions in inner parts of typical Sbc galaxies can be of
the order of 100 $kms^{-1}$ (e.g. Lewis et Freeman 1989; Bottema 1993),
showing that disks can be hot there.

In the last decade, theoretical arguments, as well as new numerical
experiments and deductions from the observations of the luminosity
profile of disks, have contributed to modify the point of view about the
role played by halos on global stability. The stabilization of disks
against non-axisymmetric perturbations, if ensured by halos, should only
depend on the halo mass located in the inner region of the disc, where a
bar usually develops; it should not depend on the halo mass in the outer
regions, where the rotation curve remains flat. Kalnajs (1987) disturbed
minds by claiming 1) that the ``luminous ''  rotation curves, calculated
from the observed exponential stellar disk luminosity profiles, would
fit very well the observed rotation curves in the optical region of
galaxies, and 2) that, as a consequence, dark halos would be useless to
stabilize against a bar.

Today, the global stability in the disks against bar-like perturbations
is no longer the acute problem that it used to be in the seventies. As
mentioned in section 2, at least two thirds of observed spirals are
recognized to be barred or to present an oval structure in the inner
regions (de Vaucouleurs 1963). The remaining third includes edge-on and
dusty galaxies in which a bar can hardly be detected. In many cases, IR
photometry, less affected by dust, reveals barred structures invisible
in the B band. Further on, extensive numerical simulations by Athanassoula
and Sellwood (1986) have shown that the growth of a bar may be prevented
if the central part of the disk is hot enough. $ Q\simeq $ 2.5 at all radii
seems to be a sufficient criterion for global stability against
non-axisymmetric perturbations in all disks. However, this result
concerns pure stellar disks.

Coming back to the kinematic observations, we point out that spiral HI
rotation curves, which extend typically up to 2 - 3 $ R_{25} $, are well
known to stay in general sufficiently flat over this range for requiring
an important presence of mass in the outer parts of spirals. Some
authors tried to solve the question of the contribution of various
components (disc, bulge halo) to the observed rotation curves produced
by HI data (see e.g. Sancisi and van Albada 1987). Among the proposed
solutions, the ``maximum disc''  solution consists in maximizing the
contribution of the luminous matter to the observed rotation curve. This
yields to a conservative lower limit for the amount of dark matter and
an estimate of the $ M/L $ ratio for the disk. The maximum disk solution
received recently a strong support from an extensive work on kinematical
data of $ \sim $ 500 galaxies; it reproduces well the fine wiggles of
the light, confirming that the fraction of dark halo within the stellar
optical parts of the galaxies must not be important (Freeman 1993).
Athanassoula et al. (1987) have had recourse to the swing amplification
theory to better constrain the contribution of the halo and the disk to
the rotation curve. Considering that a lot of galaxies display a 2-arm
spiral structure, they were able to estimate the disk mass required to
kill the mode 1 and to preserve the mode 2. This interesting approach is
worth to be pursued, based on more extensive data.

{}From the present discussion, we can conclude that the only really
obvious dark matter problem subsisting in spirals concerns the outer
regions beyond the optical disk. The fact that the disk plus the bulge
on one hand, and the dark halo on the other hand, bring an essential and
about equal contribution to the flat rotation curve in two distinct
regions, namely in the inner luminous region and in the outer 21 cm
emitting HI region, has suggested a physical coupling between these
components, also called disc-halo conspiracy (Bahcall and Casertano
1985; van Albada and Sancisi 1986). No convincing explanation of it has
yet been given.

Furthermore new HI observations have restricted the range of the
conspiracy: Casertano and van Gorkom (1991) published HI rotation curves
characterized by a large decrease between 1 and 3 $ R_{25} $ and have
found a clear correlation between the peak circular velocity, its
central brightness and the slope of the rotation curve in the outer
parts of the disks (see section 2, remark 7). This result leads these
authors to suggest that the ratio of the dark to luminous matter might
be the critical parameter controlling the Hubble sequence. It is crucial
to determine if really late-type galaxies contain more dark matter than
early-type ones, or if the importance of dark matter decreases with
luminosity. These questions introduced by Tinsley (1981) have been
recently discussed again (see for example Salucci et al. 1991) and
references therein).

A curious coincidence (Bosma 1981), until now unexplained, has been
recently recalled by Freeman (1993, and references therein) and better
documented by Broeils (1992): the surface density ratio of dark matter
and HI gas in a sample of galaxies remains constant outside the optical
disc, around 10 - 30. If dark matter tends to follow HI with a constant
ratio of about 20, we have the very curious situation that dark matter
is concomitant to HI radially, yet dark matter does not prevent HI to
flare out the plane just beyond the optical disk. The flaring shows at
least that dark matter cannot be much flatter than HI.

It can be assumed that if the dark matter is radially proportional to
HI, it should be also proportional perpendicularly. We quote here
the arguments given  by Pfenniger et al. (1994) and Pfenniger and Combes
(1994) to suggest that dark matter in spiral galaxies is essentially
made of cold hydrogen in a disc supported mostly by rotation.

Different observational constraints can be advanced on the presumed form
of hydrogen. A massive amount of hot or warm form of hydrogen can be
ruled out, essentially because hot and warm gas already fills most of
the interstellar volume at a much too low density to contribute to the
mass in an appreciable amount. On the contrary, although cold gas fills
a small part of the volume, its density is large enough to encompass
most of the ISM mass. Moreover, this condensed structure explains in
part the invisible character of the medium, its low cross-section for
absorption studies, and its transparency to external radiation. Even if
the medium is essentially molecular, the CO molecule can hardly serve as
a tracer, because of the low metallicity of this quasi-primordial gas. A
fraction could also be in atomic form. but HI emission cannot be
detected in a medium at 3 $K$. Absorption studies of the outer parts of
galaxies could probe this medium. Cold gas is observed to be fractal
over several decades of length and density. Details are given in
Pfenniger and Combes (1994). It turns out that both the problem of mass
underestimate in HI disks and the problem of star non-formation in outer
disks is closely linked to the fractal structure. The physical state of
this gas must be high density and cold temperature. Since no significant
heating sources in the outer disks presumably exist, it  can be assumed
that the gas is bathing in the cosmological background, and that its
temperature is about 3 $K$. In these nearly isothermal conditions,
clouds can fragment until they reach small clump units, where the
cooling time becomes comparable to the free-fall time. The average
typical density of these elementary clouds   lets, called
``clumpuscules''  is 10$^{10}$ cm$^{-3}$ , column density 10$^{24} $
cm$^{-2}$, size 30 AU, and mass 10$^{-3}$M$_\odot$. These small units
are the building blocks of a fractal structure, that ranges upwards over
4 to 6 orders of magnitude in scales. They are gravitationally bound,
and the corresponding thermal width along the line of sight for
molecular hydrogen at $T =$ 3$K$ is about 0.1 $kms^{-1}$.

We will come back later (section 10) to this suggestion since it could
certainly have consequences on the secular evolution of disks and
particularly on the morphological changes that galaxies can suffer with
time.

If the dark matter in disk galaxies is assumed to be made of cold gas,
the question of stability of such disks must be discussed. In fact, HI
disks are far from being perfectly axisymmetric, smooth and thin:
large asymmetries, warps, spiral arms, massive HI complexes are
observed. The gas is unsteady. $Q$ could be subcritical. But the
classical Toomre formula is less obvious in this context. An extensive
discussion in Pfenniger et al. (1994) shows that the stability and
self-consistency of cold gaseous disks is a less severe problem than
commonly believed.

\newpage
\section{Interaction between components}

\subsection{Disk-halo interactions}

a) \u{Adiabatic invariance of actions for spheroid stars}

When the actions $\ub{J} (J_r, L_z, J_\theta)$ characterizing stellar
orbits in a galactic potential are well defined (for stars on regular
orbits), they can be considered as adiabatic invariants if the
potential is varying slowly (see section 3). Such is the case of the
actions of stars belonging to a spheroid in the frame of the scenario
(Fall and Efstathiou, 1980) according to which the disk gradually
forms in the potential well of the spheroid. Information on the
original orbits of these stars can be inferred owing to this property.
The distribution function $F(\ub{J})$ is also an invariant.

In axisymmetric models of our Galaxy, the most important resonance in
the meridian plan ($R,z$) is 1/1. The corresponding resonant orbits
occupy a small fraction of phase space ($\leq$ 8\%). Chaotic orbits only
concern stars with very small angular momentum ($\leq$ 200 $kms^{-1}
kpc$) (Cretton and Martinet, 1994). It is shown that actions could be
defined for many halo stars in the solar neighborhood even if the
$z$-velocity is relatively large ( $\approx$ 100  $kms^{-1}$).

Binney and May (1986) have studied the response of a galactic spheroid
to the slow accumulation of a massive disk. It appears that models which
start from a spherical distribution cannot reproduce the presently
observed solar neighborhood because the  final state in this case would
have $\sigma_\theta > \sigma_r$  or  $\sigma_\theta > \sigma_\phi$ or
both, which are in contradiction with the kinematical properties of halo
stars. Models starting from flattened initial conditions (axis ratio
$q_0 =$ 0.7) can yield velocity dispersions   $\sigma_\phi / \sigma_r$ or
$\sigma_\phi / \sigma_\theta$ in agreement with observations and the
final axis ratio is $q_F = $ 0.33. The slow accretion of a disk must
progressively flatten the spheroid.

b) \u{Transfer of angular momentum from disk to the halo in $N$-body
simulations}

In many  $N$-body simulations of disk galaxy evolution, a current
approximation is to consider the halo as a rigid body. It is necessary
to know whether simulations in which disk and halo stars are treated
self-consistently give the same results. Sellwood (1980) has shown that
rigid approximation is sufficient for studies of global stability.
However it is expected that the disk loses angular momentum to the halo.
In fact a significant transfer of angular momentum from the disk to the
halo is observed only after the formation of a strong bar, which appears
in fact as a tool of this transfer. In this work, the evolution was
followed for at most 1200 $Myr$ and the long term consequences of the
interaction were not evaluated. The problem has been recently restarted
by Little and Carlberg (1991) and Hernquist and Weinberg (1992) (see the
next subsections).

\subsection{Dynamical friction and bar-disk interactions}

Let us consider a set of stars (field stars) of individual mass $m$ having
a velocity distribution  $F(v_m)$, and a test body of mass $M$ moving through
the stellar background with an initial velocity $ V_o $ relatively to
the center of mass of the field stars. As a result, $ M $ undergoes a
purely gravitational  retarding force due to the density enhancement,
behind it, of stars deviated by its passage. The net  effect, termed
``dynamical friction'', is a motion deceleration of $M$, which has been
evaluated, fifty years ago, by Chandrasekhar (1943) for the case of a
rigid body passing through an infinite homogeneous medium (see also
H\'enon, 1973).

For instance if $F(v_m)$ is Maxwellian, the deceleration is

$$
\f{d \vect{V}\!_M}{dt}  = - \f{4 \pi \mbox{ln}\Lambda G^2 (M+ m)}{v^3_M} \rho
\left[\mbox{erf}(X) - \f{2X}{\sqrt{\pi}} e^{-X^2}\right] \vect{V}\!_M
$$
where $X = v_m/2\sigma $, ln$\Lambda$ is the Coulomb logarithm with
$\Lambda = b_{max} / b_{min} = $ ratio of the maximum to the minimum
impact parameter, $\rho$ is the field stellar density. So, the
deceleration is proportional to $M\cdot \rho/ V^2_M$ and only stars with
velocity $v_m < V_M$ contribute to it.

Dynamical friction is potentially important in various problems of
galactic dynamics: the motion of globular clusters or dwarf galaxies
around parent galaxies, the motion of bars within galactic disks or
the interaction of disk galaxies. A number of questions arise about
the application of the Chandrasekhar formula in such contexts: Is the
self-gravity of the field population negligible? Is the local
treatment adequate? What about the hypothesis of rigidity? Is the
dynamical friction dominant with regard to resonance effects occurring
in disk galaxy interactions? Furthermore, we must recall shortcomings
such as the arbitrary choice of $b_{max}$ or the use of keplerian
hyperbolae for orbits in 2-body encounters (on which the calculation
of dynamical friction is based). Especially a controversy recently
arised concerning the importance of the self-gravity in dynamical
friction (see for instance Combes (1992) and references therein). In
our context of interactions between components of disk galaxies, we
will here be concerned by the self-gravity problem in the frame of the
bar-disk angular momentum exchange. We will come back in section 8 on
the dynamical friction in galaxy-galaxy interactions.

A 2-D analytic calculation of the friction between a bar and a disk has
been made in the angle-action formalism by Little and Carlberg (1991) in
the absence of self gravity. The chosen unperturbed potential $\Phi_0$
is that of a Kuzmin-Toomre disc

$$
\Phi_0 = - \f{G M_{tot}}{\sqrt{r^2 + r^2_h}} = - \f{1}{\sqrt{r^2 + 1}}
$$

In presence of a bar, the Hamiltonian of a disk star is

$$ H' = H_0 + \epsilon H_1 $$

\noindent
where $H_1$ includes a barlike perturbing potential

$$
\Phi_1(r, \varphi, t) = \Psi(r) \cos [2( \varphi - \Omega_b t)]\sin^2\nu
t \; \; ,  \; \; \; 0 \leq \nu t \leq \pi
$$

\noindent
with

$$
\Psi(r) = Q_{bar} \f{(r/b)^2}{1+(r/b)^5}
$$
with $ Q_{bar} = $ constant controlling the strength of the bar, $b$
measuring the radial size of the bar. The cosine term indicates the
bipolar nature of $\Phi_1$, $\Omega_b$ is the rotation frequency of the
bar, the sin$^2\nu t$ term is there to gradually turn the bar on at $t =
0$ and off at $t = t_f$.

Introducing the angle-action variables corresponding to the perturbed
Hamiltonian:

$$
J_i' = J_i + \sum \epsilon^n \Delta_n J  \ \ \ ,\ \ \ w_i' = w_i +  \sum
 \epsilon^n \Delta w\ \
$$

\noindent
the authors use the classical theory of perturbations to evaluate the
first and second order change $ \Delta_1 J_2$ and  $\Delta_2 J_2$ in the
star's angular momentum $J_2$. $\Phi_1$ being periodic in $w_1$, $w_2$ is
expanded as usually in Fourier series

$$
\Phi_1(\ub{J}, \ub{w}, t) = \f{1}{4\pi^2} \sum \Psi_{lm}
(\ub{J}, t) e^{i(lw_1 + mw_2)}
$$

In view of comparing the analytic treatment with $N-$body simulations,
it is necessary to average  $ \Delta_1 J_2$ and  $\Delta_2 J_2$ over
($w_1, w_2$) for many stars. The contributions to the leading term
$\langle  \Delta_2 J_2 \rangle$  can be calculated in the epicyclic
approximation ($\langle  \Delta_1 J_2 \rangle = 0$). They are essentially
resonant ($l = 0$ for the corotation, $l = \pm 1$ for the  ILR and OLR).
Their sum approximately gives the true phase average angular momentum
change per star by unit mass. It is proportional to $(Q_{bar})^2$.

Little and Carlberg undertaken numerical simulations of the bar-disc
system described above to evaluate the role of the self-gravity which is
parameterized by  $s_g  = M_{disc}/M_{tot}$ $(0 \leq s_g \leq 1)$. The
axisymmetric galaxy model consists of a $N$-body disk and a rigid halo.
An artifice permits to vary the relative masses of the disk and halo
without changing the overall potential set equivalent to Kuzmin disk in
the plane (with the object of comparing with the theoretical results). As
an example, fig. 10 shows the angular momentum change $\langle  \Delta J
\rangle$ for a case of sufficiently weak bar having a high pattern speed
($s_g = 0$). The agreement between theoretical prediction and numerical
simulation is very good. The angular momentum exchange occurs at the
corotation ($r_0 \sim 2.3$) and at the OLR  ($r_0 \sim 3.2$). In another
example (low pattern speed of the bar)  an absorption of angular
momentum near the corotation located just beyond the edge of the disk is
observed as well as an emission near the outer ILR. These behaviors are
consistent with the detailed and very clear discussion by Lynden-Bell
and Kalnajs (1972).

As for the self-gravity effect, studied by varying $s_g$ in the
simulation, it appears that it tends to increase the absolute angular
momentum transfer for high $\Omega_b$ and inversely to decrease it for low
$\Omega_b$. In fact the present situation is quite similar to the
familiar problem of an harmonic oscillator driven at steady frequency $2(
\Omega -  \Omega_b)$ lower or higher than its natural frequency
($\kappa$) as underlined by Little and Carlberg. As the case may be,
this  oscillator responds in phase or in antiphase with the imposed
perturbation. Self-gravity amplifies the process significantly but by
less than 60\%.

The authors mention that the present results on the role of the
self-gravity cannot be compared with others connected with the decay of
a satellite around a parent galaxy for which case the physics is very
different. Therefore it is not surprising that the respective
conclusions be different.

\subsection{Bar-spheroid interactions}

Bars also can interact with spheroidal components of galaxies such as
bulges, luminous or dark halos. By using the angle-action formalism
introduced in section 3 and by restricting oneself to a linear
approximation of the perturbation theory, it is possible to estimate the
torque exerted by a bar on a spheroid. The usual  starting point is the
collisionless Boltzmann equation formally written

$$
\f{\d F}{\d t} + \f{\d F}{\d \ub{w}} \cdot  \f{\d H}{\d \ub{I}} -
 \f{\d F}{\d \ub{I}} \cdot  \f{\d \Phi}{\d \ub{w}} = 0
$$

\noindent
where $H$ is the Hamiltonian and $\Phi$, the potential. Introducing a
small potential perturbation $\Phi_1$ uniformly rotating with a rotation
frequency $\omega$, and with $F = F_0 + F_1 + ...$, $\Phi = \Phi_0 +
\Phi_1 + ...$, the Fourier-transformed linearized Boltzmann equation
gives a response

$$ \tilde{F}_n = \f{\ub{n} \cdot \d F_0/ \d \ub{I}}{\ub{n}\cdot \ub{\Omega}
-\omega } \; \tilde{\Phi}_{1n}
$$

\noindent
where $\Omega =\ub{\nabla}_I H_0(\ub{I})$ and $\ub{n} = (n_1, n_2, n_3)$. This
classical result has been reproduced by Hernquist and Weinberg (1992) in
the context of the interaction between a rigid bar with an initially
non-rotating spheroid. It shows that large variations of density occur
when $\ub{n}\cdot \ub{\Omega} -\omega \simeq 0$ that is to say at and near
to resonances. In the limited frame of the non-self-gravitating problem
and if the bar rotation axis is parallel to the $z$-axis of the
spheroid, Weinberg (1985) had calculated that the torque exerted by the
bar, given by the time derivative of the angular momentum $J_z$, depends
on the square of the perturbed amplitude

$$
\f{dJ_z}{dt} = 4\pi^4 \int\int\int d\ub{I} \sum_n n_3 \Omega_b \ub{n}
\f{\d F_0}{\d I} \mid \Phi_{1n} \mid^2 \delta (\ub{n}\cdot \ub{\Omega} -
n_3\Omega_b)
$$

\noindent
where $ \Omega_b = \omega/2 $ is now the angular velocity of the bar.

Numerical simulations allow to study more thoroughly the effects of the
interaction, in particular how the angular momentum transfer depends on
various parameters implied in the problem as  $ \Omega_b$, the bar to
spheroid mass ratio, or the choice of the model for both.

For their simulations, Hernquist and Weinberg (1992) chosen for the bar
the Ferrer's density profile, often used to describe the barred
component in galaxies (see for ex. Athanassoula and al. (1983) for
details)

$$
\rho(\mu^2) = \rho_0(1-\mu^2)^2 \; \; \mbox{if} \; \; \mu^2 \leq 1   \; \;
\mbox{with} \; \;
\mu^2 = \f{x^2}{a^2} +  \f{y^2}{b^2} +  \f{z^2}{c^2}
$$

\noindent
and for the spheroid, various models, the more realistic of which having
a density profile

$$
\rho(r) = \f{M}{2\pi} \cdot \f{a_s}{r} \cdot \f{1}{(r+a_s)^3}
$$

\noindent
associated to the potential

$$
\Phi_s = -\f{G M}{r+a_s}
$$

A general result is that most of angular momentum exchange is resonant
confirming the prediction of the linear theory. The quantity of angular
momentum $\Delta J$ transferred to the spheroid depends on the presence
of low order resonances which are the most important and on the location
and the number of resonances. In a typical simulation, with $M_b/M_s =
.3$ inside corotation and a bar rotation period of 10$^8$ years, the bar
loses its angular momentum in less than 10$^9$ years. Furthermore, a
change of structure is observed in the spheroid: the density near the
center decreases with time, the density in the outer parts increases
with time. The spheroid becomes rotationally flattened. A pic value of
the ratio of the systematic to the random velocity in the azimuthal
direction is 0.2 - 0.25 which corresponds for an oblate spheroid to an
ellipticity $\sim$ 0.9. The physical process which triggers the angular
momentum transfer is qualitatively well described by using the
quadrupole component of the bar potential only.

It seems that we have here a new indication that bars are responsible of
secular dynamical evolution in galaxies. However, in the mentioned
experiments, the spheroid is initially non-rotating, the bar is rigid and
does not respond to the spheroid. Now we know that bars can evolve for
various reasons as seen in the others subsections. These limitation can
restrict the bearing of the results mentioned above, at least
quantitatively.

\subsection{Stellar motions perpendicular to a disk perturbed by a bar:
Resonant excitations.}

Resonant excitation of motion perpendicular to the galactic plane
($z$-direction) can concern stars on quasi-circular orbits in a
flattened slightly non-axisymmetric rotating potential. Theoretical
prediction about it have been given by Binney (1981) and the results are
clearly confirmed by 3-D numerical simulations which will be described
in the next subsection.

Binney starts from the equation of $z$-motion for a star in a rotating
potential $\Phi(R, \phi, z) = \Phi_0(R, z) + \Phi_1(R, z)  \cos(2\phi)
+  \Phi_2(R, z)  \cos(4\phi)  + ...$

$$
\ddot{z} + \{ \nu_o^2 + 2 q_A' \cos (\kappa t + \phi_0) + 2 q_B' \cos[
2t(\Omega - \Omega_p)]\}z = 0
$$

\noindent
with

 $$
\phi = (\Omega - \Omega_p)\ , \ \ \  q_A'=\f{1}{2}A \f{\d \nu_0^2}{\d
R}\ , \ \ \   \nu_0^2 =  \f{\d ^2\Phi_0}{\d z^2}
$$

\noindent
and

$$
 q_B'=\f{1}{2}\left\{ \nu_1^2 - \f{2\Omega \Phi_1 \d \nu_0^2/\d
R}{R_0(\Omega - \Omega_p)[\kappa^2 -4(\Omega - \Omega_p)^2] }\right\}
$$

\noindent
with
 $  \nu_1^2 =  \f{\d ^2\Phi_1}{\d z^2}$.

In the case $ q_A'= 0$, this equation of motion is reduced to the
Mathieu equation

$$
\f{d^2z}{d\tau^2} + [a + 2q\cos 2\tau]z = 0
$$

\noindent
with

$$
 a = \f{\nu_0^2}{(\Omega - \Omega_p)^2} = n^2
$$

\noindent
and

$$
q= \f{1}{2} \left[ \nu_1^2 -  \f{2\Omega \Phi_1 \d \nu_0^2/\d
R}{R_0(\Omega - \Omega_p)[\kappa^2 -4(\Omega - \Omega_p)^2] }\right] / (\Omega
-
\Omega_p)^2
$$

Instability strips able to trigger important perpendicular motions are
characterized (for the most important) by

$$ -q -\f{1}{8}q^2 < a-1 < q -\f{1}{8}q^2 $$

$$  -\f{1}{12}q^2 < a-4 < \f{5}{12}q^2 $$

Between $R = 0$ and $R = R_{corotation}$, $a$ increases from
$-\nu_0^2/\Omega^2$ to $\infty$. $\nu_0^2/\Omega^2$  is in principle
larger than 1 for a flattened galaxy. For  $R > R_{corotation}$, $a$
decreases as $\sim R^{-2}$ at large $R$ for a flat rotation curve.

Amongst the resonance conditions, $n =$ 2 is the most important for direct
orbits. If the potential is not strongly barred, the first term of $q$
in the bracket is negligible and $q$ would be large only if $\kappa \sim
2(\Omega - \Omega_p)$ or $\Omega  \sim \Omega_p$ corresponding to the
classical resonance regions. In fact, only the Lindblad resonance case
has to be considered, because if $\Omega = \Omega_p$, $a \gg $ 4.
For a strongly barred potential, $\nu_1/\Omega$ is high and $q$ can be
large. Outside the outer Lindblad resonance, $a$ may go through 1.

For a retrograde orbit, $\Omega_p < 0$, $a$ decreases from $\sim
(\nu_0/\Omega)^2$ at $R_0 \approx 0$ to zero at large $R$. Then if $\nu_0
> \Omega$, $\nu_0/\Omega-\Omega_p$ can be $<$ 1 or go through 1. Then $n
= $ 1 is the most interesting resonance in this case and large
$z$-motions can also be developed.

In such an investigation, the potential is given. If numerous stars
develop such $z$-oscillations, the potential will be modified.
Consequently rigorous treatment of the problem requires to take
collective effects into consideration. 3-D $N$-body simulations such as
those reported by Combes at al. (1990) confirm the importance of
perpendicular resonances in the secular evolution of barred galaxies
(see next subsection).

\subsection{3-D simulations of bar-disk interactions: box and
peanut-shapes}

Bars are often strong non-axisymmetric perturbations in regions of the
disks where the coupling of motions in the plane of the disk and
perpendicular to it could be important. Instabilities perpendicular to
the plane (often called ``vertical'' instabilities) must be not neglected
a priori. Therefore a realistic treatment of barred galaxies needs a
full 3-D dynamics. As we are going to see, many vertical instability
strips occur in bars even if they have no important thickness in the
$z$-direction and peanut-shaped bulges are formed in 3D $N$-body bars as a
consequence of resonant coupling between relative plane motions and
perpendicular oscillations. That confirms the predictions of the previous
subsection.

A detailed analysis of the 3-D problem can be found in Combes et al.
(1990) and Pfenniger and Friedli (1991) where the authors use the
``purely stellar dynamics'' part of the Geneva 3-D fully consistent
numerical code with gas and stars (PMSPH). As the aim is to study the
structure of the bar as well as the orbital behavior of stars in the
perturbed potential, the initial conditions are chosen so that the
disc be initially globally unstable. For a representative model
characterized by bulge scale length 0.14 and disk scale length 3, with
a mass ratio $M_B/M_D =$ .18, a bar is formed in 3 - 4 dynamical time
$\simeq$ 3 to 4$\cdot$10$^8$ years. The angular momentum is ejected by
strong two armed spiral patterns. A secular evolution of the bar is
observed in $z$ and a box or a peanut-shaped structure is reached
after 2 $\times$ 10$^9$ years, through a short phase of symmetry
breaking in $z$ (Fig.  11). The bar is quasi stationary after 2.5
$\times$ 10$^9$ years. Its pattern speed slowly decreases (
$\Omega_p(15 \tau_{dyn}) = $ 0.035 and $\Omega_p(50 \tau_{dyn}) = $
0.028 in the units of the code).

The surface density profile of the disk outside the bar region is
exponential along the major axis of the bar and follows a  $R^{1/4}$ law
along its minor axis. It results from some relaxation process produced
by the formation and the evolution of the bar. The velocity ellipsoid is
characterized by $\sigma_R : \sigma_\phi : \sigma_z =$ 1. : 0.85 : 0.75
in the bar and 1. : 0.55 : 0.45 in the disk. The radial
behavior of the velocity
dispersions is shown in fig. 12, displaying in particular a significant
vertical heating.

Repeatedly, the bar is a very efficient engine for
transferring the angular momentum outwards: the total angular momentum
inside the corotation radius decreases by more than 50\% between the
initial time and 5 $\times$ 10$^9$  years.

To the first order, $N$-body bars are similar to observed stellar bars.
The morphology and the cylindrical rotation agree with observations by
Jarvis (1990) in NGC 128 for instance, which is the prototype of peanut
galaxies. The actual fraction of box-peanut bulges is $\sim$ 20\%. Taking
account  of the time scale for forming this structure and the selection
effects due to the difficulty of their detection according to the
inclination of the galaxies, Combes and al. (1990) suggested that any
observed box- or peanut-shaped bulge could be the signature of a strong
bar inside the galaxy.

To the second order, the central parts of $N$-body and observed bars are
different: $N$-body bars show similar ellipticities in the isophotes
whereas most of observed SB0's for instance have round isophotes. An
explanation of the disagreement could be found in the possibility of gas
accretion towards the central parts. In fact an essential ingredient is
missing in the simulations described above: the gaseous component, and
consequently the star formation, which imply energy and angular momentum
dissipation, as well as self-regulating processes which control the
balance between cooling and heating. This problem will be pointed out in
sections 7 and 9.

Let us come to the connection between the box- or peanut-shape and
resonances. It can be understood by studying orbits in the $N$-body
bars, which have the advantage on existing analytical models to be
self-consistent, to have a measurable $\Omega_b$ and to permit to
follow the time evolution. The 2-D orbital structure of $N$-body bars
has been examined by Sparke and Sellwood (1987). For the extension to
3-D structure, Pfenniger and Friedli (1991) used the Geneva code
already mentioned. The position of the resonances can be estimated by
averaging on azimuthal angles the generalized frequencies defined by
Pfenniger (1990)

$$
\Omega' = \f{1}{R} \f{\d \Phi}{\d R}\ , \ \ \
  \kappa'^2 =  \f{\d ^2 \Phi}{\d x^2} +  \f{\d ^2 \Phi}{\d y^2} +
2\Omega'^2 \ , \ \ \ \nu^2_z =  \f{\d ^2 \Phi}{\d z^2}
$$

\noindent
at any point ($x$, $y$, $z =$ 0).

The main new result of the 3-D orbital structure is the existence of
vertical bifurcations, in particular:

\begin{itemize}

\item[-] The direct $x_1$   family of periodic orbits which essentially
sustains the bar has a vertical instability strip (2/1 resonance)
responsible for the box-peanut shape. Examples of periodic and
quasi-periodic orbits supporting the peanut-shape are elongated orbits
along the bar with  $\Omega/\kappa/\nu_z = $ 1/2/2 or eventually 1/4/4
(Fig 13).

\item[-] The usual retrograde family  $x_4$  has a vertical instability
strip (1/1 resonance). From the boundaries of the strip, anomalous
inclined orbits bifurcate. They were previously mentioned (section 4) as
one of the important class of orbits in systems with 3 degrees of freedom.
\end{itemize}

We note that the  $x_1$   and  $x_4$    families are connected by two resonant
families 3/1 and 4/1.

\newpage
\section{Dissipation in disks}

Until recently, apart from studies of the gas response to perturbations
(bar-like for ex.), the dynamical importance of gas on large galactic
scale has not drawn the attention very much, partially because it was
considered as a minor contributor to the total mass of galaxies.
However, because of its dissipative nature, its behavior is
fundamentally different from the stellar one. In the disk evolution, the
role of this tracer of spiral patterns as coupling element with stars
and as fuel for nuclear activity is essential.

In the present section, we begin with a description of various methods
used to introduce dissipation in evolution calculations. Then we will
examine some of the key numerical experiments which allow to understand
a) the spiral structure evolution and the various Hubble types of
spirals, b) the effect of gas adjunction on the disk stability against
the bar formation, c) the gas inflow towards the central region of
galaxies and its consequences. Recent 3-D fully consistent simulations
represent decisive progress about these questions.

One of the key physical processes in disk evolution is the angular
momentum redistribution between the components (see also section 6).
Probably the most effective mechanisms responsible of the loss of angular
momentum by the gas are a) the collisional viscosity due to highly
dissipative cloud-cloud collisions and supersonic velocity, b) the
dynamical friction on massive clouds produced by the background stars,
c) the gravitational  torques  exerted by non-axisymmetric perturbations in
the stellar component (bars or spiral arms) on the gas. Shlosman and
Noguchi (1994) recently have evaluated the relative importance of these
mechanisms and concluded that the dynamical friction can cause a rapid
inflow of gas if much of it is in massive clumps. The efficiency of the
bar driven inflow depends on the bar strength and on the gas mass
whereas the collisional viscosity seems to be unimportant in general. The
development of bar-like structures and their interaction with the gas has
been extensively studied by several groups in recent years and we will
report on some of works in this section.

\subsection{Modelisation of the gas behavior}

Until now, essentially two different types of approaches have been
proposed to follow the gas behavior in numerical simulations of disc
galaxies in evolution:

1) \underline{Ballistic particles}

Cold dense clouds have been considered as ballistic particles with a
finite cross section (Schwarz, 1981 and 1984). The basic principle is to
take a rectangular grid with specified box size overlaid on the disk.
Two gas particles (clouds) collide if they are in the same box and have
a component of relative velocity towards each other. The collision
reverses the relative velocities of the gas particles and reduces them
by some ``restitution coefficient''. Typically the size of the box is of
the order of 0.5\%  of the disk radius. The ``particle graininess''
is a non negligible source of small amplitude initial perturbation: cold
disk are strong amplifiers of small disturbances. But in a real galaxy,
a significant fraction of gas is considered into a large number of
molecular clouds which display a similar level of graininess. Therefore
this approach has been often considered as sufficiently realistic in
spite of the rather ad hoc estimation of energy dissipation in clouds.
It has been used for instance by Carlberg and Freedman (1985), Combes
and Elmegreen (1993) amongst others.

2) \underline{Smooth particle hydrodynamics}

A collection of clouds is considered as a fluid with a sound speed of
the order of  the velocity dispersion of the clouds
(typically 5 - 10 $kms^{-1}$) At present time, for simulations of long-time
evolution, the smooth particle hydrodynamics (SPH) method is often used:
a set of smoothed-out quasi-particles is considered so that the fluid
density at any point is obtained by summation on all particles at that
point. SPH is based on the  Lagrangian description. For a detailed
review on this technique, see for ex. Monaghan (1992) and Benz (1990).
Here a stumbling-block is the perhaps excessive shear viscosity derived
from an artificial viscosity term introduced in the hydrodynamical
equations to simulate the dissipation in shocks. Various solutions have
been proposed to try to control this effect (Friedli and Benz, 1993) by
reducing the artificial viscosity in largely shear flows and leaving it
unchanged in strong shocks.

\subsection{Spiral activity in disks}

A first example will be given here to underline the importance of gas
relatively to the large frequency of observed spiral structures. A
purely stellar disk displays a strong spiral activity which quickly
raises the velocity dispersion of stars and the disk becomes less
responsive to further disturbances. It is a general result of purely
$N$-body simulations. Dissipation and (or) accretion can act as gas
cooling factors. That could be a way to maintain some fraction of the
disk mass at a low velocity dispersion resulting in a quasi-permanent
sensitivity to the growth of spiral disturbances.

This process has been investigated by Carlberg and Freedman (1985) by
adding the Schwarz dissipative collision scheme to the Sellwood $N$-body
disk + halo code (Sellwood, 1981). $N$ = 40000 and a 3:1 stellar to gas
mass ratio were adopted. The initial values of $Q_{star}$ and $Q_{gas}$
were respectively 1.33 and 0.

After 5 revolutions ( $\sim$ 10$^9$ years), it clearly appears that the spiral
structure is much more prominent  in the gas than in the star component
because of the low velocity dispersion of the gas. The morphology
depends on the relative mass of the disk in the sense that the number of
arms is a strong function of the disc-halo mass ratio, as expected from
the swing amplification theory (Toomre, 1981). For a shearing disc, the
critical wavelength  $\lambda_c$  for long wave instabilities is

$$
\lambda_c = \f{4\pi^2 G \Sigma(R)}{\kappa^2}
$$

Let us assume the rotation velocity $V_0 = cte$, then $\kappa
=\f{\sqrt{2}V_0}{R}$, $\Sigma(R) =\f{f V_0^2}{2\pi G R}$ where $f$ is the
mass fraction in the disk. The number of arms $m$ in the spiral pattern
is connected to  the azimuthal wavelength   $\lambda_\phi$ by

$$
\lambda_\phi = \f{2\pi R}{m}
$$

The peak amplification corresponds to $\lambda_\phi \sim 2\lambda_c$
(Toomre, 1981).

{}From these relations, it comes

$$ m = \f{1}{f} $$

In a pure stellar disc, the trailing spirals transfer angular momentum
outwards, according to Lynden-Bell and Kalnajs (1972). In the
simulations here described, the gas systematically loses angular
momentum to the stars and a radial inflow is settling with a time scale
of the order of an Hubble time. We will find back later this key problem
of angular momentum transfer in experiments dealing with bars.

\subsection{Bar-disk evolution in early and late-type spirals}

In the past, $N$-body simulations have been essentially devoted to
studies of disk stability and growth of modes without consideration for
the observed distinct morphology of Sa to Sd galaxies. However, one of
the aims of galactic studies is to explain the different Hubble types
and eventually the evolution  from one type to another. For example
differences between early and late-type barred spirals, such as
emphasized by Elmegreen and Elmegreen (1985, 1989) will have to be taken
into account. Some properties of bars in these different types are
summarized in the table 2. These are general tendencies that nevertheless admit
exceptions.

\begin{table}[h]
\begin{center}
\caption{Properties of bars in early- and late-type galaxies}

\vspace{4mm}

\begin{tabular}{|l|c|c|}
\hline
                            &  Early-type        & Late-type \\
\hline
Bar perturbation             &   strong           &   weak    \\
Length of the bar/$R_{25}$   &   long             &  short     \\
Extension of bars           & corotation ?        & well inside
corotation(ILR?)\\
Density of bars             &  exponential profile    & exponential profile\\
                            &   flatter near the center     & \\
Spiral morphology          & grand design, 2 arms& multiarms or
flocculent structure\\
Spiral amplitude            &radially decreasing &radially increasing\\
\hline
\end{tabular}
\end{center}
\end{table}

In order to understand the origin of the differences mentioned in the
table, Combes and Elmegreen (1993) studied by numerical simulations the
formation and the evolution of bars in $N$-body systems having
characteristics of early and late-type galaxies (essentially large and
small bulges respectively). They also examined the gas behavior. The
stars are initially in a Toomre disk ($Q = const.$) and the gas distributed in
an exponential disk. The bulge-spheroid component is a Plummer sphere.
The mass ratio $M_B/M_D = $ 2 for early-type system and 1/10 to 1/5 in
late-type ones. The main evolution phases observed in the case of purely
stellar system are the following:

\underline{For late-type systems}, at the beginning,  development of a
small-amplitude $m = $2 wave with $\Omega_p \gg \Omega -
\f{\kappa}{2}$ which transfers angular momentum from the stars in the
inner disk to the stars in the outer disk; then growth of a bar from
the center which traps particles (stars) at larger and larger radii.
After $\sim$ 10$^9$ years, the bar ceases from growing because
corotation overtakes regions with not enough particles to receive
angular momentum. The length of the bar is constrained by the scale
length of the disk. The value of $\Omega_p$ ($\sim$ 10 $kms^{-1}
kpc^{-1}$) is of the order of $\Omega - \f{\kappa}{2}$.  In this case,
the bar seems to end inside corotation. This limitation comes from the
fact that the bar is unable to transfer angular momentum in low
density region.

\underline{For early-type systems} we expect that the disk is more
stable. The evolution begins similarly to the late-type case, but
$\Omega_p$ is initially much larger ($\Omega - \f{\kappa}{2}$ is
larger for galaxies with a large bulge). The corotation radius is
sufficiently short so that angular momentum from the inner disk finds
a reservoir for transfer up to corotation during the slow growth in
length and strength of the bar, for over a Hubble time. An other
difference obtained between late and early type systems is the density
distribution in the bars: the density profiles are exponential but,
for early types, they are flatter near the center.

The gas is expected to be important in the evolution of spiral waves and
bars: it dissipates the wave energy and amplifies the spirals through
gaseous self-gravity. Particularly in late-type galaxies, the mass as
well as the angular momentum fraction of gas is relatively high. This
gas is a reservoir for the angular momentum transferred out by the wave.
Combes and Elmegreen (1993) suggest that this angular momentum absorption is
responsible for a prolonged bar evolution by comparison with the case of
a purely stellar disk. The growth phase for the bar is also larger.

The main result concerning the late-type simulation is the appearance of
a very thin bar in the gas with the same length as the fat stellar bar.
More structure is present in the gas than in the stars, with transient
waves. The gas is strongly driven inwards by the gravitational torque
from the bar. The end of the bar coincides with one disk exponential
scale length (Fig. 14). In the case of early-type simulation, the
stellar bar extends nearly up to corotation and a thin bar develops in
the gas as well as a nuclear ring at the ILR.

An important difference between simulations with and without gas appears
in the time scale for the setting-up of the bar: 2 $\cdot$ 10$^8$ years
for a late-type galaxy and  5 $\cdot$ 10$^8$ years for an early type
one. That is much faster than for purely stellar cases which can be
stable with $Q = $ 1.5 for a Hubble time. This result is in agreement
with predictions by Jog and Solomon (1984) mentioned in sect. 5,
according to which a gas-star system can be unstable even if one of both
components is stable by itself. The fact that the dissipation maintains
$Q =$ 1 for the gas explains the evolution described above.

\subsection{Bar driven gas fueling of galactic nuclei}

As indicated in 7.1, another effective method to describe the behavior
of gas in galaxies is based, via the classical hydrodynamic equations,
on smooth particle hydrodynamics (SPH) codes. Such a code has been
used by Friedli and Benz (1993) (see also Friedli et al.  (1991)).
These authors investigated the ability of bars to transport gas
towards the central parts of galaxies. In particular, they looked for
what parameters at work are the most important in this context. A
fully consistent 3-D simulation of stars/gas systems consists in 3
main steps:

\begin{enumerate}

\item Given the mass distribution for gas and stars, to compute the
gravitational potential $\Phi$ by using the part of the code consisting
in a particle-mesh (PM) algorithm in Fourier space on a polar grid. This
fast method allows to include a large number of particles (1 to 5 $\cdot$
10$^5$), a fraction of them being gaseous ($\sim$ 10\%).

\item Using the SPH approach, to compute the hydrodynamical forces
acting on the gaseous component only. At the present time, the
modelisation of this component does neither take account of the variety
of the interstellar matter composition (diffuse, atomic, warm and
cloudy, molecular, cold components) nor of its fractal structure. It
would be illusory to introduce complexity at this level as far as the
permitted numerical resolution prevents to describe the ISM in fine
details.

\item To move all the particles by using for instance a second order
Runge-Kutta-Fehlberg integrator.
\end{enumerate}

The main results from this fully 3-D self-consistent evolution confirm
the fact previously suggested by Shlosman et al. (1989) that a bar
acting on the gaseous component of an isolated galaxy allows a transport
of mass up to several 10$^9 M_\odot$ towards the center with a mean rate
of some $ M_\odot yr^{-1}$ but reaching roughly 200 $ M_\odot yr^{-1}$
in extreme situations. On average more than 1\% of the total mass of the
galaxy in the form of gas could be brought near the nucleus. The mass
accretion is increased  1) if the bulge/disk mass ratio is lowered and
2) if the axis ratio of the stellar bar $a/b$ is larger. The accretion
is smaller by a factor 4 for gas in retrograde motion compared with
direct motion. Finally the gas self-gravitation reinforces the accretion
only when the gas density becomes high enough.

Different experiments  by Friedli and Benz (1993) lead to the conclusion
that the gaseous bars
are always induced by barred stellar potentials. No gaseous bar is
formed spontaneously in models with fixed axisymmetric stellar
potential. Moreover the gaseous bars are always shorter than their
stellar counterparts and their axis ratio $b/a$ is smaller by a factor 2
- 3. Fig. 15 indicates the time evolution of the central mass concentration
inside the radius  $R_L = $ 0.2, 0.5 and 1 $kpc$ for the gas and the stars, for
one of their models.

The mass accreted can represent a small but non-negligible part of the galactic
mass (1 - 2\%). Consequently a strong ILR can settle, which causes the
main elongated plane orbits which support the bar to be replaced by
orbits perpendicular to the bar which is no longer sustained. In a few
Gyr, the initial bar can be transformed into a triaxial bulge such as
observed in many spirals. These bulges could be the relics of old bars
(see also section 8).

The dynamical effects of star formation are not included in this
scenario. It is a new scaling in complexity. At present time, star
formation cannot be rigorously modeled as an additional equation in the
present context. We will come back in section 9 on recipes which allow
to mimic this part of the global evolution scenario.

\subsection{Bars within bars and their effects}

The reality of the peculiar phenomenon of bars within bars has emerged
from the following three complementary approaches:

1) Secondary nuclear bars or triaxial bulges have been observed in
visible light by different authors in several barred spiral or
lenticular galaxies, for example NGC 1291 (de Vaucouleurs 1974), or
NGC 3945 (Kormendy 1979, 1981, 1982); in addition, recent CO
observations have also revealed central molecular bars or rings not
aligned with the stellar bars (Kenney 1991; Devereux et al. 1992, see
also Lo et al. 1984). Galaxies with a nuclear ring eventually crossed
by a secondary bar were studied by Buta (1990). The prototype could be
NGC 3081.

2) Theoreticians have proposed bars within bars as a possible mechanism
of gas fueling in active galactic nuclei (Shlosman et al. 1989, 1990).
The possible existence of self-consistent models of periodically
time-dependent stellar systems has been confirmed by Louis and Gerhard
(1988) and Sridhar (1989).

3) Self-consistent 3D $N$-body simulations with gas and stars
(Friedli, 1992) can lead to the formation of systems with two bars
rotating with two different pattern speeds provided that the initial
gas mass amounts to about 10\% of the total mass (Friedli and
Martinet, 1993). This emphasizes the key role of dissipation. The
secondary bar results from a decoupling between the nearly
self-gravitating central part and the outer part of the primary bar.
The decoupling seems to be sensitive to mass accumulation onto the
$x_2$ orbit families of the primary bar resulting from the presence of
a moderate ILR. Either the primary bar first appears followed by the
secondary one, or the two bars almost simultaneously develop. In the
latter case, a nuclear gaseous ring is in general formed near the
secondary bar end.

The best model has shown that these double-barred systems are stable
over more than 5 turns of the secondary bar. It has the following
characteristics:  $\alpha = \Omega_s/ \Omega_p \approx $ 79.0/25.5
$kms^{-1} kpc^{-1} \approx $ 3.1, $\beta = l_p/l_s \approx $ 9.5/2.5 $kpc$
$\approx$ 3.8, and the axis-ratios are $a_s \approx $ 0.6 and $a_p \approx
$ 0.7 (Fig. 16).

Bar-driven gas fueling can be significant and can compress gas inside
the bar to about 10 - 20\% of the bar length. The system of nested bars
thus transports amounts of gas much closer to the galactic center and
could be invoked as a possible mechanism to fuel AGN. The gas accretion
rate can be very high and could be related to some starburst mechanism.
However, the efficiency of star formation must be explored before
definite conclusions can be drawn.

The two bar phase is followed by the dissolution of the secondary bar
(or even the two bars depending on the model) which is induced by a
broad and strong ILR. The final central galaxy shape resulting from the
two bar destruction is similar to triaxial observed bulges, suggesting
that some of them are relics of destroyed bars.

The above scenario is different from that proposed by Shaw and al.
(1993). Both scenarios need the presence of a massive and strongly
dissipative component but in the one described above it is supposed to
initiate the central dynamical decoupling while in Shaw et al., it
is only supposed to gravitationally deform the stellar component. As a
major consequence, in our scenario, the two stellar bars have different
pattern speeds whereas in theirs they have the same. Since theoretical,
numerical as well as observational studies seem to indicate that single
and multiple pattern speeds can occur in galaxy dynamics, it is
difficult to dismiss either of these scenarios, and consequently to
determine if only one (and which one?) or both are in action in real
double-barred galaxies. However, the existence of secondary
 trailing stellar bars seems difficult to explain with Shaw and al.'s
scenario. A mixing up can also eventually follow from the existence of
both central twisted isophotes (resulting either from projection effects
onto triaxial bodies or from torques initiated by strong spiral arms)
and two distinct bars. Shaw et al.'s scenario may eventually better
correspond to the case of twists than the one of two distinct and
persisting bars. Clearly, more studies are necessary to assess the
respective merit of these two scenarios as well as to determine the
essential parameters  which initiate the models to be channeled or not
into the way of the multiple pattern speeds.

\newpage
\section{Environment effects. Accretion of satellite \protect\newline
galaxies}

The present review essentially turns on the internal dynamical
evolution of galaxies. The previous section displayed important
effects able to produce morphological modifications of galaxy
structure only due to interactions between components of the system
considered, including the gas. In the discussion of the processes at
work, it was implicitly admitted that the systems were isolated.
However, at present time, nobody contests that most galaxies must
suffer more or less strong interactions with their neighbours during
their lifetimes. The proceedings edited by Wielen (1990) reviewed
various problems concerning these phenomena. Visible effects of
interactions such as bridges, tails, warps, shells, ripples, dust
rings, multiple nuclei, more or less advanced stages of merging are
the subject of continuous attention. The interest for these questions
is strengthened by recent qualitative evidences of a connection
between interactions and bursts of star formation (see section 9).

In our own context, we will be here interested in the morphology
transformations of disks and bulges resulting from accretion of small
satellite galaxies. Beforehand, we will briefly mention the case of
strong destroying collisions between two similar disk galaxies, the
result of which could account for starbursts which will be discussed in
section 9.

\subsection{An example of strong interaction between galaxies}

Strong interactions can be defined as those in which galaxy number is not
conserved. They have been recently discussed especially by Barnes and Hernquist
(1991, 1992) and Barnes (1992). In particular these authors presented a
simulation of merger of two self-consistent bulge/disk/halo galaxies of
equal mass including 10\% of gas. The mass ratio of the components is
1:3:16 in both galaxies. Two fundamental questions about such simulated
parabolic collision are: 1) What is the morphology of the remnant? 2)
What is the fate of gas?

Using a SPH algorithm to describe the gas behavior, Barnes and
Hernquist (1991) concluded the following:

\begin{itemize}
\item[-] Mergers such as considered in the frame of the chosen peculiar
initial conditions produce a triaxial slowly rotating remnant with a
$r^{1/4}$ luminosity profile.\\[-7mm]

\item[-] The interaction transfers a large fraction of the energy of the
relative motion to the internal degrees of freedom. The ``passage''
leads to a more tightly bound orbit, then to a complete merger in $\sim$
750 millions years.\\[-7mm]

\item[-] Gas and disk stars have a similar behavior on large scale but not
on small scale: the strong tidal field triggers a strong stellar bar.
The gas response is similar but dissipative shocks appear. The gaseous
bar is narrower and shorter.\\[-7mm]

\item[-] Half the gas initially in the disk is transferred toward the center
of the merger by purely gravitational torque (5 $\cdot$ 10$^9 M_\odot$
within $\sim$ 200 $pcs$).\\[-7mm]

\item[-] Strong  shocks  and radiative cooling are a necessary condition for
such a
process.\\[-6mm]

\end{itemize}

We refer the reader to the review mentioned above for the
details. An analogous mechanism has been described in section 7 in
connection with the bar-driven gas fueling of nucleus in an isolated
galaxy. In both cases the final evolution is uncertain due to the fact
that star formation and supernova heating have been ignored in the
experiments.

Let us add that some peculiar initial conditions in other simulations
(Pfenniger, 1994) are able to produce counter-rotating disks similar to
that observed in NGC 4550 for ex. (Rix et al., 1992)

\subsection{Coming back to dynamical friction}

Dynamical friction mechanism has been described in section 6. Orbits of
satellites around massives galaxies decay due to such an effect so that
galaxies could have accreted some mass in the form of dwarf galaxies or
globular clusters over a Hubble time. Even if the implied mass quantities
are not so important as in collisions between giant galaxies, the
resulting deformation of the target can be not negligible as we will see
in the next subsection.

Three fundamental physical aspects of the problem must be mentioned to
recall that the well known simplifying Chandrasekhar formula cannot claim
to give a quantitatively correct estimate of the deceleration due to
dynamical friction in the present context:

1) The self-gravity of the target.

2) The global nature of the target tidal deformations.

3) The non-rigidity of the satellite.

Combes (1992) draws attention to the dissociation of two effects
connected with the self-gravity: the barycenter displacement of the
``target'' galaxy and its true deformation. Conflicting results as
regards to the role of the self-gravity in the satellite braking come
from the different way of calculating or interpreting it. If the true
deformation, which is the only deceleration factor, is computed in the
center of mass frame, it appears that the self-gravity is negligible for
a satellite-target mass ratio $\sim$ 0.1  (Prugniel and Combes, 1992). If
the target is fixed in an inertial frame (Weinberg, 1989), taking the
center of mass shift as a self-gravity factor, it is found that
self-gravity slowers the decay! The Chandrasekhar formula is also based
on a fixed  target and also overestimates the friction: its local nature
is unrealistic in the present context since the tidal deformation of the
target which induces the dynamical friction may be considered as global.

Moreover, the tidal deformation of the satellite which is able to take
away some orbital energy and angular momentum of the relative motion
must not be neglected: the decay time scale is appreciably reduced with
respect to the case of a rigid satellite as seen  in
 Prugniel and Combes (1992).

These considerations have been inferred from experiments on an
elliptical galaxy and its companion. Combes (1992) also mentioned a
possible complication with spiral galaxies coming from resonance effects
which could remove the satellite instead of braking it. In fact
simulations show that the friction remains dominant in this case but
self-gravity typically reduces the decay time by a factor of two.

\subsection{Heating of the disks by  satellite mergers. An explanation
for thick disk?}

This mechanism has already been mentioned in section 5 amongst those able
to heat the disk in our Galaxy. Certainly it is quite general and we
have discussed it in more details by basing ourselves on the deepest
treatment of the problem nowadays (Quinn et al. 1993), in spite of some
limitation given below.

The used $N$-body models consists of self-consistent bulge-less
exponential disks having constant vertical scale length $h_z$.

These disks are imbedded in rigid modified isothermal halos, the density
of which is

$$
\rho(r) = \f{\rho_0}{1+ (r/\gamma)^2}
$$

The density distribution in satellite galaxies is given by the spherical
Jaff\'e models

$$
\rho(r) = \f{M}{4\pi r_0^3} \left( \f{r}{r_0} \right)^{-2} \left( 1+ \f{r}{r_0}
\right)^{-2}
$$

\noindent
with the mass ratio $M_{sat}/M_{disk} = $ 4 to 20\%.

The satellites are initially placed on circular prograde orbits at a
radius $6 h_R$ ($h_R =$ exponential disk length scale), with
various inclinations relative to the disk plane. Friction due to the
halo is ignored.

The main limitations of such experiments, which could lessen
the bearing of the results are the following: 1) the rigidity of the
halo does  not allow to give quantitatively reliable sinking rates, 2)
only prograde encounters are considered, 3) the gas effects are not
included, 4) only 32768 particles in the disk and 4096 in the satellite
are used.

The different mergers obtained are characterized by various common
effects:

\begin{itemize}

\item[-] The dynamical coupling between disks and satellites leads to satellite
orbital decay, at first by sinking into the plane of the disk, then
radially into the center.\\[-7mm]

\item[-] A strong two-armed spiral pattern is generally induced in the disk
which
transports angular momentum outwards.\\[-7mm]

\item[-] The disk radially spreads and perpendicularly inflates.\\[-7mm]

\item[-] As the satellite sinks into the central region of the galaxy, the disk
is able to find back a new axisymmetric equilibrium.\\[-7mm]

\item[-] The disk is thickened, warped and flared by the merger. The
quantitative
effect depends on the initial conditions, in particular the mass ratio
and the initial inclination of the satellite orbit. Typically for a mass
ratio of 1/10 and an inclination of 30$^\circ$, the vertical scale height
$h_z$ inside $h_R = $ 3.5 $kpc$  has increased by 50\%.\\[-7mm]

\item[-] Moreover, the mergers perturb the velocity distribution of stars in
the
disks due to three effects: a) the deposit of kinetic energy and
angular momentum from satellites, b) the action of spiral pattern and c)
the accretion of satellite debris.\\[-5mm]

\end{itemize}

In the typical case mentioned above, the ratio of azimuthal to radial
velocity dispersion $\sigma_{\phi}/\sigma_R$ varies from 1 in the center to
0.7 between 10 and 20  $kpc$.  $\sigma_z/\sigma_R$  varies from 0.75 in
the center to values larger than 1 for $R >$ 15 $kpc$. The initial range
of $Q$ is 1 $< Q <$ 2 for  0 $< R <$ 15 $kpc$ which allows a response to
the mode $m = 2$ (spiral arms). At the end of the simulation ($T = $ 5.4
$\cdot$ 10$^9$ years), $Q(R < 2h) \sim $ 1.5 and, $Q(2h < R < 4h) \sim $
2-4 which is able to stabilize the disk against the mode $m = $ 2.

The recurrence of mergers shows less spectacular effects: the disk is
hotter and more diffuse than the original thin disk! The evolution of
the satellite when it is represented by a realistic self-consistent
$N$-body model is characterized by a stripping due to the tidal field and
an impulsive heating.

There is competition between internal heating of the satellite and
external heating of the disk. It appears that if the satellite is denser
than the disk, the disk is heated impulsively whereas if the satellite
is less dense, it is heated by the disk. Of course the stripping of a
self-consistent satellite and the disk tidal field control the
efficiency of the disk heating by the satellite. Typically, in the model
considered as standard by Quinn and al. (mass ratio = 1/10,
30$^\circ$ inclination) a significant $z$-heating is obtained with 60\% of the
satellite kinetic energy transferred into $z$ motions of disk stars.

On the other hand, in the same experiment, the kinetic energy of the
disk increases by 30 - 40\% of the satellite orbital energy through the
spiral response of the disk to the satellite perturbation.

The flare in the disk is formed by satellite debris and stars from the
disk. It seems to be due to the fact that the velocity impulse is especially
in the vertical direction at large radii, resulting there in a more
efficient absorbtion of vertical kinetic energy than of azimuthal kinetic
energy by the disk.

In brief, a thin disk having suffered satellite impacts must present
the observed features already mentioned: radial spreading, thickening,
vertical flaring and warping. Many galaxies display these features. See,
for instances Shaw and Gilmore (1989) for thickening, Bosma and Freeman
(1991) for flaring, Sanchez-Saavedra and al. (1990) for warping.

Although the final structures obtained by the mechanism here described
are  similar to those observed in numerous galaxies, we must
remember that other mechanisms of disk heating could be efficient, in
particular the bars. The relative quantitative contribution of all
mechanisms put forward here and in section 6 to the evolution towards the
observed structures of disk is difficult to estimate: the simulations
which reveal the dynamical processes at work suffer from limitations
mentioned repeatedly and sufficiently detailed observations of external
galaxies concerning particularly kinematics and multi-band photometry
are still lacking.

It is interesting to note that the accretion of satellites by spiral galaxies
is
a process similar to that suggested by Searle and Zinn (1978) for the
halo formation from subsystems such as dwarf spheroidal galaxies.

\subsection{Bars induced by companion passages  and dissolved \protect\newline
by
  satellite mergers}

a) \u{Influence of interaction on stellar bar evolution}

In section 5, the conditions of disk instabilities to bar formation were
examined in the case of isolated galaxies. Since the seminal paper by
Toomre and Toomre (1972), there are numerous evidences that tidal
interactions can have an effect on the developments of substructures
inside the galaxies. Therefore it is important to bring out the
dominating features in the interplay of bar formation and interactions.
To date, the most complete study was undertaken by Gerin and al. (1990)
who simulated  close 2-D and 3-D encounters between a disk galaxy
formed of a Toomre disk and a rigid Plummer halo and a perturber
(Plummer sphere) on direct or retrograde orbit (in the sense of the
galaxy rotation or in the opposite sense). Various perturbing masses,
pericenter distance and mass distribution of the galaxy were tried as
determinant parameters. Two different situations were considered: a)
the ``target'' galaxy is initially axisymmetric and b) it is initially
already strongly barred.

Amongst the results, we point out the following facts:

\begin{enumerate}

\item A direct tidal interaction destabilizes a disk which otherwise does
not develop a bar for more than 1 Gyr. An interaction reduces the time scale
 of bar formation. Retrograde encounters are not so effective. The
process obviously depends on the mass ratio  $M_{halo}/M_{disk}$.

\item In the perturbed case (compared to the isolated case), the
interaction triggers a spiral wave in the outer parts of the disk, which
propagates towards the central regions, strengthening the bar formation.

\item In the case of an already initially barred galaxy, the phase of the
perturber acts on the strength and the pattern speed of the bar,
practically without change of the axisymmetric part of the galaxy
potential. Consequently an interaction may shift the Lindblad Resonances
in a stellar disk.

\item Bending and fattening of the disk are essentially produced by
interactions with a companion perpendicular to the plane of the galaxy.
The increase of the bar strength is less pronounced than in the other
geometries due to the shorter time during which the axisymmetric
perturbing force acts.
\end{enumerate}
b) \u{Destruction of bars by satellite mergers}

The influence of the mass concentration growth on the dynamics of disks
in isolated galaxies has been discussed in section 4 (effect of a
blackhole on orbits plunging towards the center) and in section 7 (gas
inflow towards central regions induced by a bar). In both cases mentioned,
the process implies the creation of an ILR and the enlargement of the
influence zone of the resonance with the growth of the compact mass. The
response of stellar orbits in the plane is perpendicular to the bar.
Consequently the bar shape is no longer supported. Starting from the
$N$-body simulation of a robust bar, Friedli and Pfenniger (1991) and Friedli
(1994)
 confirmed the effect mentioned. Artificially increasing slowly
the central mass of the disk, they observed that the addition of 1-2\%
of the total mass in the region inside the ILR is enough to destroy the
strong bar.

But a bar can also be dissolved after being induced as a consequence of
an environment effect in the form of satellite merger. The experiment
described below is complementary to the Quinn et al. (1993) study
presented in the previous subsection, which did not imply a bar
evolution.

Pfenniger (1991a) showed the dynamical evolution of a disk perturbed by a
compact satellite, initially on an inclined nearly circular direct
orbit, with a mass equal to 10\% of the system total mass. Induced by
the tidal action of the perturber, a bar is formed in 100 Myr, and
produces spiral arms and disk heating. Simultaneously the satellite
spirals by resonance and dynamical friction coupling effect. It reaches
the bar Lagrangian points respectively at $T =$ 1200 Myr, then at $T \simeq
$ 1300 Myr. Afterwards it plunges toward the center in 50 Myr and the
bar is destroyed according to the process above mentioned (see fig 1. in
Pfenniger (1991a)).
The final structure is a small spheroidal bulge. During the spiraling
of the satellite towards the center, there is transfer of its orbital
angular momentum to the disk which gets warped.

Inasmuch as the bar formation is easy and their observed frequency is
high, it is not impossible that the barred state be a compelled step
in the galaxy life. There is apparently no problem to explain that
$1/3$ of disks are unbarred since efficient mechanisms seem to exist
to destroy the bars!

\subsection{Bulge growth induced by satellite}
The bulge-to-disk ratio is a qualitative as well as a quantitative
classification criterion of galaxies along the Hubble sequence (cf.
Simien and de Vaucouleurs, 1985). In spite of a large dispersion in the
relation between this ratio and the Hubble type, early-type galaxies
generally have the largest bulges. Bulge-less galaxies are also
observed. Is this diversity a signature of initial conditions or the
clue of secular evolution? The simulations presented elsewhere in this
review (6.5, 8.4) supply evidences in favor of the latter possibility.
Small bulges could result from disk material accretion through the bar
dissolution process. There we present an additional argument for secular
evolution, which concerns the galaxies with big bulges, the well known
prototype of which is the Sombrero galaxy.

A possible explanation for the existence of these bulges has been
recently proposed by Pfenniger (1993). Considering a model which
consists in an initial disk with a scale length $h_R = $ 4 $kpc$ and a
scale height $h_z = 1 kpc$, as already described in 6.5. The disk was
initially slightly unstable to bar formation and a peanut-shaped
bulge-bar structure was formed. A 3-D simulation has been started in
which ten point mass satellites initially at the outskirts of the galaxy
are slowly accreted by dynamical friction. When the ratio of satellite
mass to galaxy mass $M_s/M_G$ is of the order of 4\%, the satellites modify
the structure of the bar which becomes more oval. If this ratio is
increased to 10\%, the bar is destroyed and the disk is so much inflated
by heating that the remaining structure is dominated by a big bulge
where all mass comes from the disk. If $M_s/M_G >$ 10\%, the disk is
destroyed, the remnant is a purely spheroidal system.

We must note that these simulations as well as those of the previous
subsection suppose that the satellites are rigid (point masses). It has
been mentioned in 8.3 that there could exist a competition between
internal heating of the satellites and external heating of the disk, when
the satellites are presented by realistic self-consisting models.
Therefore the mechanisms described above and in 8.4 can be really
effective only if the satellites are denser than the disk.

\newpage
\section{Coupling of dynamics and star formation}

As seen in section 7, effective numerical codes now exist which allow
to follow the dynamical evolution of galaxies consisting of stars and
gas.  However an essential ingredient is still lacking: star
formation. Theoretical as well observational primordial questions
about it are: Where does it occur? What is its rate? What are the main
parameters which govern the star formation efficiency?

At present time, only partial answers can be given, inasmuch as the
complexity and the diversity of processes at work is high. Silk (1992)
draws up an impressive list of mechanisms at work in star formation
which ought to be ideally taken into account. I quote: `` Ionization
and heating by X-rays, cosmic rays and UV-radiation, growth and
destruction of dust grains, absorption and photo-desorption of
molecules from grain surfaces, gas phase and surface molecular
chemistry, cooling transitions by molecular and atomic species,
molecular photodissociation, recombination of electrons and ions,
radiative transfer of molecular lines important for cooling, heat
input via external shocks, turbulence dissipation, Alfven wave
dissipation, internally and externally generated radiation fields and
feedback from proto-stars via bipolar flows, winds and flares,
dynamics of cloud contraction, collapse and fragmentation, role of
magnetic fields in cloud support and ambipolar diffusion, without
speaking of initial mass function origin''.

In view of the extremely complicated nature of local star formation
problems, global approaches involving a not too large number of
parameters must be adopted. A minimum task should be: 1) to find a
collapse criterion in order to determine where the stars are formed,
2) to introduce a parameter connected with the star formation
efficiency and 3) to introduce other parameters which allow to take
account of thermal and mechanical energies injected in the medium
through supernovae explosions and stellar winds. The present
discussion will focus on some recipes recently introduced into
simulations of galactic evolution for lack of a detailed theory of
star formation. Simple models will allow to tackle the problem of the
gas fate in central galactic regions and the connection between galaxy
interactions and observed big starbursts, for instance.  Proceeding
step by step, it will be possible in a near future to introduce in the
models more detailed physics specifying the conditions transforming
gas into stars inside molecular clouds.

\subsection{Observational tracers of star formation}

The observational tracers generally used to infer star formation rates
(SFR) are $H_\alpha$, far-UV and far-IR fluxes. The calculation of SFR
results from the luminosity $L_i$ in the band $i$ by

$$
L_i = SFR \int t_i(m) L(m)\phi (m) dm / \int m \phi(m) dm
$$

\noindent
where $L(m)$ is the luminosity of a star of mass $m$, $\phi(m)$ is the
initial mass function (IMF) in the range of mass 0.1 - 100 \Mo and
$t_i(m)$ is the characteristic time scale over which a star of mass
$m$ emits radiation in the band $i$.

The main source of systematic errors in the interpretation of the
$H_\alpha$ emission is extinction (Kennicutt, 1989). Typical
extinctions of HII regions in galaxies are of the order of 0.5 to 2
magnitudes. In so far as corrections for this effect can be applied,
the fluxes can be converted to current star formation rates.
Uncertainties on the Initial Mass Function are another origin of
errors. According to Kennicutt (1983)

\vspace{3mm}

$$
SFR_{H_\alpha}(\gsim 10 \Mo\!\!) = 1.42 \times 10^{-42} L(H_\alpha)\Mo\!yr^{-1}
$$

\noindent
and

$$
SFR_{H_\alpha}(\mbox{total}) = 8.9 \times 10^{-42} L(H_\alpha) \Mo\!yr^{-1}
$$

\vspace{3mm}

\noindent
where $L(H_\alpha)$ is the $H_\alpha$ luminosity obtained from the
$H_\alpha$ flux $ F(H_\alpha) $ [$erg \ cm^{-2}s^{-1}$] by

\begin{center}

$L(H_\alpha)$ = 3   $\times$ 10$^{16} F(H_\alpha)  D^2 [L_\odot]$
\end{center}
$D$ being the galaxy distance in unit of $Mpc$. The total SFR is a
strong function of Hubble type as mentioned in section 2. Integrated
over all stellar masses it was estimated by Kennicutt (1983) and
Caldwell et al. (1991) from $H_\alpha$ data: From 0.1 - 1
\Mo$\!\!yr^{-1}$ in SO/a galaxies to $\sim$ 10 \Mo$\!\!yr^{-1}$ in
Sc-Irr galaxies. Overall it ranges from less than 0.001
\Mo$\!\!yr^{-1}$ in extreme dwarf galaxies to more than 100
\Mo$\!\!yr^{-1}$ in luminous starbursts.

The same kind of difficulty connected with the extinction is present
in tentatives to estimate SFR from UV datas (Donas et al. (1987),
Bersier et al. (1994)). The current star formation rate inferred from
UV data (2000 \AA), obtained with the SCAP balloon experiment, is


$$
SFR = 2.3  \times 10^{-40} L(\lambda 2000) \Mo\!yr^{-1}
$$


\noindent
which represents a recent star formation rate on a period of time
($\sim$ 100 Myr), about 50 times longer than SFR based on $H_\alpha$.

{}From the appendix of Catalogued Galaxies and Quasars in the IRAS
Survey (1985), the IRAS 40 - 120 $\mu$m luminosities are, according to
Devereux and Young (1990)


$$
L(40-120\mu m) = 3.65 \times 10^5 (2.58 S_{60} + S_{100}) D^2 [L_\odot]$$

\noindent
$ S_{60}$ and $S_{100}$ are the IRAS 60 and 100 $\mu$m flux density in
unit of $J_y$. The resulting star formation rates given by these
authors are

$$
SFR_{IR}(\geq 10 \Mo\!\!) = 9.2 \times 10^{-11} L_{IR} \Mo\!yr^{-1}$$


$$SFR_{IR}(\geq 0.1 \Mo\!\!) = 6.3 \times 10^{-10} L_{IR} \Mo\!yr^{-1}$$

However Sauvage and Thuan (1994) discuss the ambiguity of the
information on star formation inferred from far-IR measurements in
particular through the 4 IRAS pass-bands (12, 25, 60 and 100$\mu$m).
The problem rests on the fact that the far-IR emission can have a
multiple origin: 1) Dust associated to molecular clouds and
star-forming regions heated by the radiation of newly formed OB stars,
2) dust in the interstellar medium heated by an interstellar radiation
field (cirrus), 3) photospheres or circumstellar envelopes of evolved
stars, 4) dust heated by a compact source inside an active galactic
nucleus. The present and past SFR, the chemical composition and the
size distribution with respect to the heating sources, the optical
depth of the dust play a significant role in the emission mechanisms
mentioned above.

In the extensively used FIR color-color diagram ($\log(S_{60}/S_{100})
$ versus $\log (S_{12}/S_{25})$ where $S_i$ represents the flux in the
4 IRAS bands), galaxies populate a relatively narrow sequence from the
high $S_{12}/S_{25}$ and low $S_{60}/S_{100}$ end to the low
$S_{12}/S_{25}$ and high $S_{60}/S_{100}$ end. Typical representatives
of quiescent spiral disks and starbursts would be respectively in the
lower right end and in the upper left end of the sequence. According
to Sauvage and Thuan (1994), the FIR colors of galaxies along the
Hubble sequence are a combined effect of the star formation efficiency
(SFE) and variations in dust distribution and composition. Only for
Hubble types from Sbc to Sdm-Im, the FIR colors seem to be essentially
controlled by an increasing SFE from quiescent to starburst objects.
If the major part of the 60 - 100 $\mu$ emission is associated with
regions of current star formation for these types, the IR-to-blue flux
ratio can be used as an indicator of relative present (IR) to past
(measured by B passband) star formation rate. According to Keel
(1993),

$$L_{FIR}/L_B = 4.16 \times 10^{-7} \cdot 10^{0.4B} (2.58 S_{60} +
S_{100})$$

\noindent
for blue B magnitude.

\subsection{Star formation rate and gas density}

The Schmidt law (Schmidt, 1959) $\dot{M}_s \div \rho^n_g$, where $M_s$
is the stellar mass and $\rho_g$ the gas density, has abundantly been
used in the past to modelize the average star formation rate in
galaxies. The observational fits to this law have displayed a large
dispersion (0$ < n < $4), also present in the applications of the law
to different regions of a same galaxy. Clearly, a part of the physics
associated to the star formation process is neglected. Taking account
of the numerous processes at work, one could have formalized SFR as
$SFR(\rho_g, c_s, w_s, V_T/c_s, \Omega, A, |\vect{B}|, Z, \rho_s)$
(Lynden-Bell, 1977) where $\rho_g = $ gas density, $c_s = $ gas sound
speed, $w_s = $shock frequency, $V_T/c_s = $ shock strength, $ \Omega
=$ gas rotation, $A = $ shearing rate, $|\vect{B}|$ magnetic field
strength, $Z = $ gas metal abundance, $\rho_s = $ star density.
Unfortunately, it is difficult to make this dependence quantitatively
explicit!

The tracers introduced in the previous subsection can be used with
recent data on HI and H$_2$ galaxy contents to refine the relation
between SFR and the gas density. However, we must pay attention to the
fact that larger and more luminous galaxies may have more of
everything: more blue stars, more gas, more dust. For instance, if
$H_\alpha$ is used as a star formation tracer, SFR being proportional
to $L(H_\alpha)$, a linear relation between total SFR and total gas
mass, such as found by several authors could be due to purely scaling
effect and not to a true physical dependence. The conclusion by
Devereux and Young (1991) that the $L_{FIR}/M(H_2)$ ratio is constant
along the Hubble sequence could mean that $L_{FIR}$ is simply
proportional to molecular hydrogen mass without necessarily implying
the constancy of the star formation efficiency (SFE) from Sa to Sc.
Sauvage and Thuan (1992) showed that in fact $L_{FIR}/L_{H_\alpha}$
decreases towards later types. Since $L_{H_\alpha}$ is proportional to
SFR (M $\gsim$ 10 \Mo$\!\!\!$), $L_{FIR}/L_{H_\alpha}$ is proportional
to the ratio of gas mass to SFR $\div $(SFE)$^{-1}$: The star
formation efficiency increases by a factor of $\sim$ 7 from Sa to Sd.

{}From $H_\alpha$ and UV data, it clearly appears that the SFR
indicators are not well correlated with surface density of molecular
gas. Average SFR is most strongly correlated either with HI or with HI
+ H$_2$, (Kennicutt, 1989). The important scatter must be attributed
to the extinction problem as seen above but also to the uncertainty on
H$_2$ masses and densities extrapolated from CO measurements. Large
variations in the conversion factor between CO intensity and H$_2$
density from galaxy to galaxy have not to be excluded.

Other informations on the star formation law can be inferred from
comparison of SFR and gas density within individual galaxies. The
relationship between $H_\alpha$ surface brightness and the total gas
density within field and Virgo spirals which is shown in fig. 17 is
due to Kennicutt (1993). The figure suggests three distinct regimes
for the star formation law: At high gas densities, the SFR follows a
Schmidt law with $n \sim $ 1.3. Below a critical density (between 1
and 10 \Mo$\!\!pc^{-2}$) the relation is much steeper. Well below this
critical threshold, SFR is very low if not zero. A more detailed study
shows that, for early type galaxies, the upper right quasi-linear part
of the law is absent contrary to late types.

\subsection{Dynamical indicators of star formation}

Cloud complexes and new stars are thought to be produced by the
conjugate action of gravity, stirring from massive stars, energy
dissipation and magnetic forces. Spontaneous cloud formation results
from combination of three instability types: gravitational, Parker and
thermal instabilities as described in details by Elmegreen (1992). In
the calculations reported by this author, the rate at which mass is
processed through the instability is proportional to $\rho^{1.3} -
\rho^{1.5}$. The characteristic wavelength of the instability is
$\sim$ 2.5 $kpc$ and the mass of the condensation $\sim$ 10$^7$
\Mo$\!\!$. The formation of molecular clouds ($\sim$ 10$^6$
\Mo$\!\!\!\!$) can occur through condensation and dissipation inside
the core of these complexes or fragmentation of giant shells.
Nevertheless, the details of the scenario as well as the connection
between the instabilities responsible of cloud formation and the
efficiency of star formation are still too speculative to be taken
into account.

Twenty years ago, Quirk (1972) had suggested that the star formation
threshold was associated with gravitational instabilities in the gas.
Guiderdoni (1987) has shown that the observed threshold corresponds to
the critical density for stability given by the Toomre criterion

$$
\Sigma_c = \alpha \f{\kappa c}{3.36 G}
$$

\noindent
where $c$ is the velocity dispersion of the gas (see also Kennicutt,
1989). So star formation onset could be characterized by $Q < 1$.
Recently, several models of large-scale star formation resulting from
gravitational instability in disks have been proposed with the least
number of free parameters (Silk (1992), Wyse and Silk (1989), Wang and
Silk (1994)).

The starting point is the equation for the star formation rate

$$
SFR = \epsilon \f{\Sigma_g}{t_s}
$$

\noindent
where $\epsilon = $SFE, $\Sigma_g$ is the gas surface density and
$t_s$ is the time-scale of star formation. Wang and Silk (1994) assume
that $1/t_s$ is related to the growth rate of the gravitational
instability of the gas disk. Identifying $1/t_s$ with the maximum
growth rate of the instability inferred from a local linear analysis,
they show that

\vspace{3mm}
$$
1/t_s \simeq \f{\kappa (1 - Q^2)^{1/2}}{Q}
$$

Then

$$
SFR =  \f{\epsilon \kappa \Sigma_g  (1 - Q^2)^{1/2}}{Q} \ \ \ \ \ \ (Q<1)
$$

\vspace{3mm}

This result indicates that 1) star formation occurs when $Q<1$, 2)
$t_s$ is related to $\kappa^{-1}$ or $\Omega^{-1}$, 3) the smaller
$Q$, the more rapidly stars formed, 4) SFR depends not only on
$\Sigma_g$ but also on the galaxy rotation rate. If $Q$ and $c$ are
approximatively constant, $\kappa \div \Sigma_g$ so that SFR $\div
\Sigma_g^2$ . Point 1) is equivalent to Quirk's proposition mentioned
above. Let us note with Wyse and Silk (1989) that a SFR proportional
to $\Sigma_g\Omega$ is able to reproduce color and metallicity
gradients in galaxy disks.

If the energy dissipation is rapid and the gas gravitational
instability triggers star formation, a regulation process could be
started: when $Q$ is below the threshold, SFR increases, the formed
stars stir up the interstellar medium, $Q$ increases, star formation
is stopped, $Q$ decreases etc...

Friedli and Benz (1993) tested different prescriptions relative to SFR
in view of using the most suitable in global evolution simulation of
disks. It is a question of comparing the predicted SFR sites in their
star + gas evolution models (described in section 7) with the usually
observed ones. Four criteria for star formation were considered a
priori: 1) $\Sigma_g > 4$ \Mo$\!\!pc^{-2}$, 2) $J_g < 1$ (Jeans
criterion), 3) $Q_g<1$.4) $\Delta S <0$ (S=entropy).

As seen in fig. 18, this preliminary approach seems to favour the
third criterion (or perhaps the fourth one) which at best restores the
observed zones of strong shocks obtained in the simulations (where
star formation is effectively expected). Other experiments based on
more elaborate physics are planned by the same authors.

\subsection{Star formation and dynamical perturbations in galaxies}

Enhanced rates of star formation (starbursts) have been often
associated in recent past with perturbations occurring in the host
systems such as bars and interactions. In fact both are able, in
principle and with more or less high efficiency, to trigger inflows of
gas and accumulation of matter towards the central regions as
previously explained.

Enhanced star formation in the cores of SB galaxies must depend on the
strength of the bar (mass and (or) axis ratio) and on the available
gas quantity. The fact that for instance only 40\% of early-type SB
galaxies exhibit excess 10$\mu$m central emission (Devereux, 1987)
indicates that the {\it presence} of the bar is only a necessary
condition to create a starburst.

Using IRAS galaxies of Hubble type Sbc to Sdm-Im from samples selected
by Devereux (1987) and Young et al. (1989), we observe that SA
galaxies rather lie in the lower right part of the color-color
sequence defined in subsection 9.1 whereas SB galaxies are found
everywhere along the sequence: the starburst region is essentially
populated by SB galaxies but some SB's seem quiescent (see Martinet,
1994 for details). It would be important to quantitatively estimate
the relation between the starburst intensity and the structural
parameters of bars concerned. Work in progress including star
formation processes into the evolution scheme of barred systems was
planned to this effect (Friedli and Benz, in preparation).

Evidences that galaxy interactions can trigger starbursts have been
given by various authors (table 3).

\begin{table}[h]
\begin{center}
\caption{Observational evidences of enhanced star formation in interacting
galaxies}

\vspace{4mm}

\begin{tabular}{|c|l|}
\hline
    Pass-band                        &   Reference        \\
\hline
UV       &  Larson and Tinsley (1978)       \\
Near-IR  &   Joseph and Wright (1985)        \\
Optical emission line strength &  Kennicutt and Keel (1984)       \\
Radio-emission  &  Hummel (1981)       \\
IRAS pass-bands & Sanders et al. (1988)        \\
\hline
\end{tabular}
\end{center}
\end{table}

Starbursts in interacting galaxies must depend on the encounter
orbital parameters and also on the gas available in the pre-encounter
galaxies. Advanced mergers for instance seem to be also only a
necessary condition for an enhanced IR luminosity. Whereas bright
normal isolated spirals have a range of 10 $\mu$m luminosity between
10$^5$ and 7 $\times$ 10$^8 L_\odot$, for interacting galaxies, the
range is 4 $\times$ 10$^7$ to 7 $\times$ 10$^9 L_\odot$ and for
mergers 4 $\times$ 10$^9$ to 5 $\times$ 10$^{10} L_\odot$. These
typical values correspond to a conversion of gas into early type stars
in the central few arc seconds at rates of 1 to 100 \Mo$\!\!yr^{-1}$.
By comparison, for our Galaxy, SFR in a region of 1 $kpc$ diameter is
$\sim$ 0.003 \Mo$\!\!yr^{-1}$. Large starbursts are expected in
stellar systems that readily form molecular gas out of atomic gas
(Mirabel, 1992). The most luminous IR galaxies have larger
$L_{FIR}/(M_{H_2}+M_{HI})$ ratio than isolated galaxies. For Arp 220,
this ratio is $\sim$ 20 times the Milky Way value (1-3
$L_\odot$\Mo$\!\!\!^{-1}$) and $\sim$ 5 times the M82-typical
starburst value (5-10 $L_\odot$\Mo$\!\!\!^{-1}$). However some
strongly interacting and merging systems are not ultra-luminous in
infra-red. Here also it will be important to quantitatively estimate
the relation between some features of the encounter and the intensity
of starburst which could be produced. The recent approach of the
problem by Mihos et al. (1992, 1993) can be used as an example:

The authors carried out 3 types of merger simulations: 1)
Prograde-prograde (PP) halo-disk galaxy interaction (the disks rotate
in the same sense as their initially parabolic orbital motion, 2)
retrograde-retrograde (RR) interaction (disk rotate in the opposite
sense to the orbital motion, 3) each disk plane is orthogonal to the
orbital plane and to the companion galaxy (O). They used Hernquist's
tree code (Hernquist, 1987) to calculate the gravitational forces
acting on stars and gas, modifying it to model star formation and
interactions between gas clouds. The gas consists of discrete clouds
interacting with one another through the merging of colliding clouds
and fragmentation of massive clouds into smaller ones: if two clouds
are at a distance less than the size of the larger cloud, they collide
and merge. A continued coalescence leading to gas locked into a too
small number of too massive clouds (unobserved!) is avoided by
introduction of an exponential lifetime $\tau_i \div (\log M_i)^{-1}$.
In fact the real process explaining the lack of too massive clouds
(heating process by star formation and supernovae) is here explicitly
burked.

Star formation rate in each cloud is given by

$$
\dot{M}_i = C M_i \rho_i
$$

\noindent
where $M_i$ is the cloud mass, $\rho_i$ the local density in a
spherical region surrounding the cloud.

Let us note that this law corresponds to a Schmidt law index $n
\approx $ 1.8. The initial gas depletion time is chosen so that SFR$_0
\sim $ 1-5 \Mo$\!\!yr^{-1}$.  In order to appraise the only
interaction effect, the considered initial conditions imply neither
diffusion of gas toward the center nor starburst development.

Essential of a typical PP merger is the quick increase of SFR to 5
times the pre-interaction rate followed by a decline as tidal tails
develop. Because of angular momentum loss, the galaxies fall back upon
one another and merge by $T =$ 165 $Myr$. A dramatic increase of star
formation (factor $>$ 20) is due to a large quantity of gas falling
into the center of the remnant ($\sim$ 50\% into the inner $kpc$)
(Fig. 19). The lifetime of the starburst is $\sim$ 200 $Myrs$.

Actually a competition settles between the gravitational forces which
draw the gas towards the center and the energy input from the
starburst which tends to disperse the gas. A very important question
is: what level of concentration could be reached? It is impossible to
answer as long as one has not a detailed understanding of star
formation. The central density able to trigger a starburst in the
simulation described here is far from that suspected in Arp 220 for
instance (5 \Mo$\!\!pc^{-3}$ instead of 60 \Mo$\!\!pc^{-3}$!). Is the
used star formation law correct in so extreme conditions?

The result of RR merger is qualitatively similar to the PP case, but
the massive tails are prevented by the lack of resonances between the
inner rotational and orbital motions. More gas remain for a starburst
which lasts 300 $Myr$, consuming 70\% of the gas.

The orthogonal merger (O) takes longer time to reach completion (400
$Myr$) due to the drag minimisation on the galaxies. Less gas is
channelled into the central region since the clouds suffer a
perturbation out of the disk plane. SFR is 10 times that of an
isolated disk. Only 25\% of gas was consumed by the starburst.

Fly-by interactions have been also studied by Mihos et al. (PP, RR, O,
PR). The only encounters which produce a significant increase of SFR
(factor of 5 typically) are PP. The formation of a bar triggers a flow
of gas towards the center. Enhanced star formation is not only found
in the central region but also in the outer disk. However 70\% of the
star formation activity occurs in the central $kpc$. The ``RR'' and
``O'' fly-by interactions are very ineffective. Finally, in the PR
interaction, 12\% of gas from the prograde disk is transferred to the
retrograde one. Accreted disk matter is captured on counterrotating
orbits. Very effective collisions in the retrograde disk lead to flows
of gas towards the center developing a strong starburst (SFR three
time greater than in the isolated galaxies whereas the SFR increase in
the prograde disk is only 50\%.

These examples suggest the tight dependence of the starburst intensity
on the various orbital features of the encounter.

The results can be compared to observational data suggesting
starbursts in various types of interactions.  Using information on the
star formation efficiency (SFE = $L_{IR}/ L_{CO}$) given by Solomon
and Sage (1998), Tinney et al. (1990) and Sanders et al.(1991), we can
observe that only some galaxies in strong interactions (in process of
merger) show a spectacular SFE of the order of 100 with respect to
those with companions without morphological disturbances or with
disturbances but no tidal tails or bridges (flybies!), for which
$L_{IR}/ L_{CO}$ is of the order of 10-30. Star formation rates from
FIR data for the same galaxies display a large range of values from
several \Mo$\!\!\!yr^{-1}$ to more than 200 \Mo$\!\!\!yr^{-1}$. The
upper values are obtained only for galaxies in strong interaction. But
it is not easy to establish one-to-one correspondance between
morphological features and starburst activity. Experiments similar to
those described above including a realistic treatment of star
formation will certainly allow to better understand the strong
dispersion observed in the SFR's of weakly and strongly interacting
galaxies, confirming the dependence of enhanced star formation rate on
the parameters characterizing the encounters. To go thoroughly into
the relation between enhancements of star formation and galaxy
interaction, Keel (1993) analysed kinematical data concerning direct
and retrograde pairs of galaxies. Disturbances in rotation curves can
be indicators of interactions with companions. Effectively in the
mentioned work, high rates of star formation inferred from $H_\alpha$
data are associated with large relative amplitudes of velocity
disturbances for nuclei and integrated measures. But statistically the
level of star formation is found not to be dependent on orbital
direction whereas models described before predict that fly-by
encounters must be direct to produce star formation enhancements.

Independently of questions of orbits and inclinations, detailed
studies of physical conditions in the gaseous medium can also bring
some enlightments on the fact that all interacting galaxies do not
show evidence for starburst. Jog and Das (1992) suggest that if the
central molecular pressure is less than the pressure in a giant
molecular cloud (GMC), an incoming GMC does not suffer the necessary
overpressure and a starburst will not be triggered despite gas infall
due to the interaction (or to a bar!) (see also Jog and Solomon,
1992).

Let us note in addition that the Mihos merger models seem to be unable
to produce starbursts which could power the ultra-luminous infrared
galaxies with $L_{FIR} > $ 10$^{12} L_\odot$. The implied star
formation rates would be one order greater than those predicted by the
models. There exists at least three possible sources for this
disagreement: 1) the protogalaxies are actually more gas rich than
assumed 2) the star formation criterion based on the Schmidt law is
not correct (see the previous subsection) and 3) the cloud-collision
model is too crude.

An other divergence between the present models and the observations
concerns the sites of star formation. Mihos et al. (1993) point out
various cases of pairs, for instance NGC 6872 / IC 4970 the $H_\alpha$
maps of which show star formation along the tidal arms whereas the
model indicates the bar and the nuclear region as the only privileged
sites. Probably a more rigourous approach of gas dynamics, for
instance concerning the viscosity, could clear up the question.


\newpage
\section{Does a spiral galaxy change its Hubble type \protect\newline  during
evolution?}

At length of the sections 2 and 4 to 9 of this review, a set of
processes have been discussed which are so many clues of galaxy
evolution {\it on time scales equal or smaller than the Hubble time}.
Let us summarize the essential points

\begin{enumerate}

\item From studies on orbital behavior, we learn for instance that too
highly triaxial spheroids might not exist because real equilibrium
systems would not tolerate too much chaos. The conjugate effects of a
bar and a compact central mass cause bulges to secularly grow through
enhanced stochastic orbital diffusion. The width of instability strips
on the main periodic orbits, which trigger radial or vertical
diffusion, depends on the geometry of the system which can be modified
in time.

\item In systems consisting of stars and gas, recurrent growth of
non-axisymmetric perturbations must be the rule to explain the high
frequency of spirals observed.

\item The shape of the spheroid (halo) can become flatter during the
slow setting up of a disk in the potential well of the spheroid.

\item The friction effect between a bar and a disk or a bulge produces a
transfer of angular momentum from the bar to the disk or to the bulge.
Bars evolve in response to these interactions.

\item The structure of a barred galaxy perpendicularly to the disk is
changed by resonant excitations: observed box or peanut shapes can be
the signature of a bar.

\item Satellites which merge into a disk cause thickening, flattening
and warping of the disk.

\item Bulges can grow by satellite accretions.

\item Bars easily form in a disk, they are robust, unless a strong ILR
be created by satellite accretion, compact mass or gas inflow towards
the center, what contributes to destroy the bars.

\item The dissipative nature of gas plays an important role in spiral
activity development and in fueling the galactic nuclei, the gas
systematically losing angular momentum to the stars.

\item Differences in the structure of early and late-type galaxies
cannot be explained without reference to the gaseous component the mass
of which can vary in time due to either accretion or star
formation.

\item Some observed phenomena such as bars within bars are episodic.

\item Weak interactions between galaxies trigger spiral or bar modes.
Strong interactions generate more drastic transformations. Mergers of
spirals can produce ellipticals.

\item It is possible that the local star formation be self-regulated by
the global dynamics, then the galaxy morphology must change over
time-scales of the order of gas consumption time scale which is shorter
than the Hubble time.

\item Under certain conditions, galaxy interactions enhance star
formation which can in its turn influence the subsequent dynamical
evolution.

\end{enumerate}

Jeans (1929) postulated that galaxies evolve from elliptical to
spiral, and then to irregular galaxies: this idea, which influenced
Hubble himself, has been hard to kill. In the seventies, the initial
conditions were considered as the key factor for creating the various
observed morphologies (elliptical and spirals). Larson (1976) and Gott
and Thuan (1976) introduced the ratio of the star formation to the
collapse time scales $\tau_{sf}/\tau_{coll}$ as the parameter which
determines the initial evolution: early-types (ellipticals) would
result if $\tau_{sf}/\tau_{coll} < $ 1, late-types if
$\tau_{sf}/\tau_{coll} > $ 1.

In view of the processes mentioned above, it is difficult to still
claim that only the initial conditions are responsible of the
diversity of the observed Hubble types. Here we present a quintessence
of arguments recently put forward in favor of secular evolution
through the Hubble sequence for spirals from Sd to Sa. Some ideas in
this connection had been already expressed by Kormendy (1982),
particularly concerning the fate of bars.

\noindent Larson (1993) recalled some theoretical predictions useful for the
discussion:
\begin{itemize}
\item[-] The gas depletion time scale $\tau$ is expected to increase by
a factor of 3 along the Hubble sequence.

\item[-] The rotation period could be the basic clock for the evolution
(if $Q = $ ct., $\tau \div \kappa^{-1}$).

\item[-] The spacing of arms depends on $\lambda_{crit} = 4\pi^2
G\mu/\kappa^2$. A more open spiral pattern is expected if a galaxy has
either $\mu$ higher or $\kappa$ lower.
\end{itemize}

Thus the variation of the gas content as well as the star formation and
the winding of arms must be mainly influenced by the mass distribution
and the rotation rate.

Now the Hubble sequence from Sd to Sa is a sequence of increasing
bulge-to-disk ratio, rotation, arm winding, metallicity, ratio of
current to average past star formation rate. Which regard to the list
of processes drawn up at the beginning of this section, something
irreversible appears in this sequence: a possible scenario could be
the following (Pfenniger et al. (1994) : Cooling in a mostly gaseous
disk $\longrightarrow$ gravitational instability $\longrightarrow$
formation of a bar $\longrightarrow$ secular formation of a small
bulge $\longrightarrow$ accumulation of mass towards the center
$\longrightarrow$ dissolution of the bar into a spheroidal component.
Growth of bulge can be produced by satellite mergers. Moreover
referring to the virial theorem and taking account of dissipation
(Pfenniger, 1991b), it becomes clear that a galaxy could be at first a
slowly rotating weakly concentrated object, then dynamically evolves
into a rapidly rotating and more condensed one. The outer parts are
progressively symmetrized and arms tightly wind.

As already emphasized, the role of bars seems to be here essential. A
possible evolution sequence resulting from our discussion could be
(Friedli and Martinet, 1993): Sd $\longrightarrow$ Sc
$\longrightarrow$ Sbc $\longrightarrow$ SBb $\longrightarrow$ Sb ....
or $\longrightarrow$ Sc $\longrightarrow$ SBb $\longrightarrow$ SBa
$\longrightarrow$ SBO $\longrightarrow$ SO, perhaps refined in S
$\longrightarrow$ SB $\longrightarrow$ S2B $\longrightarrow$ SB
$\longrightarrow$ S by the ``bars within bars'' episode. Such
scenarios could be either strengthened or modified by the star
formation contribution for which quantitative estimates are urgently
requested in the frame of evolution simulations.  Furthermore, the
scenario proposed above concerns essentially isolated galaxies.
Interactions can perturb it. It could be possible that mass inflows in
interacting systems change a late unbarred galaxy into an early barred
one as suggested by Elmegreen et al. (1990). Their statistics
indicates a simultaneous change of Hubble type and structure (SA
$\longrightarrow$ SB). But the size of the sample does not allow a
really decisive conclusion.

Including Ellipticals, SO and early-type spirals in the discussion,
Schweizer (1993) presented some statistical arguments according to
which many of these galaxies could be the product of collisions and
mergers .  He suggested that ``the Hubble sequence may rank galaxies
mainly by the number and the vehemence of mergers in their past
history''. Progressive mergers depend on the density. In the context
of hierarchy formation of galaxies, Larson (1993) distinguished
Ellipticals and SO as formed from subunits of stars and spirals as
formed from subunits of gas.

Study by Combes and Elmegreen (1993) already mentioned in section 7.3
is a starting point to compare in details the evolution inside the
early and late-type spirals. Whereas their early-type models evolve
slowly, late-type models evolve quickly with gas inflow towards the
center which leads to a more condensed inner region. Consequently,
these late-type models evolve towards a early-type. What is not under
control is the possible reduction of gas fraction by star bursts. But
with continuous accretion process from various origins (tidal
interactions for instance), evolution towards earlier type can
continue. In fact as we know, without infall, a Sc galaxy would
exhaust its gas in a small fraction of the Hubble time.

Kennicutt (1990) claimed that more than one evolutionary path could
lead to a particular type and that many scenarios could lead identical
protogalaxies to evolve into very different types. Taking account of
the internal processes described in this review, the interaction
effects and the changeable star formation efficiency, still badly
understood, we must obviously not exclude these possibilities.

Finally, Pfenniger et al. (1994) have discussed the implications of
the proposed evolution on the nature of dark matter. We must take into
consideration the following facts:

1) The ratio of dark matter to stellar mass decreases from Sd's to
Sa's by a factor of 10$^2$. In comparison, the ratio of dark to HI
mass is nearly constant ($\sim$ 10-30).

2) The early types, gas depleted, have a declining star formation rate
(SFR). Sb's and Sc's have a SFR = ct. The latest and irregular types
are the site of starbursts. But average time scale for gas consuming
is of the order of 4 Gyr! Either gas infall representing some percents
of total mass and (or) delayed gas recycling or stellar winds can
solve this gas consumption problem.

The solution suggested by Pfenniger et al. (1994) in order to collect
these facts in agreement with the evolution scenario from Sd to Sa is
to admit that dark matter be in a form able to create stars, i.e.
fresh diluted hydrogen. A model of fractal cold gas, clumpy down to
very small scales has been proposed by Pfenniger and Combes (1994).

The problem of dark matter in disk galaxies could be solved to a large
extent as far as it can be proved that gas mass given by present data
is underestimated by a factor of 5 to 10.


\newpage

\subsection*{Acknowledgements:}
This review is the fruit of prolonged collaboration within the
galactic dynamics group at Geneva Observatory. I would particularly
like to thank Daniel Pfenniger and Daniel Friedli for their support as
well as St\'ephane Udry and Roger Fux. Many contacts with colleagues
at this Observatory and elsewhere have stimulated our interest for the
problems raised. I am also especially grateful to Fran\c coise Combes
who has given our group her unfailing support and Prof. A. Maeder for
encouraging me to write this review. I extend my thanks to Prof. A.
Omont and to Brigitte Rocca for their kind hospitality at the
Institute of Astrophysics in Paris.

\newpage
\addcontentsline{toc}{section}{\protect \numberline{}{References}}


\newpage
\section*{Figure captions}

\begin{itemize}

\item[Fig. 1] Position of galaxies with different shape of HI
rotation velocity curve in the plane $V_{max}$ versus $h = $ radial
optical scale length. Thick part of curve corresponds to the region
outside two-thirds of optical radius. (From Casertano and van Gorkom,
1991).

 \item[Fig. 2] Resonant periodic orbits inside corotation in a barred
galaxy.  (From Contopoulos and Grosb\o l, 1989).

 \item[Fig. 3] Example of surface of section and variation of the
rotation number $Rot(x)$ in a triaxial model with axis ratio $(1 :
0.625 : 0.5)$. (From Martinet and Udry, 1990).

 \item[Fig. 4] Prediction of axis ratios of real triaxial systems
(elliptical and bulges) inferred from observations, compared with
analytical models used by Martinet and Udry (1990) (I to IV). (From
Udry and Martinet, 1994).

 \item[Fig. 5] Surface of section in a noisy triaxial potential $\Phi
= \Phi_0 + \epsilon \Phi_1 (k_i x_i)$ for different values of the
amplitude $\epsilon$ and frequency $k_i$ of noise. (From Udry, 1991).

 \item[Fig. 6] Diffusion of stochastic orbits close to the $z$-axis
orbit in galaxies with a central compact Plummer mass $M_p$: a) $GM_p
= 0$, b) $GM_p = 10^{-3}$.  (From Martinet and Pfenniger, 1987).

 \item[Fig. 7] Behaviour of the part of $Q$ containing $R$ for two
theoretical values of $R_m$, radius from which the rotation curve
becomes approximately flat.  $h$ is the radial scale length.  (From
Martinet, 1988).

 \item[Fig. 8] Theoretical relation between radial velocity dispersion
taken at $R = h$ and the maximum rotational velocity a) for different
$(M/L)_B$, b) for different $Q$'s (see the text), in comparison with
observed values. (From Bottema, 1993).

 \item[Fig. 9] Variation in the ratio of the perturbation surface
density to the unperturbed surface density with time for star (S) and
gas (G) showing the swing amplification at $\tau \geq 0$. (From Jog,
1992).

 \item[Fig. 10] Theoretical and experimental predictions of the
phase-averaged angular momentum change $\langle \Delta J \rangle$
between a bar and a disk described in the text. Case of a weak fast
rotating bar. Disk without self-gravity. (From Little and Carlberg,
1991).

 \item[Fig. 11] Evolution of the projected density in a 3-D $N$-body
model as seen disk face-on, bar end-on and bar edge-on, at T = 5 $Gyr$
compared with the initial configuration.  (From Pfenniger and
Friedli,1991).

 \item[Fig. 12] Velocity dispersions in a 3-D $N$-body simulation at T
= 5 $Gyr$ compared with the initial isotropic velocity dispersion
(From Pfenniger and Friedli, 1991).

 \item[Fig. 13] Projections of 3-D 4/4/1 resonant orbits in a strongly
barred galactic potential in the ($y, x$) (upper left), ($z, x$)
(upper right) and ($y, z$) (lower left) planes. (From Pfenniger,
1985).

 \item[Fig. 14] Evolution of star+gas model simulating a) a late-type
galaxy after 2.4 (top) and 4.8 $\times$ 10$^8 yr$ (bottom) and b) an
early-type one, after 8.4 (top) and 10.8 $\times$ 10$^8 yr$ (bottom).
Stars (left) and gas (right) particles are represented. (From Combes
and Elmegreen, 1993).

 \item[Fig. 15] Time evolution of the relative central mass
concentration within 0.5, 1.0, 1.5 $kpc$ from the center for a barred
galaxy model consisting of gas (solid lines) and stars (dotted lines).
(From Friedli and Benz, 1993).

 \item[Fig. 16] Time evolution of the stellar surface density during
the two-bar phase for a galaxy model consisting of gas and stars. $t$
is the time in M$yr$ and $\theta$ the approximate angle in degree
between the primary and the secondary bar. (From Friedli and Martinet,
1993).

 \item[Fig. 17] Distinct physical regimes in field and Virgo galaxies
from $H_\alpha$ surface brightness/gas surface density. (From
Kennicutt, 1993).

 \item[Fig. 18] Sites of star formation deduced from various criteria
described in the text for a galaxy model consisting of gas and stars
(From Friedli and Benz, 1993).

 \item[Fig. 19] Star formation evolution in a prograde-prograde merger
model. Unit time is $\sim$ 50 M$yr$. (From Mihos et al., 1992).

\end{itemize}


\noindent
\parindent=0pt

\begin{thebibliography}{}

\bibitem{}van Albada T.S. and Sancisi R. 1986,  {\it Philos. Trans.
 R. Soc. London A} {\bf 320}, 447
\bibitem{}Amendt P. and Cuddeford P., 1991  \apj {\bf 368}, 79
\bibitem{}Athanassoula E., Bienaym\'e D., Martinet L., Pfenniger D. 1988,  \aa
{\bf 127}, 349
\bibitem{}Athanassoula E., Bosma A., Papaioannou S., 1987 \aa  {\bf 179},  23
\bibitem{}Athanassoula E., and Martinet L. 1980,  \aa  {\bf 87}, L10
\bibitem{}Athanassoula E., and Sellwood J. 1986,  \mn  {\bf 221}, 213
\bibitem{}Bahcall and Casertano S. 1985,  \apj {\bf 293}, L7
\bibitem{}Barnes J.E. 1992,  in {\it Morphological and physical classification
of galaxies}, G.Longo et al. eds. Reidel Dordrecht, p.277
\bibitem{}Barnes J.E. and Hernquist L. 1991,  \apj  {\bf 370}, L65
\bibitem{}Barnes J.E. and Hernquist L. 1992,  in {\it ARA\&A: Dynamics of
interacting galaxies} {\bf 30}, 705
\bibitem{}Baumgart C.W. and Peterson C.J. 1986, \pasp {\bf 98}, 56
\bibitem{}Benz W. 1990,   {\it ``SPH: a review''}  in {\it The Numerical
modelling of non-linear stellar pulsations}: NATO ASI Series C no 302, J.R.
Buchler ed. Kluwer Dordrecht, p.269
\bibitem{}Bersier D., Blecha A., Golay M., Martinet L. 1994,  \aa in press
\bibitem{}Binney J., 1981,  \mn  {\bf 196}, 455
\bibitem{}Binney J. and May A. 1986,  \mnras  {\bf 218}, 743
\bibitem{}Binney J., and Tremaine S. 1987,  {\it Galactic Dynamics}, Princeton
Univ. Press, Princeton
\bibitem{}Born, 1927,  {\it Mechanics of atoms}, republished 1960,  F. Ungar,
Publ. Co. New York
\bibitem{}Bosma A. 1981,  \aj {\bf 101}, 1971
\bibitem{}Bosma A. and Freeman K. 1991,  unpublished
\bibitem{}Bottema, 1988,  \aa {\bf 197}, 105
\bibitem{}Bottema, 1993,  \aa {\bf 275}, 16
\bibitem{}Broeils A. 1992,   {\it Thesis}, Rijksuniversiteit Groningen
\bibitem{}Broucke R.A. 1969,  {\it Am. Inst. Aeron. Astonautics J.}, {\bf 7},
1003
\bibitem{}Buta R. 1990,  \apj {\bf 351}, 62
\bibitem{}Caldwell N., Kennicutt R.C., Phillips A.C. Schommer R.A. 1991,  \apj
{\bf 370}, 526
\bibitem{}Carlberg R.G. and Freedman W.L 1985,  \apj {\bf 298}, 486
\bibitem{}Carlberg R.G., Dawson P.C., Hsu T. and Vandenberg D.A. 1985,  \apj
{\bf 294}, 674
\bibitem{}Carlberg R.G. 1987,   \apj  {\bf 322}, 59
\bibitem{}Casertano S. and van Gorkom J.H. 1991,  \aj {\bf 101}, 1231
\bibitem{}Chandrasekhar S. 1943,  \apj {\bf 97}, 251
\bibitem{}Combes F. 1992,  in {\it Morphological and physical classification of
galaxies},  G.Longo et al. eds. Reidel Dordrecht, p.265
\bibitem{}Combes F., Debbasch F., Friedli D., Pfenniger D. 1990,  \aa {\bf
233},  82
\bibitem{}Combes F. and  Elmegreen B.G. 1993,  \aa {\bf 271}, 391
\bibitem{}Contopoulos G., 1988,  in {\it Dynamical systems}, J.R. Buchler,
 J.R. Ipser, C.D. Williams eds. Ann. of N.Y. Acad. of Sciences, {\bf 536}, 1
\bibitem{}Contopoulos G., and Grosb\o l P.1989,  \aar  {\bf 1}, 261
\bibitem{}Contopoulos G., and Magnenat P. 1985,  {\it Celestial Mechanics}
{\bf 37}, 387
\bibitem{}Contopoulos G., and Papayannopoulos 1980,  \aa  {\bf 92}, 33
\bibitem{}Corradi R.  and Capaccioli M. 1991, \aasup  {\bf  90}, 121
\bibitem{}Cretton N. and Martinet L. 1994,  in preparation
\bibitem{}de Vaucouleurs G. 1963,  \apjs {\bf 8}, 31
\bibitem{}de Vaucouleurs G. 1974,  in {\it Formation of galaxies}, IAU
Symposium no 58, J.R. Shakeshaft eds.  Reidel Dordrecht, p.335
\bibitem{}Devereux N.A. 1987, \apj {\bf 323}, 91
\bibitem{}Devereux N.A. and Young J.S. 1990, \apj {\bf 350}, L25
\bibitem{}Devereux N.A. and Young J.S. 1991, \apj {\bf 371}, 515
\bibitem{}Devereux N.A., Kenney J.D.P., Young J.S. 1992,  \aj {\bf 103}, 784
\bibitem{}Donas J., Deharveng J.M., Laget M., Milliard B. and Huguenin D. 1987,
 \aa {\bf 180}, 12
\bibitem{}Dressler A., and Sandage A. 1983,  \apj {\bf 265}, 664
\bibitem{}Efstathiou G., Lake G., Negroponte J. 1982,  \mn {\bf 130},125
\bibitem{}Einasto J., Kaasik A., Saar E. 1974,  {\it Nat} {\bf 250}, 309
\bibitem{}Elmegreen B.G., and Elmegreen D.M. 1985,  \apj {\bf 288}, 438
\bibitem{}Elmegreen B.G., and Elmegreen D.M. 1989,  \apj {\bf 342}, 677
\bibitem{}Elmegreen B.G. 1992,  in {\it Interstellar Matter}, Saas Fee Courses
of SSAA, D. Pfenniger and P. Bartholdi eds. Springer-Verlag
\bibitem{}Elmegreen D.M.  1981,  \apjs {\bf 47}, 229
\bibitem{}Elmegreen D.M., Elmegreen B.G. and Bellin A.D.  1990,  \apj {\bf
364}, 415
\bibitem{}Fall S.M., Efstathiou G.  1980,  \mnras {\bf 193}, 189
\bibitem{}Freeman K. 1993,  in {\it Physics of nearby galaxies,
 nature or nurture~?},  T. X. Thuan, C. Balkowski, D. T. T. Van eds. Editions
Fronti\`eres, Gif s/ Yvette, p.201
\bibitem{}Friedli D.  1992,  {\it Thesis}, University of Geneva
\bibitem{}Friedli D. 1994, in {\it Mass-transfer induced
	activity in galaxies} I. Shlosman ed., in press
\bibitem{}Friedli D. and Benz W. 1993,  \aa {\bf 268}, 65
\bibitem{}Friedli D. and Martinet L. 1993,  \aa {\bf 277}, 27
\bibitem{}Friedli D. and Benz W. 1994,  in preparation
\bibitem{}Friedli D., Benz W., Martinet L. 1991,   in {\it Dynamics of disk
galaxies}, B. Sundelius ed. G\"{o}teborg p.181
\bibitem{}Friedli D. and Pfenniger D. 1991,  in {\it Dynamics of
galaxies and their molecular clouds distribution},
IAU Symp. no 146, Casoli F. and Combes F. eds.  Reidel Dordrecht, p.362
\bibitem{}Gavazzi G. 1993,  \apj {\bf 419}, 469
\bibitem{}Gerhard O.E. 1985,  \aa {\bf 151}, 278
\bibitem{}Gerin M., Combes F. and Athanassoula E. 1990,  \aa  {\bf 230}, 37
\bibitem{}Goldreich and Lynden-Bell D. 1965,  \mn {\bf 130}, 125
\bibitem{}Gott J.R.  and Thuan T.X. 1976,  \apj {\bf 204}, 649
\bibitem{}Guiderdoni B. 1987, \aa {\bf 172}, 27
\bibitem{}Hasan H. and Norman C. 1990,  \apj {\bf 361}, 69
\bibitem{}Heissler J., Merrit D. Schwazschild M. 1982,  \apj {\bf 258}, 490
\bibitem{}H\'enon M. 1973,  in {\it  Dynamics of stellar systems},  3rd
Saas-Fee course of the Swiss Society of Astrophysics and Astronomy  L. Martinet
and M. Mayor eds., Geneva Observatory, p.183
\bibitem{}H\'enon M. 1981,  in {\it Chaotic behaviour of deterministic
systems},  G. Ioss, R. Helleman, R. Stora eds. North Holland Publ. Amsterdam,
p.67
\bibitem{}H\'enon M., and Heiles C. 1964,  \aj {\bf 69 }, 73
\bibitem{}Hernquist L.  1987,  \apjs {\bf 64}, 715
\bibitem{}Hernquist L., and Weinberg M.D. 1992,  \apj {\bf 400}, 80
\bibitem{}Hohl F. and Hockney R.W. 1969,  {\it J. Comput. Phys.} {\bf 4}, 305
\bibitem{}Hogg D.E., Roberts M.S., and Sandage A. 1993,  \aj {\bf 106 }, 907
\bibitem{}Hummel E  1981,  \aa {\bf 93}, 93
\bibitem{}Jarvis B. 1990,  in {\it Dynamics and interactions of galaxies},
R.Wielen ed. Springer, Berlin, p.416
\bibitem{}Jeans J.H. 1929, {\it Astronomy and Cosmogony}, Cambridge University
Press
\bibitem{}Jenkins A. and Binney J. 1990,  \mn {\bf 245}, 305
\bibitem{}Jog C. 1992,  \apj {\bf 390}, 378
\bibitem{}Jog C. and Solomon P.M. 1984,  \apj {\bf 276}, 127
\bibitem{}Jog C. and Das M. 1992,  \apj {\bf 400}, 476
\bibitem{}Jog C. and Solomon P.M. 1992,  \apj {\bf 387}, 152
\bibitem{}Joseph R.D. and  Wright G.S. 1985,  \mnras {\bf 214}, 87
\bibitem{}Kalnajs A. 1987,  in {\it Dark matter in the Universe},
IAU Symp. 117, J. Kormendy and G.R Knapp eds. Reidel Dordrecht, p.289
\bibitem{}Keel W. 1993,  \aj {\bf 106}, 1771
\bibitem{}Kenney J.D.P. 1991,  in {\it Dynamics of
galaxies and their molecular clouds distribution},
IAU Symp. no 146, Casoli F. and Combes F. eds.  Reidel Dordrecht, p.265
\bibitem{}Kenney J.D.P., Carlstrom J.E., Young J. 1993,   \apj {\bf 418}, 687
\bibitem{}Kennicutt R.C. 1981,  \aj {\bf 86}, 1847
\bibitem{}Kennicutt R.C. 1983,  \apj {\bf 272}, 54
\bibitem{}Kennicutt R.C. 1989,  \apj {\bf 344}, 685
\bibitem{}Kennicutt R.C. 1990,  in {\it The interstellar medium in galaxies},
H.A Thronson and J.M. Shull eds. Kluwer, Dordrecht, p.405
\bibitem{}Kennicutt R.C. 1993,  in {\it The environment and evolution of
galaxies}, 3rd Teton Summer School,  J.M. Shull and H.A Thronson eds. Kluwer,
Dordrecht, p.533
\bibitem{}Kennicutt R.C. and Keel W. 1984,  \apj {\bf 279}, L5
\bibitem{}Kormendy J. 1979,  \apj {\bf 227}, 714
\bibitem{}Kormendy J. 1981,  in
 {\it The structure and evolution of normal galaxies},
eds. S.M. Fall and Lynden-Bell
\bibitem{}Kormendy J. 1982,   in {\it Morphology and dynamics of galaxies},
12th Saas-Fee course of the Swiss Society of Astrophysics and Astronomy   L.
Martinet and M. Mayor eds., Geneva Observatory, p.115
\bibitem{}Kormendy J. 1984,  \apj {\bf 286}, 116
\bibitem{}van der Kruit P. 1988,  \aa {\bf 192}, 117
\bibitem{}van der Kruit P. and Freeman K. 1986,  \apj {\bf 303}, 556
\bibitem{}van der Kruit P. and Searle 1981,  \aa {\bf 95}, 105
\bibitem{}Lacey  1991,  in {\it Dynamics of disk galaxies}, B. Sundelius ed.,
G\"{o}teborg, p.257
\bibitem{}Larson R.B.   1976,  \mnras {\bf 176}, 31
\bibitem{}Larson R.B.   1993,  in {\it Physics of nearby galaxies,
 nature or nurture~?}, T. X. Thuan, C. Balkowski, D. T. T. Van eds. Editions
Fronti\`eres, Gif s/ Yvette, p.487
\bibitem{}Larson R.B. and Tinsley B.    1978,  \apj {\bf 218}, 46
\bibitem{}Lewis and Freeman K. 1989,  \aj {\bf 97}, 139
\bibitem{}Lin C.C. and Shu F. 1964,  \apj {\bf 140}, 646
\bibitem{}Little B. and Carlberg R.G. 1991,  \mn {\bf 251}, 227
\bibitem{}Lo K.Y. 1984,  \apj {\bf 282}, L59
\bibitem{}Louis P.D. and Gerhard O. 1988,  \mn {\bf 233}, 337
\bibitem{}Lynden-Bell D. 1973,  in {\it Dynamics of stellar systems}, 3rd
Saas-Fee course of the Swiss Society of Astrophysics and Astronomy L. Martinet
and M. Mayor eds.,  Geneva Observatory, p.91
\bibitem{}Lynden-Bell D. 1977,  in {\it Star formation}, IAU symposium no 75,
T.J. de Jong and A. Maeder eds., Reidel Dordrecht, p.291
\bibitem{}Lynden-Bell D. 1979,  \mnras  {\bf 187}, 101
\bibitem{}Lynden-Bell D. and Kalnajs A. 1972,  \mn {\bf 157}, 1
\bibitem{}Magnenat P. 1982,  \aa {\bf 108}, 69
\bibitem{}Magnenat P. and Martinet L. 1983,   in {\it Internal kinematics and
dynamics of galaxies}, IAU symposium no. 100, E. Athanassoula ed. Reidel
Dordrecht, p.293
\bibitem{}Martinet L. 1974,  \aa {\bf 32}, 329
\bibitem{}Martinet L. 1988,  \aa {\bf 206}, 253
\bibitem{}Martinet L. 1994, in preparation
\bibitem{}Martinet L., Pfenniger D. 1987,  \aa {\bf 173}, 81
\bibitem{}Martinet L. and de Zeeuw T. 1988,  \aa {\bf 206}, 269
\bibitem{}Martinet L., Udry S. 1990,  \aa {\bf 235}, 69
\bibitem{}Mihos J.C., Richstone D.O., Bothun G.D. 1992,  \apj {\bf 400}, 153
\bibitem{}Mihos J.C., Bothun G.D., Richstone D.O.1993,  \apj {\bf 418}, 82
\bibitem{}Miller R.H., and Prendergast K.H. 1968,  \apj {\bf 151}, 699
\bibitem{}Miller R.H., Franx M., Fisher D., Illingworth G. 1992,
\bibitem{}Mirabel I.F  1992, in {\it Star formation in stellar systems}, G.
Tenorio-Tagle, M. Prieto, F. Sanchez eds. Cambridge Univ. Press., Cambridge
p.479
\bibitem{}Miralda--Escud\'e J. and Schwarzschild M. 1989,  \apj {\bf 339}, 752
\bibitem{}Monaghan J.J. 1992,  {\it ARA\&A} {\bf 30}, 543
\bibitem{}Ostriker J.P. and Peebles P.J.E. 1973,  \apj {\bf 186}, 467
\bibitem{}Percival  1989,  in {\it Non-linear dynamics and the beam-beam
interaction}, M.Mouth and J.C. Hervera eds. A.I.P. conf. Proc. {\bf 57}, 302
\bibitem{}Pfenniger D. 1984a,  \aa {\bf 134}, 373
\bibitem{}Pfenniger D. 1984b,  \aa {\bf 141}, 171
\bibitem{}Pfenniger D. 1985,  \aa {\bf 150}, 112
\bibitem{}Pfenniger D. 1990,  \aa {\bf  230}, 55
\bibitem{}Pfenniger D. 1991a,   in {\it Dynamics of disk galaxies}, B.
Sundelius ed. G\"{o}teborg, p.191
\bibitem{}Pfenniger D. 1991b,   in {\it Dynamics of disk galaxies}, B.
Sundelius ed. G\"{o}teborg, p.389
\bibitem{}Pfenniger D. 1993,  in {\it  Galactic bulges}, I.A.U. Symposium no
153. H.Dejonghe and H.J. Habing eds. Reidel Dordrecht
\bibitem{}Pfenniger D. 1994,   in preparation
\bibitem{}Pfenniger D. and Friedli D. 1991,  \aa {\bf 252}, 75
\bibitem{}Pfenniger D., Combes F. 1994,  \aa in press
\bibitem{}Pfenniger D., Combes F., Martinet L. 1994,  \aa in press
\bibitem{}Poincar\'e H. 1899, {\it Les nouvelles m\'ethodes de la m\'ecanique
c\'eleste}, Dover Pub. republications, Vol. III, p.175
\bibitem{}Prugniel P., Combes F. 1992,  \aa {\bf 259}, 25
\bibitem{}Quinn P.J., Hernquist L., Fullagar D.P. 1993,  \apj {\bf 403}, 74
\bibitem{}Quirk W.J. 1971, \apj {\bf 167}, 7
\bibitem{}Rix H.W., Franx M., Fisher D., Illingworth G. 1992,  \apj {\bf 400},
L5
\bibitem{}Roberts M.S. 1969,  \aj {\bf 74}, 859
\bibitem{}Roberts W.W., Roberts M.S. and  Shu F.  1975,  \apj {\bf 196}, 381
\bibitem{}Rubin V., Burstein D., Ford W. and Thonnard N. 1985,  \apj {\bf 289},
81
\bibitem{}Salucci P., Ashman K.M., Persic M. 1991,  \apj {\bf 379}, 89
\bibitem{}Sanchez-Saavedra M.L., Battaner E., Florido E. 1990,  \mnras {\bf
246}, 458
\bibitem{}Sancisi R. and van Albada T.S. 1987,  in {\it Dark matter in the
Universe},
IAU Symp. 117, J. Kormendy and G.R Knapp eds. Reidel Dordrecht, p. 67
\bibitem{}Sanders D.B., Scoville N.Z. and Soifer B.T. 1991,  \apj {\bf  370},
158
\bibitem{}Sauvage M. and Thuan T.X. 1992,  \apj {\bf 396}, L69
\bibitem{}Sauvage M. and Thuan T.X. 1994,  \apj in press
\bibitem{}Schmidt M. 1959, \apj {\bf 129}, 243
\bibitem{}Schwarz M.P. 1981,  \apj {\bf 247}, 77
\bibitem{}Schwarz M.P. 1984,  \mnras {\bf 209}, 93
\bibitem{}Schwarzschild M. 1979,  \apj {\bf 232}, 236
\bibitem{}Schweizer F.  in {\it Physics of nearby galaxies,
 nature or nurture~?}, T. X. Thuan, C. Balkowski, D. T. T. Van eds. Editions
Fronti\`eres, Gif s/ Yvette, p. 283
\bibitem{}Searle L. and Zinn R. 1978,  \apj {\bf 225}, 357
\bibitem{}Sellwood J. 1980,  \aa {\bf 89}, 296
\bibitem{}Sellwood J. 1981,  \aa {\bf 99}, 362
\bibitem{}Sellwood J. and Carlberg R.G. 1984,  \apj {\bf 282}, 61
\bibitem{}Sellwood J. and Lin D.N.C. 1989,  \mn {\bf 240}, 991
\bibitem{}Sellwood J. and Wilkinson A. 1993,  {\it Rep. Prog. Phys.} {\bf 56},
173
\bibitem{}Shaw M.A., Combes F., Axon D.J., Wright G.S. 1993,  \aa {\bf 273}, 31
\bibitem{}Shaw M.A., Gilmore G.F. 1989,  \mnras {\bf 237}, 903
\bibitem{}Shlosman I., Frank J., Begelmann M.C. 1989,  {\it Nat} {\bf 338}, 45
\bibitem{}Shlosman I., Begelmann M.C., Frank J.,  1990,  {\it Nat} {\bf 345},
679
\bibitem{}Shlosman I. and Noguchi 1994,  \apj {\bf 414}, 474
\bibitem{}Shridar S. 1989,  \mnras {\bf 238}, 1159
\bibitem{}Silk J. 1992,  {\it J. of Austr. Phys.} {\bf 45}, 437
\bibitem{}Simien F. and de Vaucouleurs G. 1986,  \apj {\bf 302}, 564
\bibitem{}Solomon P.M and Sage L.J.  1988,  \apj {\bf 334}, 613
\bibitem{}Sparke L., Sellwood J.A. 1987,  \mnras {\bf 225}, 653
\bibitem{}Str\"{o}mgren B. 1987,  in {\it The  Galaxy},
eds.  G. Gilmore and R. Carswell eds., Reidel Dordrecht, p.229
\bibitem{}Tagger M., Sygnet J.F. and Pellat R. 1994,  in {\it N-body problems
and gravitational dynamics}, F. Combes, L. Athanassoula eds. IAP Paris, p.55
\bibitem{}Tinney C., Scoville N., Sanders D., Soifer B. 1990,  \apj {\bf 362},
473
\bibitem{}Tinsley B. 1981,  \mn {\bf 194}, 63
\bibitem{}Toomre A. 1964,  \apj {\bf 139}, 1217
\bibitem{}Toomre A. 1974,   in  {\it Highlights of Astronomy},  I.A.U. vol. 3,
Reidel ed. Dordrecht p.457
\bibitem{}Toomre A. 1981,  in {\it The structure and evolution of normal
galaxies},
eds. S.M. Fall and Lynden-Bell, p. 111
\bibitem{}Tully  R.B. and Fisher J.R. 1977,  \aa {\bf 54}, 661
\bibitem{}Toomre A. and Toomre J. 1972,  \apj {\bf 178}, 623
\bibitem{}Udry S. 1991,  \aa {\bf 245}, 99
\bibitem{}Udry S. 1992,  \aa {\bf 268}, 35
\bibitem{}Udry S. and Martinet L. 1994,  \aa {\bf 281}, 314
\bibitem{}Vandervoort P. 1970,  \apj {\bf 161}, 87
\bibitem{}Wang B. and Silk J. 1994,  preprint
\bibitem{}Weinberg 1985,  \mn {\bf 213}, 451
\bibitem{}Weinberg 1989,  \mnras {\bf 239}, 549
\bibitem{}Whitmore B.C., Mc Elroy D.B. and Tonry J.L. 1985,  \apjs {\bf 59}, 1
\bibitem{}Wielen R. 1990,  In  {\it Dynamics and interactions of galaxies},
eds. Berlin Springer-Verlag
\bibitem{}Wielen R. 1977,  \aa {\bf 60}, 263
\bibitem{}Wyse R. and Silk J. 1989 \apj {\bf 339}, 700
\bibitem{}Young J.S., Xie S. Kenney J. Rice W. 1989  \apjs {\bf 70}, 699
\bibitem{}Zaritsky D. 1993,  \pasp {\bf 105}, 1006
\bibitem{}Zaritsky D. Kennicutt R.C., Huchra J. 1994,  \apj {\bf 420}, 87
\bibitem{}de Zeeuw T. 1988,  in {\it Integrability in dynamical systems}, J.R.
Buchler,
 J.R. Ipser, C.D. Williams eds. Ann. of N.Y. Acad. of Sciences, {\bf 536}, 1
\end{thebibliography}
\end{document}